\documentclass[aps,prl,showpacs,twoside,twocolumn,10pt,nofootinbib]{revtex4-2}
\usepackage[colorlinks=true, citecolor=blue, urlcolor=blue ]{hyperref}
\usepackage{epsfig,newlfont,amssymb,amsfonts,amsmath,bm,subfigure,palatino,mathtools,amsthm,braket, soul,enumitem,xcolor, graphics,graphicx,times,physics}
\usepackage{txfonts}
\usepackage[dvipsnames]{xcolor}
\usepackage{algorithm}
\usepackage{algpseudocode}
\usepackage{tikz}
\usetikzlibrary{quantikz}
\usepackage[normalem]{ulem}

\newtheorem{remark}{Remark}

\newtheorem{result}{Result}

\begin{document}
\let\ORIGaddcontentsline\addcontentsline
\renewcommand{\addcontentsline}[3]{}
\setstcolor{red}

\title{Quantum resources and the onset of the quantum Monte Carlo sign problem}
\date{\today}

\author{Naga Dileep Varikuti}
\email{dileep.varikuti@unitn.it}
\affiliation{Pitaevskii BEC Center, CNR-INO and Dipartimento di Fisica, Universit\`a di Trento, Via Sommarive 14, Trento, I-38123, Italy}
\affiliation{INFN-TIFPA, Trento Institute for Fundamental Physics and Applications, Via Sommarive 14, Trento, I-38123, Italy}
\author{Leela Ganesh Chandra Lakkaraju}
\email{ganesh.lakkaraju@unitn.it}
\affiliation{Pitaevskii BEC Center, CNR-INO and Dipartimento di Fisica, Universit\`a di Trento, Via Sommarive 14, Trento, I-38123, Italy}
\affiliation{INFN-TIFPA, Trento Institute for Fundamental Physics and Applications, Via Sommarive 14, Trento, I-38123, Italy}
\author{Philipp Hauke}
\email{philipp.hauke@unitn.it}
\affiliation{Pitaevskii BEC Center, CNR-INO and Dipartimento di Fisica, Universit\`a di Trento, Via Sommarive 14, Trento, I-38123, Italy}
\affiliation{INFN-TIFPA, Trento Institute for Fundamental Physics and Applications, Via Sommarive 14, Trento, I-38123, Italy}

\begin{abstract}

The Monte Carlo sign problem poses a fundamental obstacle to classical simulation of quantum many-body systems, yet its connection to quantum resources remains poorly understood. We probe this connection by systematically tuning the degree of nonstoquasticity in random $k$-local Hamiltonians.
In numerical tensor-network studies, we find the sign complexity and the quantum resources of entanglement and nonstabilizerness of the ground state to be strongly suppressed in a broad regime up to substantial nonstoquasticity. 
As nonstoquasticity is further increased, the considered quantum resources experience sharp crossovers into a regime where they are strongly enhanced.
Surprisingly, numerical quantum Monte Carlo (QMC) simulations drastically deteriorate in the same region, even though the Hamiltonian is already considerably nonstoquastic well before it, indicating that the failure of ground-state QMC is correlated more with the resource content of the state than the sign structure of the Hamiltonian.
We corroborate these findings analytically by constructing an approximate ground state of $k$-local Hamiltonians in the stoquastic limit.
Moreover, we identify a distinct behaviour in graph states. Their nonstabilizerness always vanishes, but, as we demonstrate analytically, on bipartite graphs the sign can be easily cured while in multipartite graph states it cannot.
Our results thus shed light on the highly subtle interplay between quantum resources and the computational difficulty encountered by QMC methods.

\end{abstract}
\maketitle

\textit{Introduction.---}The boundary between classically tractable and intractable quantum systems is often characterized by the efficiency of Quantum Monte Carlo (QMC) methods~\cite{Huggins_2022, King_2021, PhysRevLett.94.170201}, which have been extensively employed to study many-body systems at thermal equilibrium~\cite{Hong2019, PhysRevX.5.041041, PhysRevB.77.155102, PhysRevB.88.235115, Maik_2012, aarts2026latticefieldtheoriessign, AARTS2023104070, 10.1063/1.3518900, PhysRevB.59.R14157, PhysRevB.56.11678, annurevconmatphys031218013339}. Hamiltonians with non-positive off-diagonal elements, known as \textit{stoquastic} Hamiltonians~\cite{stoq08072689X, Bravyi2015, Bravyi2008}, admit a stochastic matrix interpretation of the imaginary-time evolution, in which every configuration carries a non-negative weight. Observables can then be estimated without the cancellations between positive
and negative contributions that, in the absence of stoquasticity, incur an exponential overhead. This renders such models efficiently simulable in many cases of interest~\cite{Bravyi2015, stoq08072689X, PhysRevLett.119.100503, Hong2019, annurevconmatphys031218013339}, though not always~\cite{FBarahona_1982}.
In addition, the Perron--Frobenius theorem ensures that their ground states can be chosen with non-negative amplitudes in the computational basis~\cite{Horn_Johnson_1985}, such that the wave function itself admits an interpretation as a probability distribution.
Generic Hamiltonians, however, are nonstoquastic and therefore suffer from the infamous sign problem, whereby destructive interference between configurations can lead to exponentially large sampling variances and severely limit the efficiency of QMC simulations~\cite{PhysRevLett.94.170201}.
Despite its importance, the extent to which the severity of the sign problem affects, or is affected by, the underlying quantum properties has largely remained an open question.

A suitable handle on the quantumness of a given system is given by quantum resources such as entanglement~\cite{Amico2008, entanglement_manybody_review_lewenstein, LAFLORENCIE20161, entanglement_manybody_review_lewenstein_DeChiara} and nonstabilizerness~\cite{Ellison2021symmetryprotected, leone2022stabilizer, Winter2022,leone2024stabilizer}, which have been identified in recent years as metrics for the complexity of quantum many-body systems~\cite{PhysRevLett.91.147902, leone2022stabilizer, ORUS2014117, Winter2022, hauke2016measuring, santra2025complexitytransitionschaoticquantum, Magni2025quantumcomplexity, tarabunga2024nonstabilizerness, santra2025quantumresourcesnonabelianlattice, tarabunga2023many, PhysRevB.106.195158, 91hd-8k7g, lakkaraju2026fermionic, piroli_longrange_magic, 3yx4-1j27, capecci2025role, robin2026quantumcomplexitynewdirections,Viscardi_2026, odavic2024stabilizer, haug2023quantifying, Varikuti2026impactofclifford, catalano2024magic,aditya2026nonlocalmagicspreadingmanybody, Tarabunga2024criticalbehaviorsof}.
Uniting them with a view on QMC is particularly important in the quest towards a practical quantum advantage:
both entanglement~\cite{Jozsa2003Aug,PhysRevLett.86.5188,ent_grover} and nonstabilizerness~\cite{Kitaev2005} serve as resources for universal quantum computation.
Regimes where both are large
and which at the same time show a severe sign problem become prime targets for quantum simulators~\cite{hauke2012can, halimeh2025cold, doi:10.1126/science.adt1732, daley2022practical}, since there several workhorses (tensor networks, Clifford calculations, QMC methods) for numerical calculations on strongly-correlated quantum many-body systems fail simultaneously.
Seminal works have shown that random sign-free wave-functions have reduced entanglement~\cite{PhysRevA.92.042308}, and that in a class of model wave-functions the entanglement content correlates with the complexity of the nodal surface~\cite{PhysRevB.95.155102}.
A recent study has also examined how tensor-network contraction complexity depends on sign complexity \cite{schuch_tensor_sign}.
Nonstoquastic catalysts moreover offer a promising route to quantum speedup in quantum annealing~\cite{Hauke_2020, PhysRevB.95.184416, nishimori2017exponential, PhysRevA.99.042334}, and recent work found they can significantly enhance the quantum resources encountered in an annealing sweep~\cite{capecci2026quantum}.
Yet, systematic studies are lacking on how quantum resources relate to the nonstoquasticity of a many-body model Hamiltonian.

In this Letter, we address (i) whether the emergence of the sign problem is accompanied by the appearance of quantum resources given by entanglement~\cite{Amico2008}, nonstabilizerness~\cite{leone2022stabilizer}, and the inverse participation ratio~\cite{RevModPhys.80.1355}, which in pure states, is related to coherence~\cite{PhysRevA.94.052324, RevModPhys.89.041003},
and (ii) whether the simulation accuracy of QMC methods bears signatures of these resources.
As a family of benchmark models, we take random $k$-local ($k\geq 2$) Hamiltonians. By fixing the signs of randomly chosen entries of the Hamiltonian matrix, we control the nonstoquasticity of the model through a tunable parameter $p_\mathrm{ns}$.
Through numerical analysis on chains up to lengths $N=50$, we find a broad regime up to rather large nonstoquasticity, where the ground-state wave function remains nearly positive. In this region, we find that all considered quantum resources are strongly suppressed.
We shed light on this regime by analytically constructing the ground state in the fully stoquastic limit.
At a $k$-dependent (but nearly size-independent) crossover scale, the signs of the numerically constructed ground-state wave function begin to fluctuate rapidly. In the same region, the model displays steep increases in the quantum resources. To assess the influence of this behavior on QMC methods, we benchmark the error of projector QMC (PQMC)~\cite{Becca_Sorella_2017} at a fixed number of walkers and time steps against results obtained via the density matrix renormalization group (DMRG)~\cite{PhysRevLett.69.2863}. Remarkably, the sampling error remains low through the initial regime, raising steeply only near the crossover, in tandem with the enhancement of quantum resources.

Our findings indicate that the generation of quantum resources in generic nonstoquastic models is closely tied to the efficiency of QMC methods.
However, the relationship is not fully straightforward; e.g., the ground states of toric codes exhibit topological order and long-range entanglement, yet they have vanishing nonstabilizerness and admit simple stoquastic parent Hamiltonians \cite{PhysRevA.71.022315, HAMMA200522}. We extend this known result by analytically analyzing the sign complexity of graph states, which have vanishing nonstabilizerness. We show that states on bipartite graphs are sign-free, or can be made so by local Hadamards, whereas on general graphs a nonzero sign complexity survives.
Thus, our findings suggest that generically ground states of nonstoquastic Hamiltonians show strong correlations between different quantum resources and the wave-function sign, but there are physically relevant classes of Hamiltonians for which this does not hold.

\emph{Benchmark model.---}In this letter, we consider an ensemble of random $k$-local Hamiltonians of the form~\footnote[1]{Note that the sign problem is basis-dependent: a Hamiltonian has an \textit{intrinsic} sign problem if no basis transformation removes it. We work in the computational basis, since deciding whether such a transformation exists is NP-hard for broad classes of local Hamiltonians~\cite{doi:10.1126/sciadv.abb8341, marvian2019computational, klassen2020hardness}, and we expect no easy solution for the generic Hamiltonians considered here.}
\begin{eqnarray}\label{ham-main}
H = \sum_{j=0}^{N-k} h_j,\;\;\mathrm{and}\; h_j\in \mathrm{GOE}(2^k) \;\forall\; 0\leq j\leq N-k ,   \end{eqnarray}
where each $h_j$ has support on $k$ nearest-neighbor lattice sites, and $\mathrm{GOE}(2^k)$ denotes the Gaussian orthogonal ensemble with dimension $2^k$. We consider open boundary conditions and one spatial dimension to enable numerics with the DMRG~\cite{PhysRevLett.69.2863}. In the above family of Hamiltonians, we introduce a tunable level of \textit{stoquasticity}~\footnote[2]{In this work, we adopt the convention that a $k$-local Hamiltonian is stoquastic if it is term-wise stoquastic, i.e., if each of its $k$-local terms is separately stoquastic, with non-positive off-diagonal matrix elements in the computational basis.
} by randomly assigning negative signs to the off-diagonal elements of each $h_j$:
given matrix element $[h_j]_{l,m}$ for some $l\neq m$, we replace it (and $[h_j]_{m,l}$) with $-\lvert [h_j]_{l,m} \rvert$, with probability $p\in [0,1]$.
The diagonal elements are left unchanged.
For $p=0$, we retain the GOE ensemble, while for $p=1$ the Hamiltonian becomes fully stoquastic. Consequently, we define $p_\mathrm{ns}=1-p$ as the degree of nonstoquasticity.

\textit{Stoquastic ground state.---}When the Hamiltonian in Eq.~(\ref{ham-main}) is fully stoquastic, i.e., $p_{\mathrm{ns}}=0$, its ground state can be approximated
by the uniform superposition of all computational basis states (see the Supplemental Material (SM)~\cite{supplemental}),
\begin{eqnarray}\label{uniform}
|\mathrm{gs}_{p_{\mathrm{ns}}=0}\rangle\approx |\bm{+}\rangle
=\dfrac{1}{\sqrt{2^N}}\sum_{j\in \{0, 1\}^{\times N}}|j\rangle.
\end{eqnarray}
We validate this approximation for a typical $k$-local stoquastic Hamiltonian $H$ by examining the fluctuations of the amplitudes $\langle i|H|\bm{+}\rangle$ for some fixed $i\in\{0,1\}^{\times N}$ over the ensemble of stoquastic Hamiltonians. We compare these against the energy gap of $\mathbb{E}[H]$, denoted $\Delta_{\mathbb{E}[H]}$, since overlap elements that are comparable with $\Delta_{\mathbb{E}[H]}$ can admix other states into $|\bm{+}\rangle$. 
In the limit $2^k\gg 1$, keeping $N$ fixed, we find that the relative fluctuations decay exponentially with $k$ as (see SM~\cite{supplemental})
\begin{eqnarray}\label{RF}
\mathrm{RF}=\dfrac{\sqrt{2^N\mathrm{Var}_{H}(\langle i|H|\bm{+}\rangle)}}
{ \Delta_{\mathbb{E}[H]}}
\approx
\sqrt{\left(\dfrac{M}{S^2}-1\right)\;
\left(\dfrac{N-k+1}{2^k}\right)},
\end{eqnarray}
where $S=\mathbb{E}([h_j]_{i, i'})$ and $M=\mathbb{E}([h_j]^2_{i,i'})$ for any $0\leq j< N$ and $0\leq i\neq i'<2^k$. Thus, $\mathrm{RF}\sim 2^{-k/2}$ at a fixed $N$, indicating that the action of a typical $H$ on $|\mathbb{+}\rangle$ becomes increasingly homogeneous across the computational basis. Consequently, the true ground state becomes progressively closer to $|\bm{+}\rangle$ as $k\rightarrow N$. In contrast, for fixed $k$, RF grows as $\sim \sqrt{N}$, indicating that $\ket{\bm{+}}$ gradually becomes a poor approximation.

\begin{figure}
\includegraphics[scale=0.325]{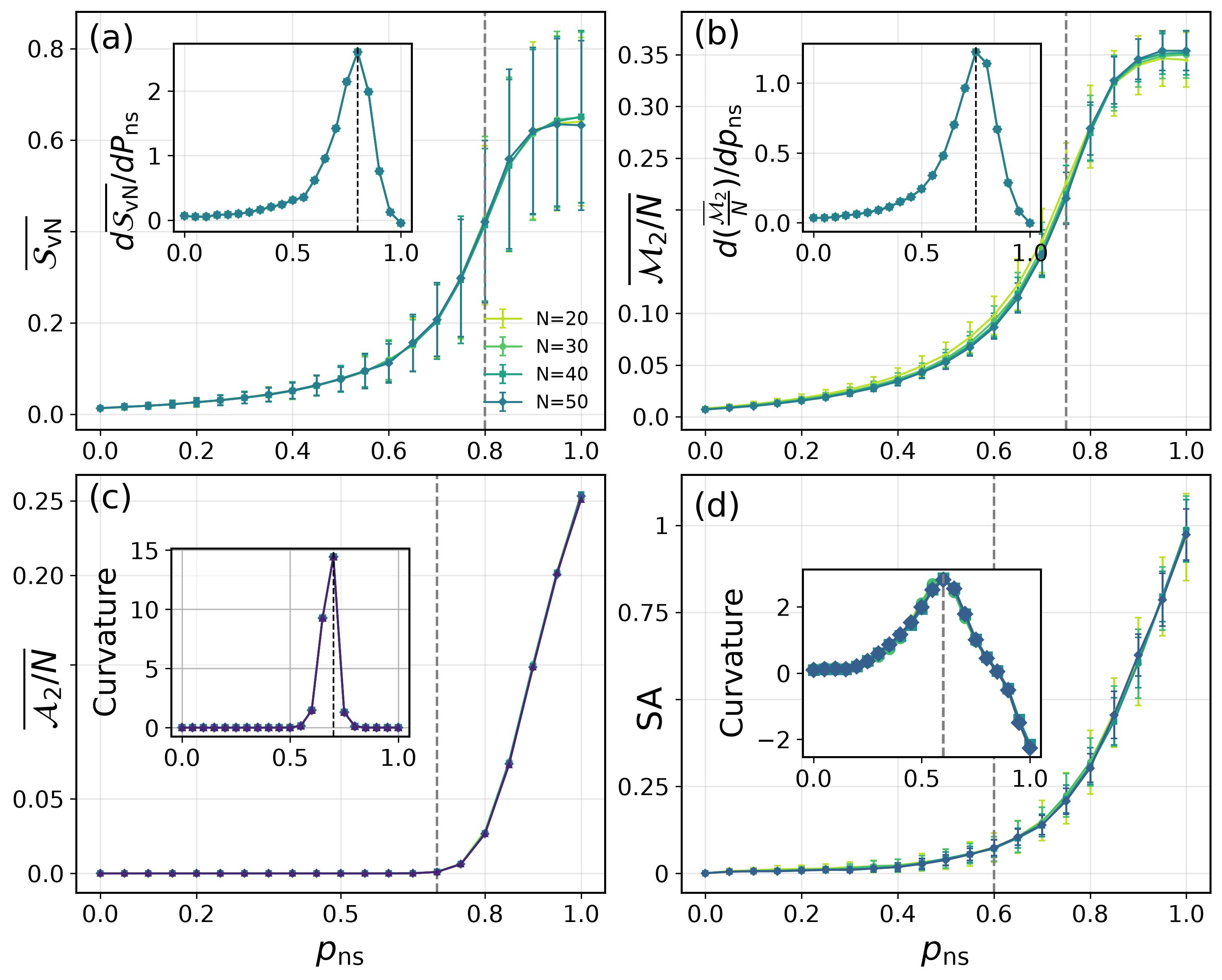}
\caption{ Quantum resources and indicators of QMC hardness of the ground state of random Hamiltonians as a function of nonstoquasticity, quantified via (a) bipartite entanglement (von Neumann entropy, $\mathcal{S}_{\mathrm{vN}}$), (b) nonstabilizerness (stabilizer Rényi entropy, $\mathcal{M}_{2}/N$), (c) the sign complexity, $\mathcal{A}_{2}/N$, and (d) QMC simulation accuracy (SA; relative deviation of PQMC ground-state energy at fixed walker number from the quasi-exact DMRG result). Data for $k=4$ and different system sizes. The insets show the gradients [(a), (b)] and the curvatures [(c), (d)] of the respective quantities. Their peaks mark crossover points, denoted as dashed vertical lines, between an almost stoquastic (low entanglement, nonstabilizerness, and sign complexity of the ground state) to a strongly nonstoquastic regime.  This crossover is further reflected in the SA: an ``easy'' regime, in which the error remains small, gives way to a ``hard'' regime in which it grows rapidly. Error bars indicate statistical standard deviations over random Hamiltonian instances.}
\label{fig:coll}
\end{figure}

Moreover, as $p_{\mathrm{ns}}$ is tuned away from $0$, the ground state acquires a sign structure. The change is initially very gradual, within an extended ``near-stoquastic'' regime, and at larger $p_{\mathrm{ns}}$ displays a sharp crossover to a ``fully nonstoquastic'' (``GOE-like'') regime. This behavior becomes apparent by tracking the ground state's sign complexity in the computational basis as well as quantum complexity resources (see Fig.~\ref{fig:coll}). What is more, this crossover marks the onset of a steep deterioration in the accuracy of ground-state sampling by basic projector QMC when the number of walkers is fixed in the numerical simulations. In what follows, we illustrate this behavior in detail by numerically probing the random instances of $H$ as a function of $p_{\mathrm{ns}}$, obtaining the ground state variationally via DMRG using a matrix product state (MPS) ansatz~\cite{PhysRevLett.69.2863, schollwock2011density}. In the main text, we illustrate our main findings for $k=4$, and other values of $k$ are discussed in the SM~\cite{supplemental}.

\textit{Bipartite entanglement.---}The sign structure of wave functions strongly influences their entanglement content: in random positive wave functions, R\'enyi entropies ($n>1$) are suppressed to an $O(1)$ constant, while the von Neumann entropy retains a volume law with a reduced prefactor~\cite{PhysRevA.92.042308}. Note however that for positive wave functions arising as ground states of general stoquastic Hamiltonians, positivity alone does not enforce low entanglement or a generalized area law~\cite{10.1063/1.4936216}.

\begin{figure}
\includegraphics[scale=0.375]{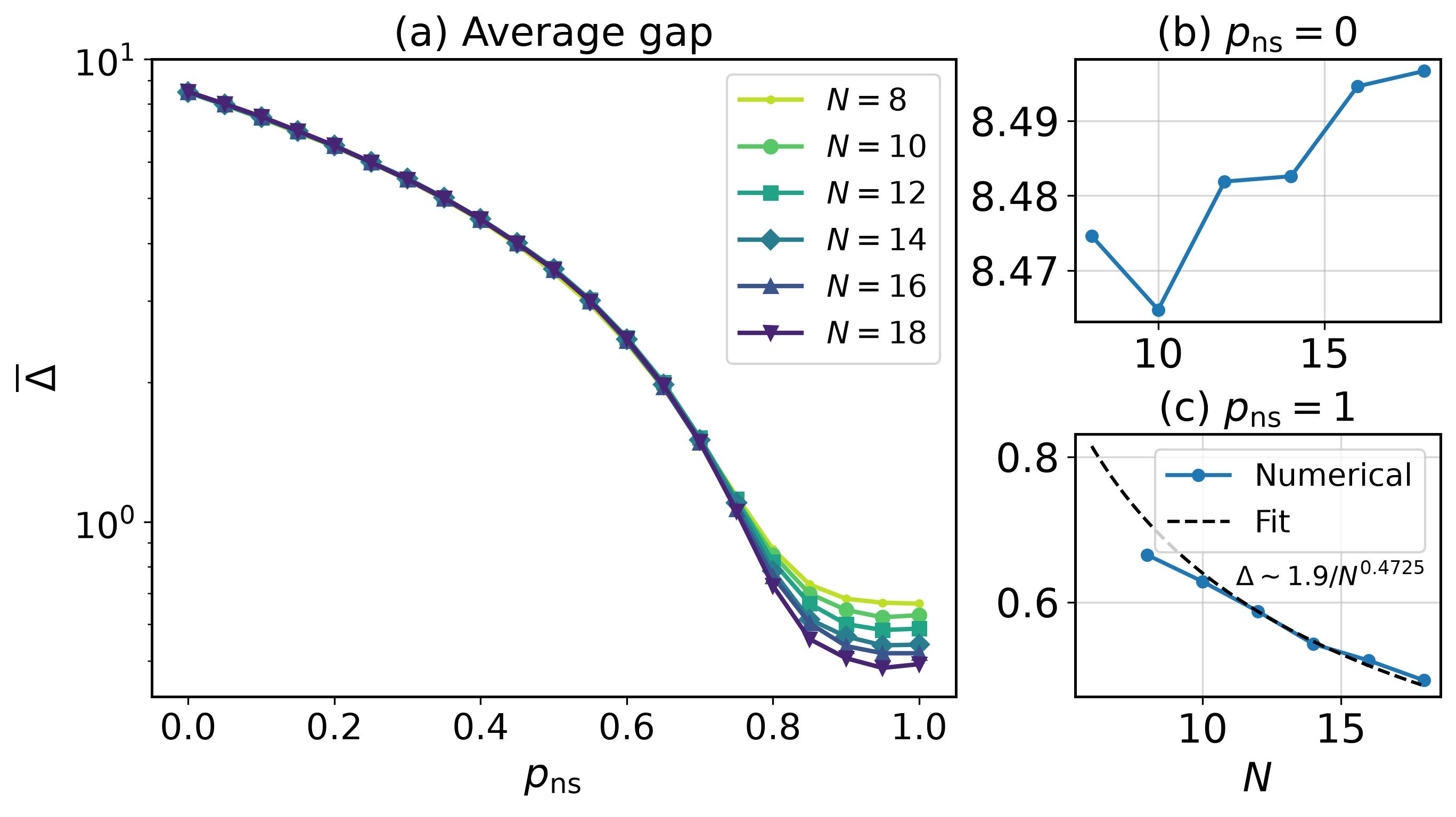}
\caption{(a) Gap between the ground and the first excited states as $p_{\mathrm{ns}}$ is varied, computed for different system sizes, averaged over several instances of the Hamiltonian with $k=4$.
(b) In the stoquastic limit, the gap stays nearly independent of system size, while (c) in the nonstoquastic limit it can be well fitted by a power-law in $N$.
}
\label{fig:GAP_main}
\end{figure}

Figure~\ref{fig:coll}(a) shows the numerical results for the half-chain von Neumann entropy, $\overline{\mathcal{S}_{\mathrm{vN}}}$, for various system sizes, where the overline indicates the average over a thousand independent random samples, and the error bars denote the corresponding statistical standard deviation.
We observe that the stoquastic limit ($p_{\mathrm{ns}}=0$) consistently displays lower entanglement than the nonstoquastic one ($p_{\mathrm{ns}}=1$).
Since the Hamiltonian is local in one dimension, its ground state is typically expected to obey an area law when gapped, with logarithmic corrections possible for gapless systems in the thermodynamic limit~\cite{Hastings_2007, Gottesman_2010, RevModPhys.82.277}.
Analytically, in the stoquastic limit, the gap of the averaged Hamiltonian $\mathbb{E}[H]$ scales like $\Delta_{\mathbb{E}[H]}=|S|2^{k}$, independent of $N$, so typical instances of $H$ are expected to be gapped; we confirm this numerically for computationally tractable system sizes in Fig.~\ref{fig:GAP_main} below and in the SM~\cite{supplemental}.

The corresponding area law scaling is clearly visible at small $p_{\mathrm{ns}}$, where the entanglement remains close to zero for all system sizes considered, and the curves coincide within their small error bars. Near $p_{\mathrm{ns}}=1$, by contrast, the entanglement shows finite-sampling fluctuations, though the data for different system sizes still agree within the error bars.
Further, near the stoquastic limit $\overline{\mathcal{S}_{\mathrm{vN}}}$ becomes increasingly suppressed as $k$ increases, while there is an opposite trend near $p_{\mathrm{ns}}=1$ (see SM \cite{supplemental}). In the limit $k=N$, the ground-state entanglement nearly vanishes in the stoquastic limit. This is in line with what one would expect for typical stoquastic Hamiltonians following Eqs.~(\ref{uniform}) and ~(\ref{RF}).

\begin{figure}
\includegraphics[scale=0.475]{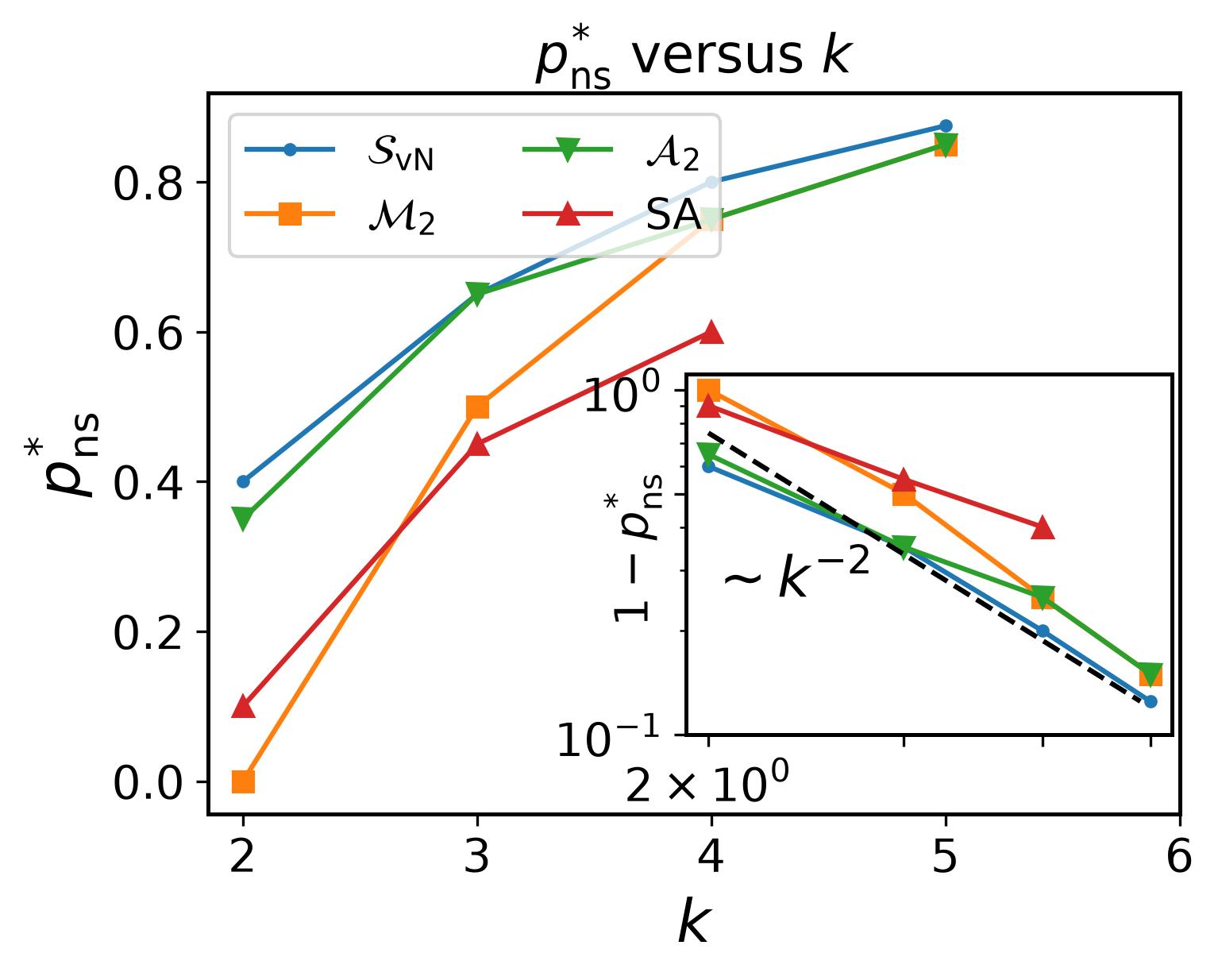}
\caption{
Crossover scale ${p}^\star_{\mathrm{ns}}$ of entanglement entropy and nonstabilizerness (peak of gradient) and of average sign and sampling accuracy (peak of curvature),
as a function of $k$. With increasing $k$, the peaks shift to higher nonstoquasticity. The numerical simulations indicate a scaling close to $1-p^{\star}_{\mathrm{ns}} \sim k^{-2}$ for the gradients. }
\label{fig:pstar_main}
\end{figure}

This qualitative separation of regimes can be sharpened by studying the gradient of $\overline{\mathcal{S}_{\mathrm{vN}}}$ with respect to $p_{\mathrm{ns}}$ [inset of Fig.~\ref{fig:coll}(a)].
For $k=4$, the gradient shows a clear peak at $p^{\star}_{\mathrm{ns}}\approx 0.8$.
As $k$ is increased, the peak shifts to larger $p_{\mathrm{ns}}$ (Fig.~\ref{fig:pstar_main} and SM~\cite{supplemental}).
We now show that such a crossover point exists consistently for different figures of merit.

\textit{Nonstabilizerness.---}The nonstabilizerness, or ``quantum magic'', of a state quantifies its distance from the set of stabilizer states, and has been identified as a key ingredient for universal quantum computation and practical quantum advantage~\cite{Kitaev2005}. As a concrete measure, we use the second stabilizer R\'enyi entropy~\cite{leone2022stabilizer, Bittel2026operational, PhysRevA.106.042426}, defined as $\mathcal{M}_2(|\psi\rangle) = -\log\left(\sum_{i=1}^{2^{2N}}\langle\psi |P_i|\psi\rangle^4/2^{N} \right)$,
where the sum runs over the $N$-qubit Pauli strings $P_i$.
The numerical results, normalized with $N$, are shown in Fig.~\ref{fig:coll}(b). We use perfect sampling of the Pauli strings to compute the non-stabilizerness of the MPS ground state~\cite{PhysRevLett.131.180401}.
Interestingly, the data corresponding to all $N$ neatly coincide, indicating that $\overline{\mathcal{M}}_{2}\sim N$ throughout. As $p_{\mathrm{ns}}$ is varied from $0$ to $1$, $\overline{\mathcal{M}}_{2}/N$ increases and displays a similar profile to the entanglement entropy. In addition, near the stoquastic limit, the nonstabilizerness gets increasingly suppressed with $k$ (see SM~\cite{supplemental}).
One can understand this behavior from the fact that the approximate ground state at low $p_\mathrm{ns}$, $|\bm{+}\rangle$, is a stabilizer state. As $p_{\mathrm{ns}}$ increases, the stabilizer entropy grows monotonically within the error bars and saturates near $p_{\mathrm{ns}}=1$, where the ordering with $k$ is reversed compared to the stoquastic limit: the
ground state now carries more nonstabilizerness at larger $k$~\cite{supplemental}. For $k=N$, the Hamiltonian becomes a fully GOE random matrix for which the eigenstates are uniformly distributed over $\mathbb{R}^{2^N}$ \cite{haake1991quantum}, and the nonstabilizerness attains the Haar value corresponding to real states.

Moreover, in good agreement with the entanglement entropy, the gradient of the nonstabilizerness [Fig.~\ref{fig:coll}(b)] peaks near $p_{\mathrm{ns}}\approx0.75$, placing the crossover in the same region. As our numerical results indicate, the gradients for nonstabilizerness across different $k$ consistently peak at a slightly smaller $p_{\mathrm{ns}}$ than that of the entanglement entropy (Fig.~\ref{fig:pstar_main} and SM~\cite{supplemental}).

\textit{Sign complexity.---}Stoquastic Hamiltonians admit for ground states with strictly non-negative amplitudes.
However, once positive off-diagonal elements are introduced, the ground-state wave function can acquire a non-trivial sign structure.
These can render fixed-node Monte Carlo approaches challenging~\cite{PhysRevLett.72.2442, PhysRevB.51.13039}, and, as we will see, correlate negatively also with the performance of imaginary-time PQMC.
To quantify the sign complexity, we adapt the measure from Ref.~\cite{PhysRevResearch.2.032060} to wave functions with real amplitudes in the computational basis,
\begin{eqnarray}
\mathcal{A}(|\psi\rangle) = 1 - \big| \sum_{i=0}^{2^N-1} |\langle i|\psi\rangle|^2 \, \mathrm{sign}(\langle i|\psi\rangle) \big|.
\end{eqnarray}
When a state is sign free, $\mathcal{A}(|\psi\rangle)=0$, whereas $\mathcal{A}(|\psi\rangle)=1$ when positive and negative amplitudes carry equal weights. Haar-random real states asymptotically approach this limit as $\mathcal{A}\approx 1-\sqrt{6/(\pi(2^N+2))}$ (see SM~\cite{supplemental}).

Since $\mathcal{A}(|\psi\rangle)$ ranges from $0$ to $1$, one may also define the associated second R\'enyi entropy $\mathcal{A}_{2}(|\psi\rangle)=-\log\left[1-\mathcal{A}(|\psi\rangle)\right]$. The corresponding results are presented in Fig.~\ref{fig:coll}(c). The data for different system sizes collapse onto a single curve once normalized by $N$, so that $\mathcal{A}_{2}\sim N$, implying that $1-\mathcal{A}$ decays exponentially with system size at any fixed $p_{\mathrm{ns}}$. Moreover, $\mathcal{A}_{2}$ crosses over from a regime of low sign complexity at small $p_{\mathrm{ns}}$ to one of high sign complexity at large $p_{\mathrm{ns}}$. Surprisingly, the ground state retains an almost vanishing sign complexity up to $p_{\mathrm{ns}}\lesssim0.75$, despite the parent Hamiltonian already containing a large fraction of positive off-diagonal elements, and only then does $\mathcal{A}_{2}$ increase sharply. The curvature of $\mathcal{A}_{2}$ peaks near $p_{\mathrm{ns}}^\star\approx0.7$, placing the crossover in the same region as those of entanglement and nonstabilizerness. The average sign therefore correlates closely with the quantum resources studied above.

\textit{Simulation accuracy of PQMC.---}The resource quantifiers examined above on the ground state serve as proxies for the cost of simulating the state classically, whether by tensor-network, stabilizer-based, or Monte Carlo methods.
In PQMC and related QMC approaches, which rely on imaginary-time evolution, the sign problem enters through negative matrix elements in the propagator, caused by a nonstoquastic Hamiltonian rather than directly through the wave function.
Here, we demonstrate that the PQMC simulation accuracy (SA) nonetheless closely mirrors the behavior of the above ground-state indicators. For the core PQMC algorithm we use and the energy estimator for generic Hamiltonians, see SM~\cite{supplemental}.

We define the SA as the relative deviation of the PQMC ground-state energy estimation at a fixed number of walkers from the quasi-exact result derived from DMRG~\footnote[3]{For the considered values of $k$, DMRG converges for bond dimensions below $\chi=200$ across the entire range of $p_{\mathrm{ns}}$ and the system sizes considered.},
\begin{eqnarray}
\mathrm{SA}=  \dfrac{\left| E_{\mathrm{PQMC}}-E_{\mathrm{DMRG}} \right|}{|E_{\mathrm{DMRG}}|}.
\end{eqnarray}
Near $p_{\mathrm{ns}}=0$, the SA ideally vanishes.
Near $p_{\mathrm{ns}}=1$, instead, one can expect it to saturate at one, since $E_{\mathrm{PQMC}}$ is reduced by the cancellation between positive and negative walkers.

The SA for $k=4$ is shown in Fig.~\ref{fig:coll}(d) as a function of $p_{\mathrm{ns}}$, evaluated with the same sampling parameters for all $N$: $200$ walkers, $500$ time steps of size $dt=0.01$, and the uniform superposition $|\bm{+}\rangle$ as the trial wave function. The markers and error bars denote the mean and standard deviation, respectively, over several random instances of the Hamiltonian.
The data roughly coincide for all $N$. Perhaps surprisingly, the SA stays nearly constant ($\sim 10^{-3}$) up to about $p_\mathrm{ns}\approx 0.5$ despite the Hamiltonian being already considerably nonstoquastic. Interestingly, within this regime the other complexity measures are also strongly suppressed [see Fig.~\ref{fig:coll}(a)--(c)].
As $p_{\mathrm{ns}}$ is further increased, the SA starts to grow uncontrollably. To roughly locate this crossover, we use the curvature of the mean SA, smoothed with a Gaussian filter of width $\sigma=0.1$. We find that the curvature peaks near $p_{\mathrm{ns}}\approx0.6$ (see inset), a little below the peaks of entanglement, nonstabilizerness, and ground-state sign. The precise crossover value may depend on the chosen numbers of walkers and time steps, but we observe the qualitative behavior to be robust~\cite{supplemental}. Moreover, the curvature peak shifts towards larger values of $p_{\mathrm{ns}}$ with increasing $k$, closely mirroring the behavior of quantum resources and sign complexity in the ground state~(see Fig.~\ref{fig:pstar_main} and SM~\cite{supplemental}). The SA therefore appears to correlate more with the resource content of the ground state than with the sign structure of the Hamiltonian itself.

\textit{Gap analysis.---}Here, we numerically probe the gap between the ground and first excited states, averaged over several instances of the random Hamiltonian, as $p_{\mathrm{ns}}$ is varied. In the stoquastic limit, the averaged Hamiltonian $\mathbb{E}[H]$ is a sum of commuting projectors, and its gap follows $\Delta_{\mathbb{E}[H]}=|S|2^{k}$, independent of $N$~\cite{supplemental}. Since $H=\mathbb{E}(H)+\delta H$, with $\mathbb{E}(\delta H)=0$, we expect that the gap averaged over several instances of $H$ follows a similar scaling. We confirm this prediction for $k=4$ across several system sizes (for other $k$ values, see SM~\cite{supplemental}). The results are presented in Fig.~\ref{fig:GAP_main}.
Importantly, the gap appears to remain of $O(1)$, almost independent of $N$ up to $p_{\mathrm{ns}}\approx0.75$, around the crossover regions identified above.
This behavior is illustrated in Fig.~\ref{fig:GAP_main}(b) for the fully stoquastic limit.
This large, $N$-independent gap ensures an area law, leading to a size-independent entanglement, consistent with the data shown in Fig.~\ref{fig:coll}(a).
For larger $p_{\mathrm{ns}}$, in contrast, the gap develops a size dependence. Figure~\ref{fig:GAP_main}(c) shows the scaling in the fully nonstoquastic regime, where the gap closing at the sizes considered is well approximated by a power law $\overline{\Delta}\sim N^{-0.4725}$.
Consistently, in this regime the observables discussed above are significantly enhanced.

\textit{Sign problem in graph states.---}While our results so far suggest a remarkable correlation between the QMC sign problem and quantum resources in ground states of random nonstoquastic Hamiltonians, this correlation need not be ubiquitous. In fact, there exist stabilizer Hamiltonians, such as the toric code, whose ground states are long-range entangled yet sign-free \cite{PhysRevA.71.022315, HAMMA200522}. In the end matter, we extend this observation to graph states, the central objects in measurement-based quantum computing \cite{briegel2009measurement}. In particular, we prove:
(a) All stabilizer states that---as the toric-code ground state---admit for a parent Hamiltonian defined on a bipartite graph are sign-free, when allowing local Hadamard operations.
(b) In contrast, stabilizer states with a parent Hamiltonian defined on a $3$- or higher-partite graph exhibit a sign problem that cannot be cured by local Hadamard operations. Corresponding details are presented in the SM~\cite{supplemental}.

As an immediate corollary, stoquastic Hamiltonians cannot have ground states that are local Hadamard equivalents of $t$-partite graph states with $t\geq 3$.
Graph states thus show a complex behavior, where entanglement may be large and nonstabilizerness always vanishes, while the sign problem is curable for $t\leq2$ and not Hadamard-curable for $t\geq 3$. This complexity highlights that the sign problem deserves a separate analysis from other quantum resources.

\textit{Discussion.} To summarize, we probed the emergence of quantum resources in the ground states of a family of random Hamiltonians and how it governs the breakdown of
ground-state quantum Monte Carlo simulations. The resources considered, entanglement and nonstabilizerness, cross over in the same region where $p_{\mathrm{ns}}$ is already sizable. The ground-state sign complexity and, as we show in the end matter, the inverse participation ratio behave in the same way. The statistical error of the Projector QMC simulations follows suit: at fixed simulation parameters, the SA for different system sizes grows slowly up to a sizable nonstoquasticity and then increases rapidly beyond a crossover scale $p_{\mathrm{ns}}^{\star}$. The crossover marks where the perturbation around $|\bm{+}\rangle$ ceases to be controlled by the gap, accounting both for the common location of the crossovers and for their shift to larger $p_{\mathrm{ns}}$ with increasing $k$. Overall, our results suggest that the PQMC method may remain robust against the Hamiltonian sign problem provided the ground state has not yet acquired the resources that may render it difficult to sample. Our analysis of graph states shows that this relation is nonetheless not one-to-one: graph states carry no nonstabilizerness yet retain a nonzero sign complexity on general graphs, further indicating that the sign structure of a wave function deserves attention in its own right.
An interesting future direction would be to test whether the results extend to more structured quantum many-body models.

\textit{Acknowledgments.---}This project has been funded by the Caritro Foundation.  This work was supported by the Swiss State Secretariat for Education, Research and Innovation (SERI) under contract number UeMO19-5.1, the Provincia Autonoma di Trento, and Q@TN, the joint lab between the University of Trento, FBK—Fondazione Bruno Kessler, INFN—National Institute for Nuclear Physics, and CNR—National Research Council, Italy.

\bibliography{main.bib}

\begin{thebibliography}{108}%
\makeatletter
\providecommand \@ifxundefined [1]{%
 \@ifx{#1\undefined}
}%
\providecommand \@ifnum [1]{%
 \ifnum #1\expandafter \@firstoftwo
 \else \expandafter \@secondoftwo
 \fi
}%
\providecommand \@ifx [1]{%
 \ifx #1\expandafter \@firstoftwo
 \else \expandafter \@secondoftwo
 \fi
}%
\providecommand \natexlab [1]{#1}%
\providecommand \enquote  [1]{``#1''}%
\providecommand \bibnamefont  [1]{#1}%
\providecommand \bibfnamefont [1]{#1}%
\providecommand \citenamefont [1]{#1}%
\providecommand \href@noop [0]{\@secondoftwo}%
\providecommand \href [0]{\begingroup \@sanitize@url \@href}%
\providecommand \@href[1]{\@@startlink{#1}\@@href}%
\providecommand \@@href[1]{\endgroup#1\@@endlink}%
\providecommand \@sanitize@url [0]{\catcode `\\12\catcode `\$12\catcode `\&12\catcode `\#12\catcode `\^12\catcode `\_12\catcode `\%12\relax}%
\providecommand \@@startlink[1]{}%
\providecommand \@@endlink[0]{}%
\providecommand \url  [0]{\begingroup\@sanitize@url \@url }%
\providecommand \@url [1]{\endgroup\@href {#1}{\urlprefix }}%
\providecommand \urlprefix  [0]{URL }%
\providecommand \Eprint [0]{\href }%
\providecommand \doibase [0]{https://doi.org/}%
\providecommand \selectlanguage [0]{\@gobble}%
\providecommand \bibinfo  [0]{\@secondoftwo}%
\providecommand \bibfield  [0]{\@secondoftwo}%
\providecommand \translation [1]{[#1]}%
\providecommand \BibitemOpen [0]{}%
\providecommand \bibitemStop [0]{}%
\providecommand \bibitemNoStop [0]{.\EOS\space}%
\providecommand \EOS [0]{\spacefactor3000\relax}%
\providecommand \BibitemShut  [1]{\csname bibitem#1\endcsname}%
\let\auto@bib@innerbib\@empty
\bibitem [{\citenamefont {Huggins}\ \emph {et~al.}(2022)\citenamefont {Huggins}, \citenamefont {O’Gorman}, \citenamefont {Rubin}, \citenamefont {Reichman}, \citenamefont {Babbush},\ and\ \citenamefont {Lee}}]{Huggins_2022}%
  \BibitemOpen
  \bibfield  {author} {\bibinfo {author} {\bibfnamefont {W.~J.}\ \bibnamefont {Huggins}}, \bibinfo {author} {\bibfnamefont {B.~A.}\ \bibnamefont {O’Gorman}}, \bibinfo {author} {\bibfnamefont {N.~C.}\ \bibnamefont {Rubin}}, \bibinfo {author} {\bibfnamefont {D.~R.}\ \bibnamefont {Reichman}}, \bibinfo {author} {\bibfnamefont {R.}~\bibnamefont {Babbush}},\ and\ \bibinfo {author} {\bibfnamefont {J.}~\bibnamefont {Lee}},\ }\bibfield  {title} {\bibinfo {title} {Unbiasing fermionic quantum monte carlo with a quantum computer},\ }\href {https://doi.org/10.1038/s41586-021-04351-z} {\bibfield  {journal} {\bibinfo  {journal} {Nature}\ }\textbf {\bibinfo {volume} {603}},\ \bibinfo {pages} {416–420} (\bibinfo {year} {2022})}\BibitemShut {NoStop}%
\bibitem [{\citenamefont {King}\ \emph {et~al.}(2021)\citenamefont {King}, \citenamefont {Raymond}, \citenamefont {Lanting}, \citenamefont {Isakov}, \citenamefont {Mohseni}, \citenamefont {Poulin-Lamarre}, \citenamefont {Ejtemaee}, \citenamefont {Bernoudy}, \citenamefont {Ozfidan}, \citenamefont {Smirnov}, \citenamefont {Reis}, \citenamefont {Altomare}, \citenamefont {Babcock}, \citenamefont {Baron}, \citenamefont {Berkley}, \citenamefont {Boothby}, \citenamefont {Bunyk}, \citenamefont {Christiani}, \citenamefont {Enderud}, \citenamefont {Evert}, \citenamefont {Harris}, \citenamefont {Hoskinson}, \citenamefont {Huang}, \citenamefont {Jooya}, \citenamefont {Khodabandelou}, \citenamefont {Ladizinsky}, \citenamefont {Li}, \citenamefont {Lott}, \citenamefont {MacDonald}, \citenamefont {Marsden}, \citenamefont {Marsden}, \citenamefont {Medina}, \citenamefont {Molavi}, \citenamefont {Neufeld}, \citenamefont {Norouzpour}, \citenamefont {Oh}, \citenamefont {Pavlov}, \citenamefont {Perminov}, \citenamefont {Prescott},
  \citenamefont {Rich}, \citenamefont {Sato}, \citenamefont {Sheldan}, \citenamefont {Sterling}, \citenamefont {Swenson}, \citenamefont {Tsai}, \citenamefont {Volkmann}, \citenamefont {Whittaker}, \citenamefont {Wilkinson}, \citenamefont {Yao}, \citenamefont {Neven}, \citenamefont {Hilton}, \citenamefont {Ladizinsky}, \citenamefont {Johnson},\ and\ \citenamefont {Amin}}]{King_2021}%
  \BibitemOpen
  \bibfield  {author} {\bibinfo {author} {\bibfnamefont {A.~D.}\ \bibnamefont {King}}, \bibinfo {author} {\bibfnamefont {J.}~\bibnamefont {Raymond}}, \bibinfo {author} {\bibfnamefont {T.}~\bibnamefont {Lanting}}, \bibinfo {author} {\bibfnamefont {S.~V.}\ \bibnamefont {Isakov}}, \bibinfo {author} {\bibfnamefont {M.}~\bibnamefont {Mohseni}}, \bibinfo {author} {\bibfnamefont {G.}~\bibnamefont {Poulin-Lamarre}}, \bibinfo {author} {\bibfnamefont {S.}~\bibnamefont {Ejtemaee}}, \bibinfo {author} {\bibfnamefont {W.}~\bibnamefont {Bernoudy}}, \bibinfo {author} {\bibfnamefont {I.}~\bibnamefont {Ozfidan}}, \bibinfo {author} {\bibfnamefont {A.~Y.}\ \bibnamefont {Smirnov}}, \bibinfo {author} {\bibfnamefont {M.}~\bibnamefont {Reis}}, \bibinfo {author} {\bibfnamefont {F.}~\bibnamefont {Altomare}}, \bibinfo {author} {\bibfnamefont {M.}~\bibnamefont {Babcock}}, \bibinfo {author} {\bibfnamefont {C.}~\bibnamefont {Baron}}, \bibinfo {author} {\bibfnamefont {A.~J.}\ \bibnamefont {Berkley}}, \bibinfo {author} {\bibfnamefont
  {K.}~\bibnamefont {Boothby}}, \bibinfo {author} {\bibfnamefont {P.~I.}\ \bibnamefont {Bunyk}}, \bibinfo {author} {\bibfnamefont {H.}~\bibnamefont {Christiani}}, \bibinfo {author} {\bibfnamefont {C.}~\bibnamefont {Enderud}}, \bibinfo {author} {\bibfnamefont {B.}~\bibnamefont {Evert}}, \bibinfo {author} {\bibfnamefont {R.}~\bibnamefont {Harris}}, \bibinfo {author} {\bibfnamefont {E.}~\bibnamefont {Hoskinson}}, \bibinfo {author} {\bibfnamefont {S.}~\bibnamefont {Huang}}, \bibinfo {author} {\bibfnamefont {K.}~\bibnamefont {Jooya}}, \bibinfo {author} {\bibfnamefont {A.}~\bibnamefont {Khodabandelou}}, \bibinfo {author} {\bibfnamefont {N.}~\bibnamefont {Ladizinsky}}, \bibinfo {author} {\bibfnamefont {R.}~\bibnamefont {Li}}, \bibinfo {author} {\bibfnamefont {P.~A.}\ \bibnamefont {Lott}}, \bibinfo {author} {\bibfnamefont {A.~J.~R.}\ \bibnamefont {MacDonald}}, \bibinfo {author} {\bibfnamefont {D.}~\bibnamefont {Marsden}}, \bibinfo {author} {\bibfnamefont {G.}~\bibnamefont {Marsden}}, \bibinfo {author} {\bibfnamefont
  {T.}~\bibnamefont {Medina}}, \bibinfo {author} {\bibfnamefont {R.}~\bibnamefont {Molavi}}, \bibinfo {author} {\bibfnamefont {R.}~\bibnamefont {Neufeld}}, \bibinfo {author} {\bibfnamefont {M.}~\bibnamefont {Norouzpour}}, \bibinfo {author} {\bibfnamefont {T.}~\bibnamefont {Oh}}, \bibinfo {author} {\bibfnamefont {I.}~\bibnamefont {Pavlov}}, \bibinfo {author} {\bibfnamefont {I.}~\bibnamefont {Perminov}}, \bibinfo {author} {\bibfnamefont {T.}~\bibnamefont {Prescott}}, \bibinfo {author} {\bibfnamefont {C.}~\bibnamefont {Rich}}, \bibinfo {author} {\bibfnamefont {Y.}~\bibnamefont {Sato}}, \bibinfo {author} {\bibfnamefont {B.}~\bibnamefont {Sheldan}}, \bibinfo {author} {\bibfnamefont {G.}~\bibnamefont {Sterling}}, \bibinfo {author} {\bibfnamefont {L.~J.}\ \bibnamefont {Swenson}}, \bibinfo {author} {\bibfnamefont {N.}~\bibnamefont {Tsai}}, \bibinfo {author} {\bibfnamefont {M.~H.}\ \bibnamefont {Volkmann}}, \bibinfo {author} {\bibfnamefont {J.~D.}\ \bibnamefont {Whittaker}}, \bibinfo {author} {\bibfnamefont
  {W.}~\bibnamefont {Wilkinson}}, \bibinfo {author} {\bibfnamefont {J.}~\bibnamefont {Yao}}, \bibinfo {author} {\bibfnamefont {H.}~\bibnamefont {Neven}}, \bibinfo {author} {\bibfnamefont {J.~P.}\ \bibnamefont {Hilton}}, \bibinfo {author} {\bibfnamefont {E.}~\bibnamefont {Ladizinsky}}, \bibinfo {author} {\bibfnamefont {M.~W.}\ \bibnamefont {Johnson}},\ and\ \bibinfo {author} {\bibfnamefont {M.~H.}\ \bibnamefont {Amin}},\ }\bibfield  {title} {\bibinfo {title} {Scaling advantage over path-integral monte carlo in quantum simulation of geometrically frustrated magnets},\ }\bibfield  {journal} {\bibinfo  {journal} {Nature Communications}\ }\textbf {\bibinfo {volume} {12}},\ \href {https://doi.org/10.1038/s41467-021-20901-5} {10.1038/s41467-021-20901-5} (\bibinfo {year} {2021})\BibitemShut {NoStop}%
\bibitem [{\citenamefont {Troyer}\ and\ \citenamefont {Wiese}(2005)}]{PhysRevLett.94.170201}%
  \BibitemOpen
  \bibfield  {author} {\bibinfo {author} {\bibfnamefont {M.}~\bibnamefont {Troyer}}\ and\ \bibinfo {author} {\bibfnamefont {U.-J.}\ \bibnamefont {Wiese}},\ }\bibfield  {title} {\bibinfo {title} {Computational complexity and fundamental limitations to fermionic quantum monte carlo simulations},\ }\href {https://doi.org/10.1103/PhysRevLett.94.170201} {\bibfield  {journal} {\bibinfo  {journal} {Phys. Rev. Lett.}\ }\textbf {\bibinfo {volume} {94}},\ \bibinfo {pages} {170201} (\bibinfo {year} {2005})}\BibitemShut {NoStop}%
\bibitem [{\citenamefont {Li}\ and\ \citenamefont {Yao}(2019)}]{Hong2019}%
  \BibitemOpen
  \bibfield  {author} {\bibinfo {author} {\bibfnamefont {Z.-X.}\ \bibnamefont {Li}}\ and\ \bibinfo {author} {\bibfnamefont {H.}~\bibnamefont {Yao}},\ }\bibfield  {title} {\bibinfo {title} {Sign-problem-free fermionic quantum monte carlo: Developments and applications},\ }\href {https://doi.org/https://doi.org/10.1146/annurev-conmatphys-033117-054307} {\bibfield  {journal} {\bibinfo  {journal} {Annual Review of Condensed Matter Physics}\ }\textbf {\bibinfo {volume} {10}},\ \bibinfo {pages} {337} (\bibinfo {year} {2019})}\BibitemShut {NoStop}%
\bibitem [{\citenamefont {LeBlanc}\ \emph {et~al.}(2015)\citenamefont {LeBlanc}, \citenamefont {Antipov}, \citenamefont {Becca}, \citenamefont {Bulik}, \citenamefont {Chan}, \citenamefont {Chung}, \citenamefont {Deng}, \citenamefont {Ferrero}, \citenamefont {Henderson}, \citenamefont {Jim\'enez-Hoyos}, \citenamefont {Kozik}, \citenamefont {Liu}, \citenamefont {Millis}, \citenamefont {Prokof'ev}, \citenamefont {Qin}, \citenamefont {Scuseria}, \citenamefont {Shi}, \citenamefont {Svistunov}, \citenamefont {Tocchio}, \citenamefont {Tupitsyn}, \citenamefont {White}, \citenamefont {Zhang}, \citenamefont {Zheng}, \citenamefont {Zhu},\ and\ \citenamefont {Gull}}]{PhysRevX.5.041041}%
  \BibitemOpen
  \bibfield  {author} {\bibinfo {author} {\bibfnamefont {J.~P.~F.}\ \bibnamefont {LeBlanc}}, \bibinfo {author} {\bibfnamefont {A.~E.}\ \bibnamefont {Antipov}}, \bibinfo {author} {\bibfnamefont {F.}~\bibnamefont {Becca}}, \bibinfo {author} {\bibfnamefont {I.~W.}\ \bibnamefont {Bulik}}, \bibinfo {author} {\bibfnamefont {G.~K.-L.}\ \bibnamefont {Chan}}, \bibinfo {author} {\bibfnamefont {C.-M.}\ \bibnamefont {Chung}}, \bibinfo {author} {\bibfnamefont {Y.}~\bibnamefont {Deng}}, \bibinfo {author} {\bibfnamefont {M.}~\bibnamefont {Ferrero}}, \bibinfo {author} {\bibfnamefont {T.~M.}\ \bibnamefont {Henderson}}, \bibinfo {author} {\bibfnamefont {C.~A.}\ \bibnamefont {Jim\'enez-Hoyos}}, \bibinfo {author} {\bibfnamefont {E.}~\bibnamefont {Kozik}}, \bibinfo {author} {\bibfnamefont {X.-W.}\ \bibnamefont {Liu}}, \bibinfo {author} {\bibfnamefont {A.~J.}\ \bibnamefont {Millis}}, \bibinfo {author} {\bibfnamefont {N.~V.}\ \bibnamefont {Prokof'ev}}, \bibinfo {author} {\bibfnamefont {M.}~\bibnamefont {Qin}}, \bibinfo {author}
  {\bibfnamefont {G.~E.}\ \bibnamefont {Scuseria}}, \bibinfo {author} {\bibfnamefont {H.}~\bibnamefont {Shi}}, \bibinfo {author} {\bibfnamefont {B.~V.}\ \bibnamefont {Svistunov}}, \bibinfo {author} {\bibfnamefont {L.~F.}\ \bibnamefont {Tocchio}}, \bibinfo {author} {\bibfnamefont {I.~S.}\ \bibnamefont {Tupitsyn}}, \bibinfo {author} {\bibfnamefont {S.~R.}\ \bibnamefont {White}}, \bibinfo {author} {\bibfnamefont {S.}~\bibnamefont {Zhang}}, \bibinfo {author} {\bibfnamefont {B.-X.}\ \bibnamefont {Zheng}}, \bibinfo {author} {\bibfnamefont {Z.}~\bibnamefont {Zhu}},\ and\ \bibinfo {author} {\bibfnamefont {E.}~\bibnamefont {Gull}} (\bibinfo {collaboration} {Simons Collaboration on the Many-Electron Problem}),\ }\bibfield  {title} {\bibinfo {title} {Solutions of the two-dimensional hubbard model: Benchmarks and results from a wide range of numerical algorithms},\ }\href {https://doi.org/10.1103/PhysRevX.5.041041} {\bibfield  {journal} {\bibinfo  {journal} {Phys. Rev. X}\ }\textbf {\bibinfo {volume} {5}},\ \bibinfo
  {pages} {041041} (\bibinfo {year} {2015})}\BibitemShut {NoStop}%
\bibitem [{\citenamefont {Wu}\ \emph {et~al.}(2008)\citenamefont {Wu}, \citenamefont {Weng},\ and\ \citenamefont {Zaanen}}]{PhysRevB.77.155102}%
  \BibitemOpen
  \bibfield  {author} {\bibinfo {author} {\bibfnamefont {K.}~\bibnamefont {Wu}}, \bibinfo {author} {\bibfnamefont {Z.~Y.}\ \bibnamefont {Weng}},\ and\ \bibinfo {author} {\bibfnamefont {J.}~\bibnamefont {Zaanen}},\ }\bibfield  {title} {\bibinfo {title} {Sign structure of the $t\text{\ensuremath{-}}j$ model},\ }\href {https://doi.org/10.1103/PhysRevB.77.155102} {\bibfield  {journal} {\bibinfo  {journal} {Phys. Rev. B}\ }\textbf {\bibinfo {volume} {77}},\ \bibinfo {pages} {155102} (\bibinfo {year} {2008})}\BibitemShut {NoStop}%
\bibitem [{\citenamefont {Rademaker}\ \emph {et~al.}(2013)\citenamefont {Rademaker}, \citenamefont {Johnston}, \citenamefont {Zaanen},\ and\ \citenamefont {van~den Brink}}]{PhysRevB.88.235115}%
  \BibitemOpen
  \bibfield  {author} {\bibinfo {author} {\bibfnamefont {L.}~\bibnamefont {Rademaker}}, \bibinfo {author} {\bibfnamefont {S.}~\bibnamefont {Johnston}}, \bibinfo {author} {\bibfnamefont {J.}~\bibnamefont {Zaanen}},\ and\ \bibinfo {author} {\bibfnamefont {J.}~\bibnamefont {van~den Brink}},\ }\bibfield  {title} {\bibinfo {title} {Determinant quantum monte carlo study of exciton condensation in the bilayer hubbard model},\ }\href {https://doi.org/10.1103/PhysRevB.88.235115} {\bibfield  {journal} {\bibinfo  {journal} {Phys. Rev. B}\ }\textbf {\bibinfo {volume} {88}},\ \bibinfo {pages} {235115} (\bibinfo {year} {2013})}\BibitemShut {NoStop}%
\bibitem [{\citenamefont {Maik}\ \emph {et~al.}(2012)\citenamefont {Maik}, \citenamefont {Hauke}, \citenamefont {Dutta}, \citenamefont {Zakrzewski},\ and\ \citenamefont {Lewenstein}}]{Maik_2012}%
  \BibitemOpen
  \bibfield  {author} {\bibinfo {author} {\bibfnamefont {M.}~\bibnamefont {Maik}}, \bibinfo {author} {\bibfnamefont {P.}~\bibnamefont {Hauke}}, \bibinfo {author} {\bibfnamefont {O.}~\bibnamefont {Dutta}}, \bibinfo {author} {\bibfnamefont {J.}~\bibnamefont {Zakrzewski}},\ and\ \bibinfo {author} {\bibfnamefont {M.}~\bibnamefont {Lewenstein}},\ }\bibfield  {title} {\bibinfo {title} {Quantum spin models with long-range interactions and tunnelings: a quantum monte carlo study},\ }\href {https://doi.org/10.1088/1367-2630/14/11/113006} {\bibfield  {journal} {\bibinfo  {journal} {New Journal of Physics}\ }\textbf {\bibinfo {volume} {14}},\ \bibinfo {pages} {113006} (\bibinfo {year} {2012})}\BibitemShut {NoStop}%
\bibitem [{\citenamefont {Aarts}\ and\ \citenamefont {Sexty}(2026)}]{aarts2026latticefieldtheoriessign}%
  \BibitemOpen
  \bibfield  {author} {\bibinfo {author} {\bibfnamefont {G.}~\bibnamefont {Aarts}}\ and\ \bibinfo {author} {\bibfnamefont {D.}~\bibnamefont {Sexty}},\ }\href {https://arxiv.org/abs/2604.24290} {\bibinfo {title} {Lattice field theories with a sign problem}} (\bibinfo {year} {2026}),\ \Eprint {https://arxiv.org/abs/2604.24290} {arXiv:2604.24290 [hep-lat]} \BibitemShut {NoStop}%
\bibitem [{\citenamefont {Aarts}\ \emph {et~al.}(2023)\citenamefont {Aarts}, \citenamefont {Aichelin}, \citenamefont {Allton}, \citenamefont {Athenodorou}, \citenamefont {Bachtis}, \citenamefont {Bonanno}, \citenamefont {Brambilla}, \citenamefont {Bratkovskaya}, \citenamefont {Bruno}, \citenamefont {Caselle}, \citenamefont {Conti}, \citenamefont {Contino}, \citenamefont {Cosmai}, \citenamefont {Cuteri}, \citenamefont {{Del Debbio}}, \citenamefont {D’Elia}, \citenamefont {Dimopoulos}, \citenamefont {{Di Renzo}}, \citenamefont {Galatyuk}, \citenamefont {Guenther}, \citenamefont {Houtz}, \citenamefont {Karsch}, \citenamefont {Kotov}, \citenamefont {Lombardo}, \citenamefont {Lucini}, \citenamefont {Maio}, \citenamefont {Panero}, \citenamefont {Pawlowski}, \citenamefont {Pelissetto}, \citenamefont {Philipsen}, \citenamefont {Rago}, \citenamefont {Ratti}, \citenamefont {Ryan}, \citenamefont {Sannino}, \citenamefont {Sasaki}, \citenamefont {Schicho}, \citenamefont {Schmidt}, \citenamefont {Sharma}, \citenamefont
  {Soloveva}, \citenamefont {Sorba},\ and\ \citenamefont {Wiese}}]{AARTS2023104070}%
  \BibitemOpen
  \bibfield  {author} {\bibinfo {author} {\bibfnamefont {G.}~\bibnamefont {Aarts}}, \bibinfo {author} {\bibfnamefont {J.}~\bibnamefont {Aichelin}}, \bibinfo {author} {\bibfnamefont {C.}~\bibnamefont {Allton}}, \bibinfo {author} {\bibfnamefont {A.}~\bibnamefont {Athenodorou}}, \bibinfo {author} {\bibfnamefont {D.}~\bibnamefont {Bachtis}}, \bibinfo {author} {\bibfnamefont {C.}~\bibnamefont {Bonanno}}, \bibinfo {author} {\bibfnamefont {N.}~\bibnamefont {Brambilla}}, \bibinfo {author} {\bibfnamefont {E.}~\bibnamefont {Bratkovskaya}}, \bibinfo {author} {\bibfnamefont {M.}~\bibnamefont {Bruno}}, \bibinfo {author} {\bibfnamefont {M.}~\bibnamefont {Caselle}}, \bibinfo {author} {\bibfnamefont {C.}~\bibnamefont {Conti}}, \bibinfo {author} {\bibfnamefont {R.}~\bibnamefont {Contino}}, \bibinfo {author} {\bibfnamefont {L.}~\bibnamefont {Cosmai}}, \bibinfo {author} {\bibfnamefont {F.}~\bibnamefont {Cuteri}}, \bibinfo {author} {\bibfnamefont {L.}~\bibnamefont {{Del Debbio}}}, \bibinfo {author} {\bibfnamefont
  {M.}~\bibnamefont {D’Elia}}, \bibinfo {author} {\bibfnamefont {P.}~\bibnamefont {Dimopoulos}}, \bibinfo {author} {\bibfnamefont {F.}~\bibnamefont {{Di Renzo}}}, \bibinfo {author} {\bibfnamefont {T.}~\bibnamefont {Galatyuk}}, \bibinfo {author} {\bibfnamefont {J.~N.}\ \bibnamefont {Guenther}}, \bibinfo {author} {\bibfnamefont {R.}~\bibnamefont {Houtz}}, \bibinfo {author} {\bibfnamefont {F.}~\bibnamefont {Karsch}}, \bibinfo {author} {\bibfnamefont {A.~Y.}\ \bibnamefont {Kotov}}, \bibinfo {author} {\bibfnamefont {M.~P.}\ \bibnamefont {Lombardo}}, \bibinfo {author} {\bibfnamefont {B.}~\bibnamefont {Lucini}}, \bibinfo {author} {\bibfnamefont {L.}~\bibnamefont {Maio}}, \bibinfo {author} {\bibfnamefont {M.}~\bibnamefont {Panero}}, \bibinfo {author} {\bibfnamefont {J.~M.}\ \bibnamefont {Pawlowski}}, \bibinfo {author} {\bibfnamefont {A.}~\bibnamefont {Pelissetto}}, \bibinfo {author} {\bibfnamefont {O.}~\bibnamefont {Philipsen}}, \bibinfo {author} {\bibfnamefont {A.}~\bibnamefont {Rago}}, \bibinfo {author}
  {\bibfnamefont {C.}~\bibnamefont {Ratti}}, \bibinfo {author} {\bibfnamefont {S.~M.}\ \bibnamefont {Ryan}}, \bibinfo {author} {\bibfnamefont {F.}~\bibnamefont {Sannino}}, \bibinfo {author} {\bibfnamefont {C.}~\bibnamefont {Sasaki}}, \bibinfo {author} {\bibfnamefont {P.}~\bibnamefont {Schicho}}, \bibinfo {author} {\bibfnamefont {C.}~\bibnamefont {Schmidt}}, \bibinfo {author} {\bibfnamefont {S.}~\bibnamefont {Sharma}}, \bibinfo {author} {\bibfnamefont {O.}~\bibnamefont {Soloveva}}, \bibinfo {author} {\bibfnamefont {M.}~\bibnamefont {Sorba}},\ and\ \bibinfo {author} {\bibfnamefont {U.-J.}\ \bibnamefont {Wiese}},\ }\bibfield  {title} {\bibinfo {title} {Phase transitions in particle physics: Results and perspectives from lattice quantum chromo-dynamics},\ }\href {https://doi.org/https://doi.org/10.1016/j.ppnp.2023.104070} {\bibfield  {journal} {\bibinfo  {journal} {Progress in Particle and Nuclear Physics}\ }\textbf {\bibinfo {volume} {133}},\ \bibinfo {pages} {104070} (\bibinfo {year} {2023})}\BibitemShut
  {NoStop}%
\bibitem [{\citenamefont {Sandvik}(2010)}]{10.1063/1.3518900}%
  \BibitemOpen
  \bibfield  {author} {\bibinfo {author} {\bibfnamefont {A.~W.}\ \bibnamefont {Sandvik}},\ }\bibfield  {title} {\bibinfo {title} {Computational studies of quantum spin systems},\ }\href {https://doi.org/10.1063/1.3518900} {\bibfield  {journal} {\bibinfo  {journal} {AIP Conference Proceedings}\ }\textbf {\bibinfo {volume} {1297}},\ \bibinfo {pages} {135} (\bibinfo {year} {2010})}\BibitemShut {NoStop}%
\bibitem [{\citenamefont {Sandvik}(1999)}]{PhysRevB.59.R14157}%
  \BibitemOpen
  \bibfield  {author} {\bibinfo {author} {\bibfnamefont {A.~W.}\ \bibnamefont {Sandvik}},\ }\bibfield  {title} {\bibinfo {title} {Stochastic series expansion method with operator-loop update},\ }\href {https://doi.org/10.1103/PhysRevB.59.R14157} {\bibfield  {journal} {\bibinfo  {journal} {Phys. Rev. B}\ }\textbf {\bibinfo {volume} {59}},\ \bibinfo {pages} {R14157(R)} (\bibinfo {year} {1999})}\BibitemShut {NoStop}%
\bibitem [{\citenamefont {Sandvik}(1997)}]{PhysRevB.56.11678}%
  \BibitemOpen
  \bibfield  {author} {\bibinfo {author} {\bibfnamefont {A.~W.}\ \bibnamefont {Sandvik}},\ }\bibfield  {title} {\bibinfo {title} {Finite-size scaling of the ground-state parameters of the two-dimensional heisenberg model},\ }\href {https://doi.org/10.1103/PhysRevB.56.11678} {\bibfield  {journal} {\bibinfo  {journal} {Phys. Rev. B}\ }\textbf {\bibinfo {volume} {56}},\ \bibinfo {pages} {11678} (\bibinfo {year} {1997})}\BibitemShut {NoStop}%
\bibitem [{\citenamefont {Berg}\ \emph {et~al.}(2019)\citenamefont {Berg}, \citenamefont {Lederer}, \citenamefont {Schattner},\ and\ \citenamefont {Trebst}}]{annurevconmatphys031218013339}%
  \BibitemOpen
  \bibfield  {author} {\bibinfo {author} {\bibfnamefont {E.}~\bibnamefont {Berg}}, \bibinfo {author} {\bibfnamefont {S.}~\bibnamefont {Lederer}}, \bibinfo {author} {\bibfnamefont {Y.}~\bibnamefont {Schattner}},\ and\ \bibinfo {author} {\bibfnamefont {S.}~\bibnamefont {Trebst}},\ }\bibfield  {title} {\bibinfo {title} {Monte carlo studies of quantum critical metals},\ }\href {https://doi.org/https://doi.org/10.1146/annurev-conmatphys-031218-013339} {\bibfield  {journal} {\bibinfo  {journal} {Annual Review of Condensed Matter Physics}\ }\textbf {\bibinfo {volume} {10}},\ \bibinfo {pages} {63} (\bibinfo {year} {2019})}\BibitemShut {NoStop}%
\bibitem [{\citenamefont {Bravyi}\ and\ \citenamefont {Terhal}(2010)}]{stoq08072689X}%
  \BibitemOpen
  \bibfield  {author} {\bibinfo {author} {\bibfnamefont {S.}~\bibnamefont {Bravyi}}\ and\ \bibinfo {author} {\bibfnamefont {B.}~\bibnamefont {Terhal}},\ }\bibfield  {title} {\bibinfo {title} {Complexity of stoquastic frustration-free hamiltonians},\ }\href {https://doi.org/10.1137/08072689X} {\bibfield  {journal} {\bibinfo  {journal} {SIAM Journal on Computing}\ }\textbf {\bibinfo {volume} {39}},\ \bibinfo {pages} {1462} (\bibinfo {year} {2010})}\BibitemShut {NoStop}%
\bibitem [{\citenamefont {Bravyi}(2015)}]{Bravyi2015}%
  \BibitemOpen
  \bibfield  {author} {\bibinfo {author} {\bibfnamefont {S.}~\bibnamefont {Bravyi}},\ }\bibfield  {title} {\bibinfo {title} {Monte carlo simulation of stoquastic hamiltonians},\ }\href {https://doi.org/10.26421/QIC15.13-14-3} {\bibfield  {journal} {\bibinfo  {journal} {Quantum Information and Computation}\ }\textbf {\bibinfo {volume} {15}},\ \bibinfo {pages} {1122} (\bibinfo {year} {2015})}\BibitemShut {NoStop}%
\bibitem [{\citenamefont {Bravyi}\ \emph {et~al.}(2008)\citenamefont {Bravyi}, \citenamefont {DiVincenzo}, \citenamefont {Oliveira},\ and\ \citenamefont {Terhal}}]{Bravyi2008}%
  \BibitemOpen
  \bibfield  {author} {\bibinfo {author} {\bibfnamefont {S.}~\bibnamefont {Bravyi}}, \bibinfo {author} {\bibfnamefont {D.~P.}\ \bibnamefont {DiVincenzo}}, \bibinfo {author} {\bibfnamefont {R.~I.}\ \bibnamefont {Oliveira}},\ and\ \bibinfo {author} {\bibfnamefont {B.~M.}\ \bibnamefont {Terhal}},\ }\bibfield  {title} {\bibinfo {title} {The complexity of stoquastic local hamiltonian problems},\ }\href {https://www.rintonpress.com/journals/doi/QIC8.5-1.html} {\bibfield  {journal} {\bibinfo  {journal} {Quantum Information and Computation}\ }\textbf {\bibinfo {volume} {8}},\ \bibinfo {pages} {361} (\bibinfo {year} {2008})}\BibitemShut {NoStop}%
\bibitem [{\citenamefont {Bravyi}\ and\ \citenamefont {Gosset}(2017)}]{PhysRevLett.119.100503}%
  \BibitemOpen
  \bibfield  {author} {\bibinfo {author} {\bibfnamefont {S.}~\bibnamefont {Bravyi}}\ and\ \bibinfo {author} {\bibfnamefont {D.}~\bibnamefont {Gosset}},\ }\bibfield  {title} {\bibinfo {title} {Polynomial-time classical simulation of quantum ferromagnets},\ }\href {https://doi.org/10.1103/PhysRevLett.119.100503} {\bibfield  {journal} {\bibinfo  {journal} {Phys. Rev. Lett.}\ }\textbf {\bibinfo {volume} {119}},\ \bibinfo {pages} {100503} (\bibinfo {year} {2017})}\BibitemShut {NoStop}%
\bibitem [{\citenamefont {Barahona}(1982)}]{FBarahona_1982}%
  \BibitemOpen
  \bibfield  {author} {\bibinfo {author} {\bibfnamefont {F.}~\bibnamefont {Barahona}},\ }\bibfield  {title} {\bibinfo {title} {On the computational complexity of ising spin glass models},\ }\href {https://doi.org/10.1088/0305-4470/15/10/028} {\bibfield  {journal} {\bibinfo  {journal} {Journal of Physics A: Mathematical and General}\ }\textbf {\bibinfo {volume} {15}},\ \bibinfo {pages} {3241} (\bibinfo {year} {1982})}\BibitemShut {NoStop}%
\bibitem [{\citenamefont {Horn}\ and\ \citenamefont {Johnson}(1985)}]{Horn_Johnson_1985}%
  \BibitemOpen
  \bibfield  {author} {\bibinfo {author} {\bibfnamefont {R.~A.}\ \bibnamefont {Horn}}\ and\ \bibinfo {author} {\bibfnamefont {C.~R.}\ \bibnamefont {Johnson}},\ }\href {https://doi.org/https://doi.org/10.1017/CBO9780511810817} {\emph {\bibinfo {title} {Matrix Analysis}}}\ (\bibinfo  {publisher} {Cambridge University Press},\ \bibinfo {year} {1985})\BibitemShut {NoStop}%
\bibitem [{\citenamefont {Amico}\ \emph {et~al.}(2008)\citenamefont {Amico}, \citenamefont {Fazio}, \citenamefont {Osterloh},\ and\ \citenamefont {Vedral}}]{Amico2008}%
  \BibitemOpen
  \bibfield  {author} {\bibinfo {author} {\bibfnamefont {L.}~\bibnamefont {Amico}}, \bibinfo {author} {\bibfnamefont {R.}~\bibnamefont {Fazio}}, \bibinfo {author} {\bibfnamefont {A.}~\bibnamefont {Osterloh}},\ and\ \bibinfo {author} {\bibfnamefont {V.}~\bibnamefont {Vedral}},\ }\bibfield  {title} {\bibinfo {title} {Entanglement in many-body systems},\ }\href {https://doi.org/10.1103/RevModPhys.80.517} {\bibfield  {journal} {\bibinfo  {journal} {Rev. Mod. Phys.}\ }\textbf {\bibinfo {volume} {80}},\ \bibinfo {pages} {517} (\bibinfo {year} {2008})}\BibitemShut {NoStop}%
\bibitem [{\citenamefont {Lewenstein}\ \emph {et~al.}(2007)\citenamefont {Lewenstein}, \citenamefont {Sanpera}, \citenamefont {Ahufinger}, \citenamefont {Damski}, \citenamefont {Sen(De)},\ and\ \citenamefont {Sen}}]{entanglement_manybody_review_lewenstein}%
  \BibitemOpen
  \bibfield  {author} {\bibinfo {author} {\bibfnamefont {M.}~\bibnamefont {Lewenstein}}, \bibinfo {author} {\bibfnamefont {A.}~\bibnamefont {Sanpera}}, \bibinfo {author} {\bibfnamefont {V.}~\bibnamefont {Ahufinger}}, \bibinfo {author} {\bibfnamefont {B.}~\bibnamefont {Damski}}, \bibinfo {author} {\bibfnamefont {A.}~\bibnamefont {Sen(De)}},\ and\ \bibinfo {author} {\bibfnamefont {U.}~\bibnamefont {Sen}},\ }\bibfield  {title} {\bibinfo {title} {Ultracold atomic gases in optical lattices: mimicking condensed matter physics and beyond},\ }\href {https://doi.org/10.1080/00018730701223200} {\bibfield  {journal} {\bibinfo  {journal} {Advances in Physics}\ }\textbf {\bibinfo {volume} {56}},\ \bibinfo {pages} {243} (\bibinfo {year} {2007})}\BibitemShut {NoStop}%
\bibitem [{\citenamefont {Laflorencie}(2016)}]{LAFLORENCIE20161}%
  \BibitemOpen
  \bibfield  {author} {\bibinfo {author} {\bibfnamefont {N.}~\bibnamefont {Laflorencie}},\ }\bibfield  {title} {\bibinfo {title} {Quantum entanglement in condensed matter systems},\ }\href {https://doi.org/https://doi.org/10.1016/j.physrep.2016.06.008} {\bibfield  {journal} {\bibinfo  {journal} {Physics Reports}\ }\textbf {\bibinfo {volume} {646}},\ \bibinfo {pages} {1} (\bibinfo {year} {2016})}\BibitemShut {NoStop}%
\bibitem [{\citenamefont {De~Chiara}\ and\ \citenamefont {Sanpera}(2018)}]{entanglement_manybody_review_lewenstein_DeChiara}%
  \BibitemOpen
  \bibfield  {author} {\bibinfo {author} {\bibfnamefont {G.}~\bibnamefont {De~Chiara}}\ and\ \bibinfo {author} {\bibfnamefont {A.}~\bibnamefont {Sanpera}},\ }\bibfield  {title} {\bibinfo {title} {{Genuine quantum correlations in quantum many-body systems: a review of recent progress}},\ }\href {https://doi.org/10.1088/1361-6633/aabf61} {\bibfield  {journal} {\bibinfo  {journal} {Rep Prog Phys}\ }\textbf {\bibinfo {volume} {81}},\ \bibinfo {pages} {074002} (\bibinfo {year} {2018})}\BibitemShut {NoStop}%
\bibitem [{\citenamefont {Ellison}\ \emph {et~al.}(2021)\citenamefont {Ellison}, \citenamefont {Kato}, \citenamefont {Liu},\ and\ \citenamefont {Hsieh}}]{Ellison2021symmetryprotected}%
  \BibitemOpen
  \bibfield  {author} {\bibinfo {author} {\bibfnamefont {T.~D.}\ \bibnamefont {Ellison}}, \bibinfo {author} {\bibfnamefont {K.}~\bibnamefont {Kato}}, \bibinfo {author} {\bibfnamefont {Z.-W.}\ \bibnamefont {Liu}},\ and\ \bibinfo {author} {\bibfnamefont {T.~H.}\ \bibnamefont {Hsieh}},\ }\bibfield  {title} {\bibinfo {title} {Symmetry-protected sign problem and magic in quantum phases of matter},\ }\href {https://doi.org/10.22331/q-2021-12-28-612} {\bibfield  {journal} {\bibinfo  {journal} {{Quantum}}\ }\textbf {\bibinfo {volume} {5}},\ \bibinfo {pages} {612} (\bibinfo {year} {2021})}\BibitemShut {NoStop}%
\bibitem [{\citenamefont {Leone}\ \emph {et~al.}(2022)\citenamefont {Leone}, \citenamefont {Oliviero},\ and\ \citenamefont {Hamma}}]{leone2022stabilizer}%
  \BibitemOpen
  \bibfield  {author} {\bibinfo {author} {\bibfnamefont {L.}~\bibnamefont {Leone}}, \bibinfo {author} {\bibfnamefont {S.~F.}\ \bibnamefont {Oliviero}},\ and\ \bibinfo {author} {\bibfnamefont {A.}~\bibnamefont {Hamma}},\ }\bibfield  {title} {\bibinfo {title} {Stabilizer r{\'e}nyi entropy},\ }\href {https://doi.org/10.1103/PhysRevLett.128.050402} {\bibfield  {journal} {\bibinfo  {journal} {Physical Review Letters}\ }\textbf {\bibinfo {volume} {128}},\ \bibinfo {pages} {050402} (\bibinfo {year} {2022})}\BibitemShut {NoStop}%
\bibitem [{\citenamefont {Liu}\ and\ \citenamefont {Winter}(2022)}]{Winter2022}%
  \BibitemOpen
  \bibfield  {author} {\bibinfo {author} {\bibfnamefont {Z.-W.}\ \bibnamefont {Liu}}\ and\ \bibinfo {author} {\bibfnamefont {A.}~\bibnamefont {Winter}},\ }\bibfield  {title} {\bibinfo {title} {Many-body quantum magic},\ }\href {https://doi.org/10.1103/PRXQuantum.3.020333} {\bibfield  {journal} {\bibinfo  {journal} {PRX Quantum}\ }\textbf {\bibinfo {volume} {3}},\ \bibinfo {pages} {020333} (\bibinfo {year} {2022})}\BibitemShut {NoStop}%
\bibitem [{\citenamefont {Leone}\ and\ \citenamefont {Bittel}(2024)}]{leone2024stabilizer}%
  \BibitemOpen
  \bibfield  {author} {\bibinfo {author} {\bibfnamefont {L.}~\bibnamefont {Leone}}\ and\ \bibinfo {author} {\bibfnamefont {L.}~\bibnamefont {Bittel}},\ }\bibfield  {title} {\bibinfo {title} {Stabilizer entropies are monotones for magic-state resource theory},\ }\href {https://doi.org/10.1103/PhysRevA.110.L040403} {\bibfield  {journal} {\bibinfo  {journal} {Physical Review A}\ }\textbf {\bibinfo {volume} {110}},\ \bibinfo {pages} {L040403} (\bibinfo {year} {2024})}\BibitemShut {NoStop}%
\bibitem [{\citenamefont {Vidal}(2003)}]{PhysRevLett.91.147902}%
  \BibitemOpen
  \bibfield  {author} {\bibinfo {author} {\bibfnamefont {G.}~\bibnamefont {Vidal}},\ }\bibfield  {title} {\bibinfo {title} {Efficient classical simulation of slightly entangled quantum computations},\ }\href {https://doi.org/10.1103/PhysRevLett.91.147902} {\bibfield  {journal} {\bibinfo  {journal} {Phys. Rev. Lett.}\ }\textbf {\bibinfo {volume} {91}},\ \bibinfo {pages} {147902} (\bibinfo {year} {2003})}\BibitemShut {NoStop}%
\bibitem [{\citenamefont {Orús}(2014)}]{ORUS2014117}%
  \BibitemOpen
  \bibfield  {author} {\bibinfo {author} {\bibfnamefont {R.}~\bibnamefont {Orús}},\ }\bibfield  {title} {\bibinfo {title} {A practical introduction to tensor networks: Matrix product states and projected entangled pair states},\ }\href {https://doi.org/https://doi.org/10.1016/j.aop.2014.06.013} {\bibfield  {journal} {\bibinfo  {journal} {Annals of Physics}\ }\textbf {\bibinfo {volume} {349}},\ \bibinfo {pages} {117} (\bibinfo {year} {2014})}\BibitemShut {NoStop}%
\bibitem [{\citenamefont {Hauke}\ \emph {et~al.}(2016)\citenamefont {Hauke}, \citenamefont {Heyl}, \citenamefont {Tagliacozzo},\ and\ \citenamefont {Zoller}}]{hauke2016measuring}%
  \BibitemOpen
  \bibfield  {author} {\bibinfo {author} {\bibfnamefont {P.}~\bibnamefont {Hauke}}, \bibinfo {author} {\bibfnamefont {M.}~\bibnamefont {Heyl}}, \bibinfo {author} {\bibfnamefont {L.}~\bibnamefont {Tagliacozzo}},\ and\ \bibinfo {author} {\bibfnamefont {P.}~\bibnamefont {Zoller}},\ }\bibfield  {title} {\bibinfo {title} {Measuring multipartite entanglement through dynamic susceptibilities},\ }\href {https://doi.org/10.1038/nphys3700} {\bibfield  {journal} {\bibinfo  {journal} {Nature Physics}\ }\textbf {\bibinfo {volume} {12}},\ \bibinfo {pages} {778} (\bibinfo {year} {2016})}\BibitemShut {NoStop}%
\bibitem [{\citenamefont {Santra}\ \emph {et~al.}(2025{\natexlab{a}})\citenamefont {Santra}, \citenamefont {Windey}, \citenamefont {Bandyopadhyay}, \citenamefont {Legramandi},\ and\ \citenamefont {Hauke}}]{santra2025complexitytransitionschaoticquantum}%
  \BibitemOpen
  \bibfield  {author} {\bibinfo {author} {\bibfnamefont {G.~C.}\ \bibnamefont {Santra}}, \bibinfo {author} {\bibfnamefont {A.}~\bibnamefont {Windey}}, \bibinfo {author} {\bibfnamefont {S.}~\bibnamefont {Bandyopadhyay}}, \bibinfo {author} {\bibfnamefont {A.}~\bibnamefont {Legramandi}},\ and\ \bibinfo {author} {\bibfnamefont {P.}~\bibnamefont {Hauke}},\ }\bibfield  {title} {\bibinfo {title} {Complexity transitions in chaotic quantum systems: Nonstabilizerness, entanglement, and fractal dimension in syk and random matrix models},\ }\href {https://arxiv.org/abs/2505.09707} {\bibfield  {journal} {\bibinfo  {journal} {arXiv preprint arXiv:2505.09707}\ } (\bibinfo {year} {2025}{\natexlab{a}})}\BibitemShut {NoStop}%
\bibitem [{\citenamefont {Magni}\ and\ \citenamefont {Turkeshi}(2025)}]{Magni2025quantumcomplexity}%
  \BibitemOpen
  \bibfield  {author} {\bibinfo {author} {\bibfnamefont {B.}~\bibnamefont {Magni}}\ and\ \bibinfo {author} {\bibfnamefont {X.}~\bibnamefont {Turkeshi}},\ }\bibfield  {title} {\bibinfo {title} {Quantum {C}omplexity and {C}haos in {M}any-{Q}udit {D}oped {C}lifford {C}ircuits},\ }\href {https://doi.org/10.22331/q-2025-12-24-1956} {\bibfield  {journal} {\bibinfo  {journal} {{Quantum}}\ }\textbf {\bibinfo {volume} {9}},\ \bibinfo {pages} {1956} (\bibinfo {year} {2025})}\BibitemShut {NoStop}%
\bibitem [{\citenamefont {Tarabunga}\ \emph {et~al.}(2024)\citenamefont {Tarabunga}, \citenamefont {Tirrito}, \citenamefont {Ba{\~n}uls},\ and\ \citenamefont {Dalmonte}}]{tarabunga2024nonstabilizerness}%
  \BibitemOpen
  \bibfield  {author} {\bibinfo {author} {\bibfnamefont {P.~S.}\ \bibnamefont {Tarabunga}}, \bibinfo {author} {\bibfnamefont {E.}~\bibnamefont {Tirrito}}, \bibinfo {author} {\bibfnamefont {M.~C.}\ \bibnamefont {Ba{\~n}uls}},\ and\ \bibinfo {author} {\bibfnamefont {M.}~\bibnamefont {Dalmonte}},\ }\bibfield  {title} {\bibinfo {title} {Nonstabilizerness via matrix product states in the pauli basis},\ }\href {https://doi.org/10.1103/PhysRevLett.133.010601} {\bibfield  {journal} {\bibinfo  {journal} {Physical Review Letters}\ }\textbf {\bibinfo {volume} {133}},\ \bibinfo {pages} {010601} (\bibinfo {year} {2024})}\BibitemShut {NoStop}%
\bibitem [{\citenamefont {Santra}\ \emph {et~al.}(2025{\natexlab{b}})\citenamefont {Santra}, \citenamefont {Mildenberger}, \citenamefont {Ballini}, \citenamefont {Bottarelli}, \citenamefont {Wauters},\ and\ \citenamefont {Hauke}}]{santra2025quantumresourcesnonabelianlattice}%
  \BibitemOpen
  \bibfield  {author} {\bibinfo {author} {\bibfnamefont {G.~C.}\ \bibnamefont {Santra}}, \bibinfo {author} {\bibfnamefont {J.}~\bibnamefont {Mildenberger}}, \bibinfo {author} {\bibfnamefont {E.}~\bibnamefont {Ballini}}, \bibinfo {author} {\bibfnamefont {A.}~\bibnamefont {Bottarelli}}, \bibinfo {author} {\bibfnamefont {M.~M.}\ \bibnamefont {Wauters}},\ and\ \bibinfo {author} {\bibfnamefont {P.}~\bibnamefont {Hauke}},\ }\href {https://arxiv.org/abs/2510.07385} {\bibinfo {title} {Quantum resources in non-abelian lattice gauge theories: Nonstabilizerness, multipartite entanglement, and fermionic non-gaussianity}} (\bibinfo {year} {2025}{\natexlab{b}}),\ \Eprint {https://arxiv.org/abs/2510.07385} {arXiv:2510.07385 [quant-ph]} \BibitemShut {NoStop}%
\bibitem [{\citenamefont {Tarabunga}\ \emph {et~al.}(2023)\citenamefont {Tarabunga}, \citenamefont {Tirrito}, \citenamefont {Chanda},\ and\ \citenamefont {Dalmonte}}]{tarabunga2023many}%
  \BibitemOpen
  \bibfield  {author} {\bibinfo {author} {\bibfnamefont {P.~S.}\ \bibnamefont {Tarabunga}}, \bibinfo {author} {\bibfnamefont {E.}~\bibnamefont {Tirrito}}, \bibinfo {author} {\bibfnamefont {T.}~\bibnamefont {Chanda}},\ and\ \bibinfo {author} {\bibfnamefont {M.}~\bibnamefont {Dalmonte}},\ }\bibfield  {title} {\bibinfo {title} {Many-body magic via pauli-markov chains—from criticality to gauge theories},\ }\href {https://doi.org/10.1103/PRXQuantum.4.040317} {\bibfield  {journal} {\bibinfo  {journal} {PRX Quantum}\ }\textbf {\bibinfo {volume} {4}},\ \bibinfo {pages} {040317} (\bibinfo {year} {2023})}\BibitemShut {NoStop}%
\bibitem [{\citenamefont {Singha~Roy}\ \emph {et~al.}(2022)\citenamefont {Singha~Roy}, \citenamefont {Carl},\ and\ \citenamefont {Hauke}}]{PhysRevB.106.195158}%
  \BibitemOpen
  \bibfield  {author} {\bibinfo {author} {\bibfnamefont {S.}~\bibnamefont {Singha~Roy}}, \bibinfo {author} {\bibfnamefont {L.}~\bibnamefont {Carl}},\ and\ \bibinfo {author} {\bibfnamefont {P.}~\bibnamefont {Hauke}},\ }\bibfield  {title} {\bibinfo {title} {Genuine multipartite entanglement in a one-dimensional bose-hubbard model with frustrated hopping},\ }\href {https://doi.org/10.1103/PhysRevB.106.195158} {\bibfield  {journal} {\bibinfo  {journal} {Phys. Rev. B}\ }\textbf {\bibinfo {volume} {106}},\ \bibinfo {pages} {195158} (\bibinfo {year} {2022})}\BibitemShut {NoStop}%
\bibitem [{\citenamefont {Varikuti}\ \emph {et~al.}(2026{\natexlab{a}})\citenamefont {Varikuti}, \citenamefont {Bandyopadhyay},\ and\ \citenamefont {Hauke}}]{91hd-8k7g}%
  \BibitemOpen
  \bibfield  {author} {\bibinfo {author} {\bibfnamefont {N.~D.}\ \bibnamefont {Varikuti}}, \bibinfo {author} {\bibfnamefont {S.}~\bibnamefont {Bandyopadhyay}},\ and\ \bibinfo {author} {\bibfnamefont {P.}~\bibnamefont {Hauke}},\ }\bibfield  {title} {\bibinfo {title} {Deep thermalization and measurements of quantum resources},\ }\href {https://doi.org/10.1103/91hd-8k7g} {\bibfield  {journal} {\bibinfo  {journal} {Phys. Rev. Lett.}\ }\textbf {\bibinfo {volume} {137}},\ \bibinfo {pages} {180401} (\bibinfo {year} {2026}{\natexlab{a}})}\BibitemShut {NoStop}%
\bibitem [{\citenamefont {Lakkaraju}\ \emph {et~al.}(2026)\citenamefont {Lakkaraju}, \citenamefont {Varikuti},\ and\ \citenamefont {Hauke}}]{lakkaraju2026fermionic}%
  \BibitemOpen
  \bibfield  {author} {\bibinfo {author} {\bibfnamefont {L.~G.~C.}\ \bibnamefont {Lakkaraju}}, \bibinfo {author} {\bibfnamefont {N.~D.}\ \bibnamefont {Varikuti}},\ and\ \bibinfo {author} {\bibfnamefont {P.}~\bibnamefont {Hauke}},\ }\bibfield  {title} {\bibinfo {title} {Fermionic non-gaussianity via computable measures and experimentally accessible bounds},\ }\href {https://arxiv.org/abs/2610.17219} {\bibfield  {journal} {\bibinfo  {journal} {arXiv preprint arXiv:2610.17219}\ } (\bibinfo {year} {2026})}\BibitemShut {NoStop}%
\bibitem [{\citenamefont {Korbany}\ \emph {et~al.}(2025)\citenamefont {Korbany}, \citenamefont {Gullans},\ and\ \citenamefont {Piroli}}]{piroli_longrange_magic}%
  \BibitemOpen
  \bibfield  {author} {\bibinfo {author} {\bibfnamefont {D.~A.}\ \bibnamefont {Korbany}}, \bibinfo {author} {\bibfnamefont {M.~J.}\ \bibnamefont {Gullans}},\ and\ \bibinfo {author} {\bibfnamefont {L.}~\bibnamefont {Piroli}},\ }\bibfield  {title} {\bibinfo {title} {Long-range nonstabilizerness and phases of matter},\ }\href {https://doi.org/10.1103/1hlj-h6t9} {\bibfield  {journal} {\bibinfo  {journal} {Phys. Rev. Lett.}\ }\textbf {\bibinfo {volume} {135}},\ \bibinfo {pages} {160404} (\bibinfo {year} {2025})}\BibitemShut {NoStop}%
\bibitem [{\citenamefont {Sierant}\ \emph {et~al.}(2026)\citenamefont {Sierant}, \citenamefont {Stornati},\ and\ \citenamefont {Turkeshi}}]{3yx4-1j27}%
  \BibitemOpen
  \bibfield  {author} {\bibinfo {author} {\bibfnamefont {P.}~\bibnamefont {Sierant}}, \bibinfo {author} {\bibfnamefont {P.}~\bibnamefont {Stornati}},\ and\ \bibinfo {author} {\bibfnamefont {X.}~\bibnamefont {Turkeshi}},\ }\bibfield  {title} {\bibinfo {title} {Fermionic magic resources of quantum many-body systems},\ }\href {https://doi.org/10.1103/3yx4-1j27} {\bibfield  {journal} {\bibinfo  {journal} {PRX Quantum}\ }\textbf {\bibinfo {volume} {7}},\ \bibinfo {pages} {010302} (\bibinfo {year} {2026})}\BibitemShut {NoStop}%
\bibitem [{\citenamefont {Capecci}\ \emph {et~al.}(2025)\citenamefont {Capecci}, \citenamefont {Santra}, \citenamefont {Bottarelli}, \citenamefont {Tirrito},\ and\ \citenamefont {Hauke}}]{capecci2025role}%
  \BibitemOpen
  \bibfield  {author} {\bibinfo {author} {\bibfnamefont {C.}~\bibnamefont {Capecci}}, \bibinfo {author} {\bibfnamefont {G.~C.}\ \bibnamefont {Santra}}, \bibinfo {author} {\bibfnamefont {A.}~\bibnamefont {Bottarelli}}, \bibinfo {author} {\bibfnamefont {E.}~\bibnamefont {Tirrito}},\ and\ \bibinfo {author} {\bibfnamefont {P.}~\bibnamefont {Hauke}},\ }\bibfield  {title} {\bibinfo {title} {Role of nonstabilizerness in quantum optimization},\ }\href {https://arxiv.org/abs/2505.17185} {\bibfield  {journal} {\bibinfo  {journal} {arXiv preprint arXiv:2505.17185}\ } (\bibinfo {year} {2025})}\BibitemShut {NoStop}%
\bibitem [{\citenamefont {Robin}\ and\ \citenamefont {Savage}(2026)}]{robin2026quantumcomplexitynewdirections}%
  \BibitemOpen
  \bibfield  {author} {\bibinfo {author} {\bibfnamefont {C.~E.~P.}\ \bibnamefont {Robin}}\ and\ \bibinfo {author} {\bibfnamefont {M.~J.}\ \bibnamefont {Savage}},\ }\href {https://arxiv.org/abs/2604.26376} {\bibinfo {title} {Quantum complexity and new directions in nuclear physics and high-energy physics phenomenology}} (\bibinfo {year} {2026}),\ \Eprint {https://arxiv.org/abs/2604.26376} {arXiv:2604.26376 [quant-ph]} \BibitemShut {NoStop}%
\bibitem [{\citenamefont {Viscardi}\ \emph {et~al.}(2026)\citenamefont {Viscardi}, \citenamefont {Dalmonte}, \citenamefont {Hamma},\ and\ \citenamefont {Tirrito}}]{Viscardi_2026}%
  \BibitemOpen
  \bibfield  {author} {\bibinfo {author} {\bibfnamefont {M.}~\bibnamefont {Viscardi}}, \bibinfo {author} {\bibfnamefont {M.}~\bibnamefont {Dalmonte}}, \bibinfo {author} {\bibfnamefont {A.}~\bibnamefont {Hamma}},\ and\ \bibinfo {author} {\bibfnamefont {E.}~\bibnamefont {Tirrito}},\ }\bibfield  {title} {\bibinfo {title} {Interplay of entanglement structures and stabilizer entropy in spin models},\ }\bibfield  {journal} {\bibinfo  {journal} {SciPost Physics Core}\ }\textbf {\bibinfo {volume} {9}},\ \href {https://doi.org/10.21468/scipostphyscore.9.1.012} {10.21468/scipostphyscore.9.1.012} (\bibinfo {year} {2026})\BibitemShut {NoStop}%
\bibitem [{\citenamefont {Odavi\ifmmode~\acute{c}\else \'{c}\fi{}}\ \emph {et~al.}(2025)\citenamefont {Odavi\ifmmode~\acute{c}\else \'{c}\fi{}}, \citenamefont {Viscardi},\ and\ \citenamefont {Hamma}}]{odavic2024stabilizer}%
  \BibitemOpen
  \bibfield  {author} {\bibinfo {author} {\bibfnamefont {J.}~\bibnamefont {Odavi\ifmmode~\acute{c}\else \'{c}\fi{}}}, \bibinfo {author} {\bibfnamefont {M.}~\bibnamefont {Viscardi}},\ and\ \bibinfo {author} {\bibfnamefont {A.}~\bibnamefont {Hamma}},\ }\bibfield  {title} {\bibinfo {title} {Stabilizer entropy in nonintegrable quantum evolutions},\ }\href {https://doi.org/10.1103/y9r6-dx7p} {\bibfield  {journal} {\bibinfo  {journal} {Phys. Rev. B}\ }\textbf {\bibinfo {volume} {112}},\ \bibinfo {pages} {104301} (\bibinfo {year} {2025})}\BibitemShut {NoStop}%
\bibitem [{\citenamefont {Haug}\ and\ \citenamefont {Piroli}(2023)}]{haug2023quantifying}%
  \BibitemOpen
  \bibfield  {author} {\bibinfo {author} {\bibfnamefont {T.}~\bibnamefont {Haug}}\ and\ \bibinfo {author} {\bibfnamefont {L.}~\bibnamefont {Piroli}},\ }\bibfield  {title} {\bibinfo {title} {Quantifying nonstabilizerness of matrix product states},\ }\href {https://doi.org/10.1103/PhysRevB.107.035148} {\bibfield  {journal} {\bibinfo  {journal} {Physical Review B}\ }\textbf {\bibinfo {volume} {107}},\ \bibinfo {pages} {035148} (\bibinfo {year} {2023})}\BibitemShut {NoStop}%
\bibitem [{\citenamefont {Varikuti}\ \emph {et~al.}(2026{\natexlab{b}})\citenamefont {Varikuti}, \citenamefont {Bandyopadhyay},\ and\ \citenamefont {Hauke}}]{Varikuti2026impactofclifford}%
  \BibitemOpen
  \bibfield  {author} {\bibinfo {author} {\bibfnamefont {N.~D.}\ \bibnamefont {Varikuti}}, \bibinfo {author} {\bibfnamefont {S.}~\bibnamefont {Bandyopadhyay}},\ and\ \bibinfo {author} {\bibfnamefont {P.}~\bibnamefont {Hauke}},\ }\bibfield  {title} {\bibinfo {title} {Impact of {C}lifford operations on non-stabilizing power and quantum chaos},\ }\href {https://doi.org/10.22331/q-2026-03-10-2017} {\bibfield  {journal} {\bibinfo  {journal} {{Quantum}}\ }\textbf {\bibinfo {volume} {10}},\ \bibinfo {pages} {2017} (\bibinfo {year} {2026}{\natexlab{b}})}\BibitemShut {NoStop}%
\bibitem [{\citenamefont {Catalano}\ \emph {et~al.}(2025)\citenamefont {Catalano}, \citenamefont {Odavi{\'c}}, \citenamefont {Torre}, \citenamefont {Hamma}, \citenamefont {Franchini},\ and\ \citenamefont {Giampaolo}}]{catalano2024magic}%
  \BibitemOpen
  \bibfield  {author} {\bibinfo {author} {\bibfnamefont {A.~G.}\ \bibnamefont {Catalano}}, \bibinfo {author} {\bibfnamefont {J.}~\bibnamefont {Odavi{\'c}}}, \bibinfo {author} {\bibfnamefont {G.}~\bibnamefont {Torre}}, \bibinfo {author} {\bibfnamefont {A.}~\bibnamefont {Hamma}}, \bibinfo {author} {\bibfnamefont {F.}~\bibnamefont {Franchini}},\ and\ \bibinfo {author} {\bibfnamefont {S.~M.}\ \bibnamefont {Giampaolo}},\ }\bibfield  {title} {\bibinfo {title} {Magic phase transition and non-local complexity in generalized w state},\ }\href {https://doi.org/10.21468/SciPostPhysCore.8.4.078} {\bibfield  {journal} {\bibinfo  {journal} {SciPost Physics Core}\ }\textbf {\bibinfo {volume} {8}},\ \bibinfo {pages} {078} (\bibinfo {year} {2025})}\BibitemShut {NoStop}%
\bibitem [{\citenamefont {Aditya}\ \emph {et~al.}(2026)\citenamefont {Aditya}, \citenamefont {Sierant},\ and\ \citenamefont {Turkeshi}}]{aditya2026nonlocalmagicspreadingmanybody}%
  \BibitemOpen
  \bibfield  {author} {\bibinfo {author} {\bibfnamefont {S.}~\bibnamefont {Aditya}}, \bibinfo {author} {\bibfnamefont {P.}~\bibnamefont {Sierant}},\ and\ \bibinfo {author} {\bibfnamefont {X.}~\bibnamefont {Turkeshi}},\ }\href {https://arxiv.org/abs/2609.20951} {\bibinfo {title} {Nonlocal magic spreading in many-body quantum dynamics: From chaotic evolution to quasi-particle picture in integrable models}} (\bibinfo {year} {2026}),\ \Eprint {https://arxiv.org/abs/2609.20951} {arXiv:2609.20951 [quant-ph]} \BibitemShut {NoStop}%
\bibitem [{\citenamefont {Tarabunga}(2024)}]{Tarabunga2024criticalbehaviorsof}%
  \BibitemOpen
  \bibfield  {author} {\bibinfo {author} {\bibfnamefont {P.~S.}\ \bibnamefont {Tarabunga}},\ }\bibfield  {title} {\bibinfo {title} {Critical behaviors of non-stabilizerness in quantum spin chains},\ }\href {https://doi.org/10.22331/q-2024-07-17-1413} {\bibfield  {journal} {\bibinfo  {journal} {{Quantum}}\ }\textbf {\bibinfo {volume} {8}},\ \bibinfo {pages} {1413} (\bibinfo {year} {2024})}\BibitemShut {NoStop}%
\bibitem [{\citenamefont {Jozsa}\ and\ \citenamefont {Linden}(2003)}]{Jozsa2003Aug}%
  \BibitemOpen
  \bibfield  {author} {\bibinfo {author} {\bibfnamefont {R.}~\bibnamefont {Jozsa}}\ and\ \bibinfo {author} {\bibfnamefont {N.}~\bibnamefont {Linden}},\ }\bibfield  {title} {\bibinfo {title} {{On the role of entanglement in quantum-computational speed-up}},\ }\href {https://doi.org/10.1098/rspa.2002.1097} {\bibfield  {journal} {\bibinfo  {journal} {Proc R Soc A}\ }\textbf {\bibinfo {volume} {459}},\ \bibinfo {pages} {2011} (\bibinfo {year} {2003})}\BibitemShut {NoStop}%
\bibitem [{\citenamefont {Raussendorf}\ and\ \citenamefont {Briegel}(2001)}]{PhysRevLett.86.5188}%
  \BibitemOpen
  \bibfield  {author} {\bibinfo {author} {\bibfnamefont {R.}~\bibnamefont {Raussendorf}}\ and\ \bibinfo {author} {\bibfnamefont {H.~J.}\ \bibnamefont {Briegel}},\ }\bibfield  {title} {\bibinfo {title} {A one-way quantum computer},\ }\href {https://doi.org/10.1103/PhysRevLett.86.5188} {\bibfield  {journal} {\bibinfo  {journal} {Phys. Rev. Lett.}\ }\textbf {\bibinfo {volume} {86}},\ \bibinfo {pages} {5188} (\bibinfo {year} {2001})}\BibitemShut {NoStop}%
\bibitem [{\citenamefont {Rossi}\ \emph {et~al.}(2013)\citenamefont {Rossi}, \citenamefont {Bru{\ss{}}},\ and\ \citenamefont {Macchiavello}}]{ent_grover}%
  \BibitemOpen
  \bibfield  {author} {\bibinfo {author} {\bibfnamefont {M.}~\bibnamefont {Rossi}}, \bibinfo {author} {\bibfnamefont {D.}~\bibnamefont {Bru{\ss{}}}},\ and\ \bibinfo {author} {\bibfnamefont {C.}~\bibnamefont {Macchiavello}},\ }\bibfield  {title} {\bibinfo {title} {Scale invariance of entanglement dynamics in grover's quantum search algorithm},\ }\href {https://doi.org/10.1103/PhysRevA.87.022331} {\bibfield  {journal} {\bibinfo  {journal} {Phys. Rev. A}\ }\textbf {\bibinfo {volume} {87}},\ \bibinfo {pages} {022331} (\bibinfo {year} {2013})}\BibitemShut {NoStop}%
\bibitem [{\citenamefont {Bravyi}\ and\ \citenamefont {Kitaev}(2005)}]{Kitaev2005}%
  \BibitemOpen
  \bibfield  {author} {\bibinfo {author} {\bibfnamefont {S.}~\bibnamefont {Bravyi}}\ and\ \bibinfo {author} {\bibfnamefont {A.}~\bibnamefont {Kitaev}},\ }\bibfield  {title} {\bibinfo {title} {Universal quantum computation with ideal clifford gates and noisy ancillas},\ }\href {https://doi.org/10.1103/PhysRevA.71.022316} {\bibfield  {journal} {\bibinfo  {journal} {Phys. Rev. A}\ }\textbf {\bibinfo {volume} {71}},\ \bibinfo {pages} {022316} (\bibinfo {year} {2005})}\BibitemShut {NoStop}%
\bibitem [{\citenamefont {Hauke}\ \emph {et~al.}(2012)\citenamefont {Hauke}, \citenamefont {Cucchietti}, \citenamefont {Tagliacozzo}, \citenamefont {Deutsch},\ and\ \citenamefont {Lewenstein}}]{hauke2012can}%
  \BibitemOpen
  \bibfield  {author} {\bibinfo {author} {\bibfnamefont {P.}~\bibnamefont {Hauke}}, \bibinfo {author} {\bibfnamefont {F.~M.}\ \bibnamefont {Cucchietti}}, \bibinfo {author} {\bibfnamefont {L.}~\bibnamefont {Tagliacozzo}}, \bibinfo {author} {\bibfnamefont {I.}~\bibnamefont {Deutsch}},\ and\ \bibinfo {author} {\bibfnamefont {M.}~\bibnamefont {Lewenstein}},\ }\bibfield  {title} {\bibinfo {title} {Can one trust quantum simulators?},\ }\href {https://doi.org/10.1088/0034-4885/75/8/082401} {\bibfield  {journal} {\bibinfo  {journal} {Reports on Progress in Physics}\ }\textbf {\bibinfo {volume} {75}},\ \bibinfo {pages} {082401} (\bibinfo {year} {2012})}\BibitemShut {NoStop}%
\bibitem [{\citenamefont {Halimeh}\ \emph {et~al.}(2025)\citenamefont {Halimeh}, \citenamefont {Aidelsburger}, \citenamefont {Grusdt}, \citenamefont {Hauke},\ and\ \citenamefont {Yang}}]{halimeh2025cold}%
  \BibitemOpen
  \bibfield  {author} {\bibinfo {author} {\bibfnamefont {J.~C.}\ \bibnamefont {Halimeh}}, \bibinfo {author} {\bibfnamefont {M.}~\bibnamefont {Aidelsburger}}, \bibinfo {author} {\bibfnamefont {F.}~\bibnamefont {Grusdt}}, \bibinfo {author} {\bibfnamefont {P.}~\bibnamefont {Hauke}},\ and\ \bibinfo {author} {\bibfnamefont {B.}~\bibnamefont {Yang}},\ }\bibfield  {title} {\bibinfo {title} {Cold-atom quantum simulators of gauge theories},\ }\href {https://www.nature.com/articles/s41567-024-02721-8} {\bibfield  {journal} {\bibinfo  {journal} {Nature Physics}\ }\textbf {\bibinfo {volume} {21}},\ \bibinfo {pages} {25} (\bibinfo {year} {2025})}\BibitemShut {NoStop}%
\bibitem [{\citenamefont {Roushan}\ and\ \citenamefont {Martin}(2025)}]{doi:10.1126/science.adt1732}%
  \BibitemOpen
  \bibfield  {author} {\bibinfo {author} {\bibfnamefont {P.}~\bibnamefont {Roushan}}\ and\ \bibinfo {author} {\bibfnamefont {L.}~\bibnamefont {Martin}},\ }\bibfield  {title} {\bibinfo {title} {Dawn of quantum simulators},\ }\href {https://doi.org/10.1126/science.adt1732} {\bibfield  {journal} {\bibinfo  {journal} {Science}\ }\textbf {\bibinfo {volume} {390}},\ \bibinfo {pages} {1000} (\bibinfo {year} {2025})}\BibitemShut {NoStop}%
\bibitem [{\citenamefont {Daley}\ \emph {et~al.}(2022)\citenamefont {Daley}, \citenamefont {Bloch}, \citenamefont {Kokail}, \citenamefont {Flannigan}, \citenamefont {Pearson}, \citenamefont {Troyer},\ and\ \citenamefont {Zoller}}]{daley2022practical}%
  \BibitemOpen
  \bibfield  {author} {\bibinfo {author} {\bibfnamefont {A.~J.}\ \bibnamefont {Daley}}, \bibinfo {author} {\bibfnamefont {I.}~\bibnamefont {Bloch}}, \bibinfo {author} {\bibfnamefont {C.}~\bibnamefont {Kokail}}, \bibinfo {author} {\bibfnamefont {S.}~\bibnamefont {Flannigan}}, \bibinfo {author} {\bibfnamefont {N.}~\bibnamefont {Pearson}}, \bibinfo {author} {\bibfnamefont {M.}~\bibnamefont {Troyer}},\ and\ \bibinfo {author} {\bibfnamefont {P.}~\bibnamefont {Zoller}},\ }\bibfield  {title} {\bibinfo {title} {Practical quantum advantage in quantum simulation},\ }\href {https://www.nature.com/articles/s41586-022-04940-6} {\bibfield  {journal} {\bibinfo  {journal} {Nature}\ }\textbf {\bibinfo {volume} {607}},\ \bibinfo {pages} {667} (\bibinfo {year} {2022})}\BibitemShut {NoStop}%
\bibitem [{\citenamefont {Grover}\ and\ \citenamefont {Fisher}(2015)}]{PhysRevA.92.042308}%
  \BibitemOpen
  \bibfield  {author} {\bibinfo {author} {\bibfnamefont {T.}~\bibnamefont {Grover}}\ and\ \bibinfo {author} {\bibfnamefont {M.~P.~A.}\ \bibnamefont {Fisher}},\ }\bibfield  {title} {\bibinfo {title} {Entanglement and the sign structure of quantum states},\ }\href {https://doi.org/10.1103/PhysRevA.92.042308} {\bibfield  {journal} {\bibinfo  {journal} {Phys. Rev. A}\ }\textbf {\bibinfo {volume} {92}},\ \bibinfo {pages} {042308} (\bibinfo {year} {2015})}\BibitemShut {NoStop}%
\bibitem [{\citenamefont {Kaplis}\ \emph {et~al.}(2017)\citenamefont {Kaplis}, \citenamefont {Kr{\"u}ger},\ and\ \citenamefont {Zaanen}}]{PhysRevB.95.155102}%
  \BibitemOpen
  \bibfield  {author} {\bibinfo {author} {\bibfnamefont {N.}~\bibnamefont {Kaplis}}, \bibinfo {author} {\bibfnamefont {F.}~\bibnamefont {Kr{\"u}ger}},\ and\ \bibinfo {author} {\bibfnamefont {J.}~\bibnamefont {Zaanen}},\ }\bibfield  {title} {\bibinfo {title} {Entanglement entropies and fermion signs of critical metals},\ }\href {https://doi.org/10.1103/PhysRevB.95.155102} {\bibfield  {journal} {\bibinfo  {journal} {Phys. Rev. B}\ }\textbf {\bibinfo {volume} {95}},\ \bibinfo {pages} {155102} (\bibinfo {year} {2017})}\BibitemShut {NoStop}%
\bibitem [{\citenamefont {Chen}\ \emph {et~al.}(2025)\citenamefont {Chen}, \citenamefont {Jiang}, \citenamefont {Hangleiter},\ and\ \citenamefont {Schuch}}]{schuch_tensor_sign}%
  \BibitemOpen
  \bibfield  {author} {\bibinfo {author} {\bibfnamefont {J.}~\bibnamefont {Chen}}, \bibinfo {author} {\bibfnamefont {J.}~\bibnamefont {Jiang}}, \bibinfo {author} {\bibfnamefont {D.}~\bibnamefont {Hangleiter}},\ and\ \bibinfo {author} {\bibfnamefont {N.}~\bibnamefont {Schuch}},\ }\bibfield  {title} {\bibinfo {title} {Sign problem in tensor-network contraction},\ }\href {https://doi.org/10.1103/PRXQuantum.6.010312} {\bibfield  {journal} {\bibinfo  {journal} {PRX Quantum}\ }\textbf {\bibinfo {volume} {6}},\ \bibinfo {pages} {010312} (\bibinfo {year} {2025})}\BibitemShut {NoStop}%
\bibitem [{\citenamefont {Hauke}\ \emph {et~al.}(2020)\citenamefont {Hauke}, \citenamefont {Katzgraber}, \citenamefont {Lechner}, \citenamefont {Nishimori},\ and\ \citenamefont {Oliver}}]{Hauke_2020}%
  \BibitemOpen
  \bibfield  {author} {\bibinfo {author} {\bibfnamefont {P.}~\bibnamefont {Hauke}}, \bibinfo {author} {\bibfnamefont {H.~G.}\ \bibnamefont {Katzgraber}}, \bibinfo {author} {\bibfnamefont {W.}~\bibnamefont {Lechner}}, \bibinfo {author} {\bibfnamefont {H.}~\bibnamefont {Nishimori}},\ and\ \bibinfo {author} {\bibfnamefont {W.~D.}\ \bibnamefont {Oliver}},\ }\bibfield  {title} {\bibinfo {title} {Perspectives of quantum annealing: methods and implementations},\ }\href {https://doi.org/10.1088/1361-6633/ab85b8} {\bibfield  {journal} {\bibinfo  {journal} {Reports on Progress in Physics}\ }\textbf {\bibinfo {volume} {83}},\ \bibinfo {pages} {054401} (\bibinfo {year} {2020})}\BibitemShut {NoStop}%
\bibitem [{\citenamefont {Hormozi}\ \emph {et~al.}(2017)\citenamefont {Hormozi}, \citenamefont {Brown}, \citenamefont {Carleo},\ and\ \citenamefont {Troyer}}]{PhysRevB.95.184416}%
  \BibitemOpen
  \bibfield  {author} {\bibinfo {author} {\bibfnamefont {L.}~\bibnamefont {Hormozi}}, \bibinfo {author} {\bibfnamefont {E.~W.}\ \bibnamefont {Brown}}, \bibinfo {author} {\bibfnamefont {G.}~\bibnamefont {Carleo}},\ and\ \bibinfo {author} {\bibfnamefont {M.}~\bibnamefont {Troyer}},\ }\bibfield  {title} {\bibinfo {title} {Nonstoquastic hamiltonians and quantum annealing of an ising spin glass},\ }\href {https://doi.org/10.1103/PhysRevB.95.184416} {\bibfield  {journal} {\bibinfo  {journal} {Phys. Rev. B}\ }\textbf {\bibinfo {volume} {95}},\ \bibinfo {pages} {184416} (\bibinfo {year} {2017})}\BibitemShut {NoStop}%
\bibitem [{\citenamefont {Nishimori}\ and\ \citenamefont {Takada}(2017)}]{nishimori2017exponential}%
  \BibitemOpen
  \bibfield  {author} {\bibinfo {author} {\bibfnamefont {H.}~\bibnamefont {Nishimori}}\ and\ \bibinfo {author} {\bibfnamefont {K.}~\bibnamefont {Takada}},\ }\bibfield  {title} {\bibinfo {title} {Exponential enhancement of the efficiency of quantum annealing by non-stoquastic hamiltonians},\ }\href {https://www.frontiersin.org/journals/ict/articles/10.3389/fict.2017.00002/full} {\bibfield  {journal} {\bibinfo  {journal} {Frontiers in ICT}\ }\textbf {\bibinfo {volume} {4}},\ \bibinfo {pages} {2} (\bibinfo {year} {2017})}\BibitemShut {NoStop}%
\bibitem [{\citenamefont {Albash}(2019)}]{PhysRevA.99.042334}%
  \BibitemOpen
  \bibfield  {author} {\bibinfo {author} {\bibfnamefont {T.}~\bibnamefont {Albash}},\ }\bibfield  {title} {\bibinfo {title} {Role of nonstoquastic catalysts in quantum adiabatic optimization},\ }\href {https://doi.org/10.1103/PhysRevA.99.042334} {\bibfield  {journal} {\bibinfo  {journal} {Phys. Rev. A}\ }\textbf {\bibinfo {volume} {99}},\ \bibinfo {pages} {042334} (\bibinfo {year} {2019})}\BibitemShut {NoStop}%
\bibitem [{\citenamefont {Capecci}\ \emph {et~al.}(2026)\citenamefont {Capecci}, \citenamefont {Nagies}, \citenamefont {Varikuti},\ and\ \citenamefont {Hauke}}]{capecci2026quantum}%
  \BibitemOpen
  \bibfield  {author} {\bibinfo {author} {\bibfnamefont {C.}~\bibnamefont {Capecci}}, \bibinfo {author} {\bibfnamefont {S.}~\bibnamefont {Nagies}}, \bibinfo {author} {\bibfnamefont {N.~D.}\ \bibnamefont {Varikuti}},\ and\ \bibinfo {author} {\bibfnamefont {P.}~\bibnamefont {Hauke}},\ }\bibfield  {title} {\bibinfo {title} {Quantum resources in non-stoquastic quantum annealing},\ }\href {https://arxiv.org/abs/2606.09995} {\bibfield  {journal} {\bibinfo  {journal} {arXiv preprint arXiv:2606.09995}\ } (\bibinfo {year} {2026})}\BibitemShut {NoStop}%
\bibitem [{\citenamefont {Evers}\ and\ \citenamefont {Mirlin}(2008)}]{RevModPhys.80.1355}%
  \BibitemOpen
  \bibfield  {author} {\bibinfo {author} {\bibfnamefont {F.}~\bibnamefont {Evers}}\ and\ \bibinfo {author} {\bibfnamefont {A.~D.}\ \bibnamefont {Mirlin}},\ }\bibfield  {title} {\bibinfo {title} {Anderson transitions},\ }\href {https://doi.org/10.1103/RevModPhys.80.1355} {\bibfield  {journal} {\bibinfo  {journal} {Rev. Mod. Phys.}\ }\textbf {\bibinfo {volume} {80}},\ \bibinfo {pages} {1355} (\bibinfo {year} {2008})}\BibitemShut {NoStop}%
\bibitem [{\citenamefont {Marvian}\ and\ \citenamefont {Spekkens}(2016)}]{PhysRevA.94.052324}%
  \BibitemOpen
  \bibfield  {author} {\bibinfo {author} {\bibfnamefont {I.}~\bibnamefont {Marvian}}\ and\ \bibinfo {author} {\bibfnamefont {R.~W.}\ \bibnamefont {Spekkens}},\ }\bibfield  {title} {\bibinfo {title} {How to quantify coherence: Distinguishing speakable and unspeakable notions},\ }\href {https://doi.org/10.1103/PhysRevA.94.052324} {\bibfield  {journal} {\bibinfo  {journal} {Phys. Rev. A}\ }\textbf {\bibinfo {volume} {94}},\ \bibinfo {pages} {052324} (\bibinfo {year} {2016})}\BibitemShut {NoStop}%
\bibitem [{\citenamefont {Streltsov}\ \emph {et~al.}(2017)\citenamefont {Streltsov}, \citenamefont {Adesso},\ and\ \citenamefont {Plenio}}]{RevModPhys.89.041003}%
  \BibitemOpen
  \bibfield  {author} {\bibinfo {author} {\bibfnamefont {A.}~\bibnamefont {Streltsov}}, \bibinfo {author} {\bibfnamefont {G.}~\bibnamefont {Adesso}},\ and\ \bibinfo {author} {\bibfnamefont {M.~B.}\ \bibnamefont {Plenio}},\ }\bibfield  {title} {\bibinfo {title} {Colloquium: Quantum coherence as a resource},\ }\href {https://doi.org/10.1103/RevModPhys.89.041003} {\bibfield  {journal} {\bibinfo  {journal} {Rev. Mod. Phys.}\ }\textbf {\bibinfo {volume} {89}},\ \bibinfo {pages} {041003} (\bibinfo {year} {2017})}\BibitemShut {NoStop}%
\bibitem [{\citenamefont {Becca}\ and\ \citenamefont {Sorella}(2017)}]{Becca_Sorella_2017}%
  \BibitemOpen
  \bibfield  {author} {\bibinfo {author} {\bibfnamefont {F.}~\bibnamefont {Becca}}\ and\ \bibinfo {author} {\bibfnamefont {S.}~\bibnamefont {Sorella}},\ }\bibinfo {title} {Green’s function monte carlo},\ in\ \href@noop {} {\emph {\bibinfo {booktitle} {Quantum Monte Carlo Approaches for Correlated Systems}}}\ (\bibinfo  {publisher} {Cambridge University Press},\ \bibinfo {year} {2017})\ p.\ \bibinfo {pages} {167–188}\BibitemShut {NoStop}%
\bibitem [{\citenamefont {White}(1992)}]{PhysRevLett.69.2863}%
  \BibitemOpen
  \bibfield  {author} {\bibinfo {author} {\bibfnamefont {S.~R.}\ \bibnamefont {White}},\ }\bibfield  {title} {\bibinfo {title} {Density matrix formulation for quantum renormalization groups},\ }\href {https://doi.org/10.1103/PhysRevLett.69.2863} {\bibfield  {journal} {\bibinfo  {journal} {Phys. Rev. Lett.}\ }\textbf {\bibinfo {volume} {69}},\ \bibinfo {pages} {2863} (\bibinfo {year} {1992})}\BibitemShut {NoStop}%
\bibitem [{\citenamefont {Hamma}\ \emph {et~al.}(2005{\natexlab{a}})\citenamefont {Hamma}, \citenamefont {Ionicioiu},\ and\ \citenamefont {Zanardi}}]{PhysRevA.71.022315}%
  \BibitemOpen
  \bibfield  {author} {\bibinfo {author} {\bibfnamefont {A.}~\bibnamefont {Hamma}}, \bibinfo {author} {\bibfnamefont {R.}~\bibnamefont {Ionicioiu}},\ and\ \bibinfo {author} {\bibfnamefont {P.}~\bibnamefont {Zanardi}},\ }\bibfield  {title} {\bibinfo {title} {Bipartite entanglement and entropic boundary law in lattice spin systems},\ }\href {https://doi.org/10.1103/PhysRevA.71.022315} {\bibfield  {journal} {\bibinfo  {journal} {Phys. Rev. A}\ }\textbf {\bibinfo {volume} {71}},\ \bibinfo {pages} {022315} (\bibinfo {year} {2005}{\natexlab{a}})}\BibitemShut {NoStop}%
\bibitem [{\citenamefont {Hamma}\ \emph {et~al.}(2005{\natexlab{b}})\citenamefont {Hamma}, \citenamefont {Ionicioiu},\ and\ \citenamefont {Zanardi}}]{HAMMA200522}%
  \BibitemOpen
  \bibfield  {author} {\bibinfo {author} {\bibfnamefont {A.}~\bibnamefont {Hamma}}, \bibinfo {author} {\bibfnamefont {R.}~\bibnamefont {Ionicioiu}},\ and\ \bibinfo {author} {\bibfnamefont {P.}~\bibnamefont {Zanardi}},\ }\bibfield  {title} {\bibinfo {title} {Ground state entanglement and geometric entropy in the kitaev model},\ }\href {https://doi.org/https://doi.org/10.1016/j.physleta.2005.01.060} {\bibfield  {journal} {\bibinfo  {journal} {Physics Letters A}\ }\textbf {\bibinfo {volume} {337}},\ \bibinfo {pages} {22} (\bibinfo {year} {2005}{\natexlab{b}})}\BibitemShut {NoStop}%
\bibitem [{\citenamefont {Hangleiter}\ \emph {et~al.}(2020)\citenamefont {Hangleiter}, \citenamefont {Roth}, \citenamefont {Nagaj},\ and\ \citenamefont {Eisert}}]{doi:10.1126/sciadv.abb8341}%
  \BibitemOpen
  \bibfield  {author} {\bibinfo {author} {\bibfnamefont {D.}~\bibnamefont {Hangleiter}}, \bibinfo {author} {\bibfnamefont {I.}~\bibnamefont {Roth}}, \bibinfo {author} {\bibfnamefont {D.}~\bibnamefont {Nagaj}},\ and\ \bibinfo {author} {\bibfnamefont {J.}~\bibnamefont {Eisert}},\ }\bibfield  {title} {\bibinfo {title} {Easing the monte carlo sign problem},\ }\href {https://doi.org/10.1126/sciadv.abb8341} {\bibfield  {journal} {\bibinfo  {journal} {Science Advances}\ }\textbf {\bibinfo {volume} {6}},\ \bibinfo {pages} {eabb8341} (\bibinfo {year} {2020})}\BibitemShut {NoStop}%
\bibitem [{\citenamefont {Marvian}\ \emph {et~al.}(2019)\citenamefont {Marvian}, \citenamefont {Lidar},\ and\ \citenamefont {Hen}}]{marvian2019computational}%
  \BibitemOpen
  \bibfield  {author} {\bibinfo {author} {\bibfnamefont {M.}~\bibnamefont {Marvian}}, \bibinfo {author} {\bibfnamefont {D.~A.}\ \bibnamefont {Lidar}},\ and\ \bibinfo {author} {\bibfnamefont {I.}~\bibnamefont {Hen}},\ }\bibfield  {title} {\bibinfo {title} {On the computational complexity of curing non-stoquastic hamiltonians},\ }\href {https://www.nature.com/articles/s41467-019-09501-6} {\bibfield  {journal} {\bibinfo  {journal} {Nature communications}\ }\textbf {\bibinfo {volume} {10}},\ \bibinfo {pages} {1571} (\bibinfo {year} {2019})}\BibitemShut {NoStop}%
\bibitem [{\citenamefont {Klassen}\ \emph {et~al.}(2020)\citenamefont {Klassen}, \citenamefont {Marvian}, \citenamefont {Piddock}, \citenamefont {Ioannou}, \citenamefont {Hen},\ and\ \citenamefont {Terhal}}]{klassen2020hardness}%
  \BibitemOpen
  \bibfield  {author} {\bibinfo {author} {\bibfnamefont {J.}~\bibnamefont {Klassen}}, \bibinfo {author} {\bibfnamefont {M.}~\bibnamefont {Marvian}}, \bibinfo {author} {\bibfnamefont {S.}~\bibnamefont {Piddock}}, \bibinfo {author} {\bibfnamefont {M.}~\bibnamefont {Ioannou}}, \bibinfo {author} {\bibfnamefont {I.}~\bibnamefont {Hen}},\ and\ \bibinfo {author} {\bibfnamefont {B.~M.}\ \bibnamefont {Terhal}},\ }\bibfield  {title} {\bibinfo {title} {Hardness and ease of curing the sign problem for two-local qubit hamiltonians},\ }\href {https://epubs.siam.org/doi/10.1137/19M1287511} {\bibfield  {journal} {\bibinfo  {journal} {SIAM Journal on Computing}\ }\textbf {\bibinfo {volume} {49}},\ \bibinfo {pages} {1332} (\bibinfo {year} {2020})}\BibitemShut {NoStop}%
\bibitem [{sup()}]{supplemental}%
  \BibitemOpen
  \href@noop {} {}\bibinfo {note} {See Supplemental Material, which includes Refs.~\cite{PhysRevA.53.2046, giovannetti2004quantum, giovannetti2006quantum, RevModPhys.89.035002, RevModPhys.90.035005, bennett1993teleporting, shor1995scheme, RevModPhys.87.307, roffe2019quantum, perez2006matrix, verstraete2008matrix, varikuti2022out, PhysRevLett.120.040402, gottesman1998heisenberg, gottesman1997stabilizercodesquantumerror, collins2006integration}}\BibitemShut {NoStop}%
\bibitem [{\citenamefont {Schollw{\"o}ck}(2011)}]{schollwock2011density}%
  \BibitemOpen
  \bibfield  {author} {\bibinfo {author} {\bibfnamefont {U.}~\bibnamefont {Schollw{\"o}ck}},\ }\bibfield  {title} {\bibinfo {title} {The density-matrix renormalization group in the age of matrix product states},\ }\href {https://doi.org/10.1016/j.aop.2010.09.012} {\bibfield  {journal} {\bibinfo  {journal} {Annals of physics}\ }\textbf {\bibinfo {volume} {326}},\ \bibinfo {pages} {96} (\bibinfo {year} {2011})}\BibitemShut {NoStop}%
\bibitem [{\citenamefont {Hastings}(2015)}]{10.1063/1.4936216}%
  \BibitemOpen
  \bibfield  {author} {\bibinfo {author} {\bibfnamefont {M.~B.}\ \bibnamefont {Hastings}},\ }\bibfield  {title} {\bibinfo {title} {How quantum are non-negative wavefunctions?},\ }\href {https://doi.org/10.1063/1.4936216} {\bibfield  {journal} {\bibinfo  {journal} {Journal of Mathematical Physics}\ }\textbf {\bibinfo {volume} {57}},\ \bibinfo {pages} {015210} (\bibinfo {year} {2015})}\BibitemShut {NoStop}%
\bibitem [{\citenamefont {Hastings}(2007)}]{Hastings_2007}%
  \BibitemOpen
  \bibfield  {author} {\bibinfo {author} {\bibfnamefont {M.~B.}\ \bibnamefont {Hastings}},\ }\bibfield  {title} {\bibinfo {title} {An area law for one-dimensional quantum systems},\ }\href {https://doi.org/10.1088/1742-5468/2007/08/P08024} {\bibfield  {journal} {\bibinfo  {journal} {Journal of Statistical Mechanics: Theory and Experiment}\ }\textbf {\bibinfo {volume} {2007}},\ \bibinfo {pages} {P08024} (\bibinfo {year} {2007})}\BibitemShut {NoStop}%
\bibitem [{\citenamefont {Gottesman}\ and\ \citenamefont {Hastings}(2010)}]{Gottesman_2010}%
  \BibitemOpen
  \bibfield  {author} {\bibinfo {author} {\bibfnamefont {D.}~\bibnamefont {Gottesman}}\ and\ \bibinfo {author} {\bibfnamefont {M.~B.}\ \bibnamefont {Hastings}},\ }\bibfield  {title} {\bibinfo {title} {Entanglement versus gap for one-dimensional spin systems},\ }\href {https://doi.org/10.1088/1367-2630/12/2/025002} {\bibfield  {journal} {\bibinfo  {journal} {New Journal of Physics}\ }\textbf {\bibinfo {volume} {12}},\ \bibinfo {pages} {025002} (\bibinfo {year} {2010})}\BibitemShut {NoStop}%
\bibitem [{\citenamefont {Eisert}\ \emph {et~al.}(2010)\citenamefont {Eisert}, \citenamefont {Cramer},\ and\ \citenamefont {Plenio}}]{RevModPhys.82.277}%
  \BibitemOpen
  \bibfield  {author} {\bibinfo {author} {\bibfnamefont {J.}~\bibnamefont {Eisert}}, \bibinfo {author} {\bibfnamefont {M.}~\bibnamefont {Cramer}},\ and\ \bibinfo {author} {\bibfnamefont {M.~B.}\ \bibnamefont {Plenio}},\ }\bibfield  {title} {\bibinfo {title} {Colloquium: Area laws for the entanglement entropy},\ }\href {https://doi.org/10.1103/RevModPhys.82.277} {\bibfield  {journal} {\bibinfo  {journal} {Rev. Mod. Phys.}\ }\textbf {\bibinfo {volume} {82}},\ \bibinfo {pages} {277} (\bibinfo {year} {2010})}\BibitemShut {NoStop}%
\bibitem [{\citenamefont {Bittel}\ and\ \citenamefont {Leone}(2026)}]{Bittel2026operational}%
  \BibitemOpen
  \bibfield  {author} {\bibinfo {author} {\bibfnamefont {L.}~\bibnamefont {Bittel}}\ and\ \bibinfo {author} {\bibfnamefont {L.}~\bibnamefont {Leone}},\ }\bibfield  {title} {\bibinfo {title} {Operational interpretation of the {S}tabilizer {E}ntropy},\ }\href {https://doi.org/10.22331/q-2026-04-15-2069} {\bibfield  {journal} {\bibinfo  {journal} {{Quantum}}\ }\textbf {\bibinfo {volume} {10}},\ \bibinfo {pages} {2069} (\bibinfo {year} {2026})}\BibitemShut {NoStop}%
\bibitem [{\citenamefont {Oliviero}\ \emph {et~al.}(2022)\citenamefont {Oliviero}, \citenamefont {Leone},\ and\ \citenamefont {Hamma}}]{PhysRevA.106.042426}%
  \BibitemOpen
  \bibfield  {author} {\bibinfo {author} {\bibfnamefont {S.~F.~E.}\ \bibnamefont {Oliviero}}, \bibinfo {author} {\bibfnamefont {L.}~\bibnamefont {Leone}},\ and\ \bibinfo {author} {\bibfnamefont {A.}~\bibnamefont {Hamma}},\ }\bibfield  {title} {\bibinfo {title} {Magic-state resource theory for the ground state of the transverse-field ising model},\ }\href {https://doi.org/10.1103/PhysRevA.106.042426} {\bibfield  {journal} {\bibinfo  {journal} {Phys. Rev. A}\ }\textbf {\bibinfo {volume} {106}},\ \bibinfo {pages} {042426} (\bibinfo {year} {2022})}\BibitemShut {NoStop}%
\bibitem [{\citenamefont {Lami}\ and\ \citenamefont {Collura}(2023)}]{PhysRevLett.131.180401}%
  \BibitemOpen
  \bibfield  {author} {\bibinfo {author} {\bibfnamefont {G.}~\bibnamefont {Lami}}\ and\ \bibinfo {author} {\bibfnamefont {M.}~\bibnamefont {Collura}},\ }\bibfield  {title} {\bibinfo {title} {Nonstabilizerness via perfect pauli sampling of matrix product states},\ }\href {https://doi.org/10.1103/PhysRevLett.131.180401} {\bibfield  {journal} {\bibinfo  {journal} {Phys. Rev. Lett.}\ }\textbf {\bibinfo {volume} {131}},\ \bibinfo {pages} {180401} (\bibinfo {year} {2023})}\BibitemShut {NoStop}%
\bibitem [{\citenamefont {Haake}\ \emph {et~al.}(2018)\citenamefont {Haake}, \citenamefont {Gnutzmann},\ and\ \citenamefont {Ku{\'s}}}]{haake1991quantum}%
  \BibitemOpen
  \bibfield  {author} {\bibinfo {author} {\bibfnamefont {F.}~\bibnamefont {Haake}}, \bibinfo {author} {\bibfnamefont {S.}~\bibnamefont {Gnutzmann}},\ and\ \bibinfo {author} {\bibfnamefont {M.}~\bibnamefont {Ku{\'s}}},\ }\href {https://doi.org/10.1007/978-3-319-97580-1} {\emph {\bibinfo {title} {Quantum Signatures of Chaos}}},\ \bibinfo {edition} {4th}\ ed.,\ Springer Series in Synergetics\ (\bibinfo  {publisher} {Springer},\ \bibinfo {year} {2018})\BibitemShut {NoStop}%
\bibitem [{\citenamefont {van Bemmel}\ \emph {et~al.}(1994)\citenamefont {van Bemmel}, \citenamefont {ten Haaf}, \citenamefont {van Saarloos}, \citenamefont {van Leeuwen},\ and\ \citenamefont {An}}]{PhysRevLett.72.2442}%
  \BibitemOpen
  \bibfield  {author} {\bibinfo {author} {\bibfnamefont {H.~J.~M.}\ \bibnamefont {van Bemmel}}, \bibinfo {author} {\bibfnamefont {D.~F.~B.}\ \bibnamefont {ten Haaf}}, \bibinfo {author} {\bibfnamefont {W.}~\bibnamefont {van Saarloos}}, \bibinfo {author} {\bibfnamefont {J.~M.~J.}\ \bibnamefont {van Leeuwen}},\ and\ \bibinfo {author} {\bibfnamefont {G.}~\bibnamefont {An}},\ }\bibfield  {title} {\bibinfo {title} {Fixed-node quantum monte carlo method for lattice fermions},\ }\href {https://doi.org/10.1103/PhysRevLett.72.2442} {\bibfield  {journal} {\bibinfo  {journal} {Phys. Rev. Lett.}\ }\textbf {\bibinfo {volume} {72}},\ \bibinfo {pages} {2442} (\bibinfo {year} {1994})}\BibitemShut {NoStop}%
\bibitem [{\citenamefont {ten Haaf}\ \emph {et~al.}(1995)\citenamefont {ten Haaf}, \citenamefont {van Bemmel}, \citenamefont {van Leeuwen}, \citenamefont {van Saarloos},\ and\ \citenamefont {Ceperley}}]{PhysRevB.51.13039}%
  \BibitemOpen
  \bibfield  {author} {\bibinfo {author} {\bibfnamefont {D.~F.~B.}\ \bibnamefont {ten Haaf}}, \bibinfo {author} {\bibfnamefont {H.~J.~M.}\ \bibnamefont {van Bemmel}}, \bibinfo {author} {\bibfnamefont {J.~M.~J.}\ \bibnamefont {van Leeuwen}}, \bibinfo {author} {\bibfnamefont {W.}~\bibnamefont {van Saarloos}},\ and\ \bibinfo {author} {\bibfnamefont {D.~M.}\ \bibnamefont {Ceperley}},\ }\bibfield  {title} {\bibinfo {title} {Proof for an upper bound in fixed-node monte carlo for lattice fermions},\ }\href {https://doi.org/10.1103/PhysRevB.51.13039} {\bibfield  {journal} {\bibinfo  {journal} {Phys. Rev. B}\ }\textbf {\bibinfo {volume} {51}},\ \bibinfo {pages} {13039} (\bibinfo {year} {1995})}\BibitemShut {NoStop}%
\bibitem [{\citenamefont {Torlai}\ \emph {et~al.}(2020)\citenamefont {Torlai}, \citenamefont {Carrasquilla}, \citenamefont {Fishman}, \citenamefont {Melko},\ and\ \citenamefont {Fisher}}]{PhysRevResearch.2.032060}%
  \BibitemOpen
  \bibfield  {author} {\bibinfo {author} {\bibfnamefont {G.}~\bibnamefont {Torlai}}, \bibinfo {author} {\bibfnamefont {J.}~\bibnamefont {Carrasquilla}}, \bibinfo {author} {\bibfnamefont {M.~T.}\ \bibnamefont {Fishman}}, \bibinfo {author} {\bibfnamefont {R.~G.}\ \bibnamefont {Melko}},\ and\ \bibinfo {author} {\bibfnamefont {M.~P.~A.}\ \bibnamefont {Fisher}},\ }\bibfield  {title} {\bibinfo {title} {Wave-function positivization via automatic differentiation},\ }\href {https://doi.org/10.1103/PhysRevResearch.2.032060} {\bibfield  {journal} {\bibinfo  {journal} {Phys. Rev. Res.}\ }\textbf {\bibinfo {volume} {2}},\ \bibinfo {pages} {032060} (\bibinfo {year} {2020})}\BibitemShut {NoStop}%
\bibitem [{\citenamefont {Briegel}\ \emph {et~al.}(2009)\citenamefont {Briegel}, \citenamefont {Browne}, \citenamefont {D{\"u}r}, \citenamefont {Raussendorf},\ and\ \citenamefont {Van~den Nest}}]{briegel2009measurement}%
  \BibitemOpen
  \bibfield  {author} {\bibinfo {author} {\bibfnamefont {H.~J.}\ \bibnamefont {Briegel}}, \bibinfo {author} {\bibfnamefont {D.~E.}\ \bibnamefont {Browne}}, \bibinfo {author} {\bibfnamefont {W.}~\bibnamefont {D{\"u}r}}, \bibinfo {author} {\bibfnamefont {R.}~\bibnamefont {Raussendorf}},\ and\ \bibinfo {author} {\bibfnamefont {M.}~\bibnamefont {Van~den Nest}},\ }\bibfield  {title} {\bibinfo {title} {Measurement-based quantum computation},\ }\href {https://www.nature.com/articles/nphys1157} {\bibfield  {journal} {\bibinfo  {journal} {Nature Physics}\ }\textbf {\bibinfo {volume} {5}},\ \bibinfo {pages} {19} (\bibinfo {year} {2009})}\BibitemShut {NoStop}%
\bibitem [{\citenamefont {Bennett}\ \emph {et~al.}(1996)\citenamefont {Bennett}, \citenamefont {Bernstein}, \citenamefont {Popescu},\ and\ \citenamefont {Schumacher}}]{PhysRevA.53.2046}%
  \BibitemOpen
  \bibfield  {author} {\bibinfo {author} {\bibfnamefont {C.~H.}\ \bibnamefont {Bennett}}, \bibinfo {author} {\bibfnamefont {H.~J.}\ \bibnamefont {Bernstein}}, \bibinfo {author} {\bibfnamefont {S.}~\bibnamefont {Popescu}},\ and\ \bibinfo {author} {\bibfnamefont {B.}~\bibnamefont {Schumacher}},\ }\bibfield  {title} {\bibinfo {title} {Concentrating partial entanglement by local operations},\ }\href {https://doi.org/10.1103/PhysRevA.53.2046} {\bibfield  {journal} {\bibinfo  {journal} {Phys. Rev. A}\ }\textbf {\bibinfo {volume} {53}},\ \bibinfo {pages} {2046} (\bibinfo {year} {1996})}\BibitemShut {NoStop}%
\bibitem [{\citenamefont {Giovannetti}\ \emph {et~al.}(2004)\citenamefont {Giovannetti}, \citenamefont {Lloyd},\ and\ \citenamefont {Maccone}}]{giovannetti2004quantum}%
  \BibitemOpen
  \bibfield  {author} {\bibinfo {author} {\bibfnamefont {V.}~\bibnamefont {Giovannetti}}, \bibinfo {author} {\bibfnamefont {S.}~\bibnamefont {Lloyd}},\ and\ \bibinfo {author} {\bibfnamefont {L.}~\bibnamefont {Maccone}},\ }\bibfield  {title} {\bibinfo {title} {Quantum-enhanced measurements: beating the standard quantum limit},\ }\href {https://doi.org/10.1126/science.1104149} {\bibfield  {journal} {\bibinfo  {journal} {Science}\ }\textbf {\bibinfo {volume} {306}},\ \bibinfo {pages} {1330} (\bibinfo {year} {2004})}\BibitemShut {NoStop}%
\bibitem [{\citenamefont {Giovannetti}\ \emph {et~al.}(2006)\citenamefont {Giovannetti}, \citenamefont {Lloyd},\ and\ \citenamefont {Maccone}}]{giovannetti2006quantum}%
  \BibitemOpen
  \bibfield  {author} {\bibinfo {author} {\bibfnamefont {V.}~\bibnamefont {Giovannetti}}, \bibinfo {author} {\bibfnamefont {S.}~\bibnamefont {Lloyd}},\ and\ \bibinfo {author} {\bibfnamefont {L.}~\bibnamefont {Maccone}},\ }\bibfield  {title} {\bibinfo {title} {Quantum metrology},\ }\href {https://doi.org/10.1103/PhysRevLett.96.010401} {\bibfield  {journal} {\bibinfo  {journal} {Physical review letters}\ }\textbf {\bibinfo {volume} {96}},\ \bibinfo {pages} {010401} (\bibinfo {year} {2006})}\BibitemShut {NoStop}%
\bibitem [{\citenamefont {Degen}\ \emph {et~al.}(2017)\citenamefont {Degen}, \citenamefont {Reinhard},\ and\ \citenamefont {Cappellaro}}]{RevModPhys.89.035002}%
  \BibitemOpen
  \bibfield  {author} {\bibinfo {author} {\bibfnamefont {C.~L.}\ \bibnamefont {Degen}}, \bibinfo {author} {\bibfnamefont {F.}~\bibnamefont {Reinhard}},\ and\ \bibinfo {author} {\bibfnamefont {P.}~\bibnamefont {Cappellaro}},\ }\bibfield  {title} {\bibinfo {title} {Quantum sensing},\ }\href {https://doi.org/10.1103/RevModPhys.89.035002} {\bibfield  {journal} {\bibinfo  {journal} {Rev. Mod. Phys.}\ }\textbf {\bibinfo {volume} {89}},\ \bibinfo {pages} {035002} (\bibinfo {year} {2017})}\BibitemShut {NoStop}%
\bibitem [{\citenamefont {Pezz\`e}\ \emph {et~al.}(2018)\citenamefont {Pezz\`e}, \citenamefont {Smerzi}, \citenamefont {Oberthaler}, \citenamefont {Schmied},\ and\ \citenamefont {Treutlein}}]{RevModPhys.90.035005}%
  \BibitemOpen
  \bibfield  {author} {\bibinfo {author} {\bibfnamefont {L.}~\bibnamefont {Pezz\`e}}, \bibinfo {author} {\bibfnamefont {A.}~\bibnamefont {Smerzi}}, \bibinfo {author} {\bibfnamefont {M.~K.}\ \bibnamefont {Oberthaler}}, \bibinfo {author} {\bibfnamefont {R.}~\bibnamefont {Schmied}},\ and\ \bibinfo {author} {\bibfnamefont {P.}~\bibnamefont {Treutlein}},\ }\bibfield  {title} {\bibinfo {title} {Quantum metrology with nonclassical states of atomic ensembles},\ }\href {https://doi.org/10.1103/RevModPhys.90.035005} {\bibfield  {journal} {\bibinfo  {journal} {Rev. Mod. Phys.}\ }\textbf {\bibinfo {volume} {90}},\ \bibinfo {pages} {035005} (\bibinfo {year} {2018})}\BibitemShut {NoStop}%
\bibitem [{\citenamefont {Bennett}\ \emph {et~al.}(1993)\citenamefont {Bennett}, \citenamefont {Brassard}, \citenamefont {Cr{\'e}peau}, \citenamefont {Jozsa}, \citenamefont {Peres},\ and\ \citenamefont {Wootters}}]{bennett1993teleporting}%
  \BibitemOpen
  \bibfield  {author} {\bibinfo {author} {\bibfnamefont {C.~H.}\ \bibnamefont {Bennett}}, \bibinfo {author} {\bibfnamefont {G.}~\bibnamefont {Brassard}}, \bibinfo {author} {\bibfnamefont {C.}~\bibnamefont {Cr{\'e}peau}}, \bibinfo {author} {\bibfnamefont {R.}~\bibnamefont {Jozsa}}, \bibinfo {author} {\bibfnamefont {A.}~\bibnamefont {Peres}},\ and\ \bibinfo {author} {\bibfnamefont {W.~K.}\ \bibnamefont {Wootters}},\ }\bibfield  {title} {\bibinfo {title} {Teleporting an unknown quantum state via dual classical and einstein-podolsky-rosen channels},\ }\href {https://doi.org/10.1103/PhysRevLett.70.1895} {\bibfield  {journal} {\bibinfo  {journal} {Phys. Rev. Lett.}\ }\textbf {\bibinfo {volume} {70}},\ \bibinfo {pages} {1895} (\bibinfo {year} {1993})}\BibitemShut {NoStop}%
\bibitem [{\citenamefont {Shor}(1995)}]{shor1995scheme}%
  \BibitemOpen
  \bibfield  {author} {\bibinfo {author} {\bibfnamefont {P.~W.}\ \bibnamefont {Shor}},\ }\bibfield  {title} {\bibinfo {title} {Scheme for reducing decoherence in quantum computer memory},\ }\href {https://doi.org/10.1103/PhysRevA.52.R2493} {\bibfield  {journal} {\bibinfo  {journal} {Phys. Rev. A}\ }\textbf {\bibinfo {volume} {52}},\ \bibinfo {pages} {R2493(R)} (\bibinfo {year} {1995})}\BibitemShut {NoStop}%
\bibitem [{\citenamefont {Terhal}(2015)}]{RevModPhys.87.307}%
  \BibitemOpen
  \bibfield  {author} {\bibinfo {author} {\bibfnamefont {B.~M.}\ \bibnamefont {Terhal}},\ }\bibfield  {title} {\bibinfo {title} {Quantum error correction for quantum memories},\ }\href {https://doi.org/10.1103/RevModPhys.87.307} {\bibfield  {journal} {\bibinfo  {journal} {Rev. Mod. Phys.}\ }\textbf {\bibinfo {volume} {87}},\ \bibinfo {pages} {307} (\bibinfo {year} {2015})}\BibitemShut {NoStop}%
\bibitem [{\citenamefont {Roffe}(2019)}]{roffe2019quantum}%
  \BibitemOpen
  \bibfield  {author} {\bibinfo {author} {\bibfnamefont {J.}~\bibnamefont {Roffe}},\ }\bibfield  {title} {\bibinfo {title} {Quantum error correction: an introductory guide},\ }\href {https://doi.org/10.1080/00107514.2019.1667078} {\bibfield  {journal} {\bibinfo  {journal} {Contemporary Physics}\ }\textbf {\bibinfo {volume} {60}},\ \bibinfo {pages} {226} (\bibinfo {year} {2019})}\BibitemShut {NoStop}%
\bibitem [{\citenamefont {Perez-Garcia}\ \emph {et~al.}(2007)\citenamefont {Perez-Garcia}, \citenamefont {Verstraete}, \citenamefont {Wolf},\ and\ \citenamefont {Cirac}}]{perez2006matrix}%
  \BibitemOpen
  \bibfield  {author} {\bibinfo {author} {\bibfnamefont {D.}~\bibnamefont {Perez-Garcia}}, \bibinfo {author} {\bibfnamefont {F.}~\bibnamefont {Verstraete}}, \bibinfo {author} {\bibfnamefont {M.~M.}\ \bibnamefont {Wolf}},\ and\ \bibinfo {author} {\bibfnamefont {J.~I.}\ \bibnamefont {Cirac}},\ }\bibfield  {title} {\bibinfo {title} {Matrix product state representations},\ }\href {https://doi.org/10.26421/QIC7.5-6-1} {\bibfield  {journal} {\bibinfo  {journal} {Quantum Info. Comput.}\ }\textbf {\bibinfo {volume} {7}},\ \bibinfo {pages} {401–430} (\bibinfo {year} {2007})}\BibitemShut {NoStop}%
\bibitem [{\citenamefont {Verstraete}\ \emph {et~al.}(2008)\citenamefont {Verstraete}, \citenamefont {Murg},\ and\ \citenamefont {Cirac}}]{verstraete2008matrix}%
  \BibitemOpen
  \bibfield  {author} {\bibinfo {author} {\bibfnamefont {F.}~\bibnamefont {Verstraete}}, \bibinfo {author} {\bibfnamefont {V.}~\bibnamefont {Murg}},\ and\ \bibinfo {author} {\bibfnamefont {J.~I.}\ \bibnamefont {Cirac}},\ }\bibfield  {title} {\bibinfo {title} {Matrix product states, projected entangled pair states, and variational renormalization group methods for quantum spin systems},\ }\href {https://doi.org/10.1080/14789940801912366} {\bibfield  {journal} {\bibinfo  {journal} {Advances in physics}\ }\textbf {\bibinfo {volume} {57}},\ \bibinfo {pages} {143} (\bibinfo {year} {2008})}\BibitemShut {NoStop}%
\bibitem [{\citenamefont {Varikuti}\ and\ \citenamefont {Madhok}(2024)}]{varikuti2022out}%
  \BibitemOpen
  \bibfield  {author} {\bibinfo {author} {\bibfnamefont {N.~D.}\ \bibnamefont {Varikuti}}\ and\ \bibinfo {author} {\bibfnamefont {V.}~\bibnamefont {Madhok}},\ }\bibfield  {title} {\bibinfo {title} {{Out-of-time ordered correlators in kicked coupled tops: Information scrambling in mixed phase space and the role of conserved quantities}},\ }\href {https://doi.org/10.1063/5.0191140} {\bibfield  {journal} {\bibinfo  {journal} {Chaos}\ }\textbf {\bibinfo {volume} {34}},\ \bibinfo {pages} {063124} (\bibinfo {year} {2024})}\BibitemShut {NoStop}%
\bibitem [{\citenamefont {G{\"a}rttner}\ \emph {et~al.}(2018)\citenamefont {G{\"a}rttner}, \citenamefont {Hauke},\ and\ \citenamefont {Rey}}]{PhysRevLett.120.040402}%
  \BibitemOpen
  \bibfield  {author} {\bibinfo {author} {\bibfnamefont {M.}~\bibnamefont {G{\"a}rttner}}, \bibinfo {author} {\bibfnamefont {P.}~\bibnamefont {Hauke}},\ and\ \bibinfo {author} {\bibfnamefont {A.~M.}\ \bibnamefont {Rey}},\ }\bibfield  {title} {\bibinfo {title} {Relating out-of-time-order correlations to entanglement via multiple-quantum coherences},\ }\href {https://doi.org/10.1103/PhysRevLett.120.040402} {\bibfield  {journal} {\bibinfo  {journal} {Phys. Rev. Lett.}\ }\textbf {\bibinfo {volume} {120}},\ \bibinfo {pages} {040402} (\bibinfo {year} {2018})}\BibitemShut {NoStop}%
\bibitem [{\citenamefont {Gottesman}(1998)}]{gottesman1998heisenberg}%
  \BibitemOpen
  \bibfield  {author} {\bibinfo {author} {\bibfnamefont {D.}~\bibnamefont {Gottesman}},\ }\bibfield  {title} {\bibinfo {title} {The heisenberg representation of quantum computers},\ }\href {https://arxiv.org/abs/quant-ph/9807006} {\bibfield  {journal} {\bibinfo  {journal} {arXiv preprint quant-ph/9807006}\ } (\bibinfo {year} {1998})}\BibitemShut {NoStop}%
\bibitem [{\citenamefont {Gottesman}(1997)}]{gottesman1997stabilizercodesquantumerror}%
  \BibitemOpen
  \bibfield  {author} {\bibinfo {author} {\bibfnamefont {D.}~\bibnamefont {Gottesman}},\ }\href {https://arxiv.org/abs/quant-ph/9705052} {\bibinfo {title} {Stabilizer codes and quantum error correction}} (\bibinfo {year} {1997}),\ \Eprint {https://arxiv.org/abs/quant-ph/9705052} {arXiv:quant-ph/9705052 [quant-ph]} \BibitemShut {NoStop}%
\bibitem [{\citenamefont {Collins}\ and\ \citenamefont {{\'S}niady}(2006)}]{collins2006integration}%
  \BibitemOpen
  \bibfield  {author} {\bibinfo {author} {\bibfnamefont {B.}~\bibnamefont {Collins}}\ and\ \bibinfo {author} {\bibfnamefont {P.}~\bibnamefont {{\'S}niady}},\ }\bibfield  {title} {\bibinfo {title} {Integration with respect to the haar measure on unitary, orthogonal and symplectic group},\ }\href {https://link.springer.com/article/10.1007/s00220-006-1554-3} {\bibfield  {journal} {\bibinfo  {journal} {Communications in Mathematical Physics}\ }\textbf {\bibinfo {volume} {264}},\ \bibinfo {pages} {773} (\bibinfo {year} {2006})}\BibitemShut {NoStop}%
\bibitem [{\citenamefont {Hein}\ \emph {et~al.}(2006)\citenamefont {Hein}, \citenamefont {Dür}, \citenamefont {Eisert}, \citenamefont {Raussendorf}, \citenamefont {den Nest},\ and\ \citenamefont {Briegel}}]{hein2006entanglementgraphstatesapplications}%
  \BibitemOpen
  \bibfield  {author} {\bibinfo {author} {\bibfnamefont {M.}~\bibnamefont {Hein}}, \bibinfo {author} {\bibfnamefont {W.}~\bibnamefont {Dür}}, \bibinfo {author} {\bibfnamefont {J.}~\bibnamefont {Eisert}}, \bibinfo {author} {\bibfnamefont {R.}~\bibnamefont {Raussendorf}}, \bibinfo {author} {\bibfnamefont {M.~V.}\ \bibnamefont {den Nest}},\ and\ \bibinfo {author} {\bibfnamefont {H.~J.}\ \bibnamefont {Briegel}},\ }\href {https://arxiv.org/abs/quant-ph/0602096} {\bibinfo {title} {Entanglement in graph states and its applications}} (\bibinfo {year} {2006}),\ \Eprint {https://arxiv.org/abs/quant-ph/0602096} {arXiv:quant-ph/0602096 [quant-ph]} \BibitemShut {NoStop}%
\bibitem [{\citenamefont {Aaronson}\ and\ \citenamefont {Gottesman}(2004)}]{aaronson2004improved}%
  \BibitemOpen
  \bibfield  {author} {\bibinfo {author} {\bibfnamefont {S.}~\bibnamefont {Aaronson}}\ and\ \bibinfo {author} {\bibfnamefont {D.}~\bibnamefont {Gottesman}},\ }\bibfield  {title} {\bibinfo {title} {Improved simulation of stabilizer circuits},\ }\href {https://doi.org/10.1103/PhysRevA.70.052328} {\bibfield  {journal} {\bibinfo  {journal} {Phys. Rev. A}\ }\textbf {\bibinfo {volume} {70}},\ \bibinfo {pages} {052328} (\bibinfo {year} {2004})}\BibitemShut {NoStop}%
\end{thebibliography}%
\section*{End Matter}


\emph{Inverse participation ratio.---}A complementary marker for the complexity of a ground state, motivated by quantum statistical physics, is the inverse participation ratio (IPR)~\cite{RevModPhys.80.1355}. In a given reference basis $\{|i\rangle\}_{i=0}^{2^{N}-1}$, it reads $ \mathrm{IPR}(|\psi\rangle)=\sum_{i=0}^{2^{N}-1}\left|\langle i|\psi\rangle\right|^{4}$,  from which one defines the second participation entropy, normalized per site,
\begin{eqnarray}
\mathrm{I}_{2}(|\psi\rangle)=-\frac{1}{N}\log\left[\mathrm{IPR}(|\psi\rangle)\right].
\end{eqnarray}
We evaluate $\mathrm{I}_{2}$ in the Hadamard basis $\{|+\rangle,|-\rangle\}^{\otimes N}$, which contains $|\bm{+}\rangle$, the approximate ground state at $p_{\mathrm{ns}}=0$. In this basis, $\mathrm{I}_{2}$ vanishes when the ground state coincides with $|\bm{+}\rangle$ and grows as it delocalizes, reaching $\log 2$ for a state spread uniformly over the basis. The IPR is closely related to the fractal dimension, which provides an information-theoretic characterization of a quantum state's delocalization with respect to a chosen reference basis. It is defined as $
\mathrm{D}_2 = \lim_{\mathcal{D}\to \infty}\mathrm{I}_2/\log \mathcal{D}
$. Here, $\mathrm{D}_2=1$ corresponds to a fully delocalized state, $0<\mathrm{D}_2 <1$ indicates a multifractal state, and $\mathrm{D}_2=0$ characterizes a localized state.

\begin{figure}
\includegraphics[scale=0.45]{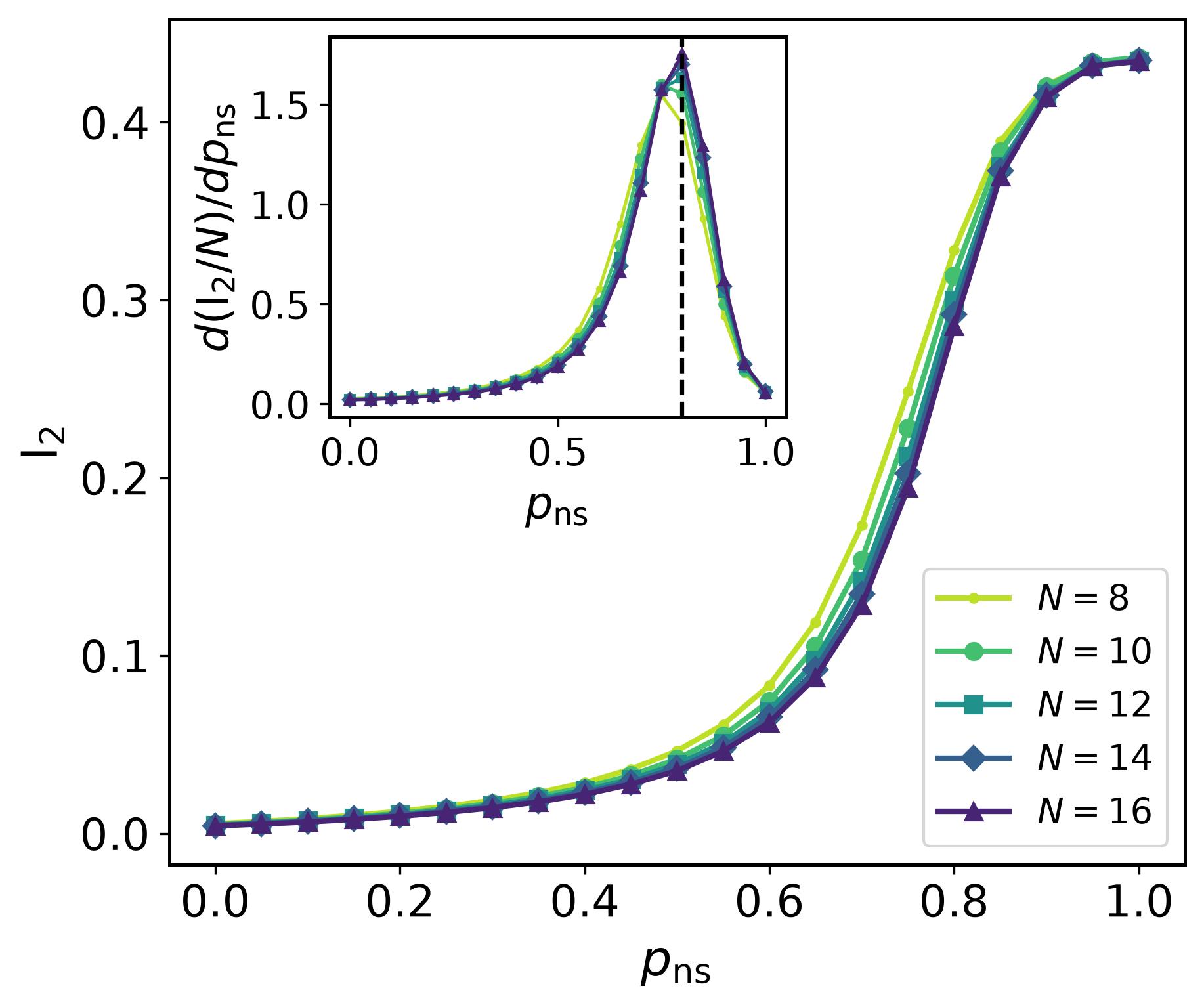}
\caption{Normalized second participation entropy, $I_2$, of the ground state of the random Hamiltonians in main-text Eq.~\eqref{ham-main} as a function of the degree of nonstoquasticity $p_{\mathrm{ns}}$, for $k=4$ and different system sizes. The reference basis is the Hadamard basis $\{\ket{+}, \ket{-}\}^{\otimes N}$. The data nearly coincide for all $N$. Near $p_{\mathrm{ns}}=0$, $I_2$ stays close to zero as the true ground state remains close to $|\bm{+}\rangle$. As $p_{\mathrm{ns}}$ is increased, $I_2$ grows. The values in the range $0<I_2<\log 2$ indicate multifractality. The inset shows the gradient of the data, which peaks near $P_{\mathrm{ns}}\approx 0.8$, indicating a crossover from localization to delocalization in the reference basis, consistent with the behavior of other resources.}
\label{fig:IPR}
\end{figure}

For the numerical simulations, reported in Fig.~\ref{fig:IPR}, we focus on $4$-local Hamiltonians, averaging over $O(10^{3})$ realizations. Near the stoquastic limit, $\mathrm{I}_{2}$ approaches zero, indicating strong localization in the Hadamard basis, consistent with $|\bm{+}\rangle$ providing a good approximation to the ground state.
It slowly increases as the Hamiltonian is tuned away from that limit, passes a sharp crossover at a sizeable value of $p_\mathrm{ns}$ (marked by a pronounced peak of the first derivative of $\mathrm{I}_{2}$, see inset), and saturates in the nonstoquastic regime. This behavior closely mirrors that of entanglement entropy, nonstabilizerness, the wave-function sign, and the projector QMC simulation accuracy reported in the main text.

This behavior can be qualitatively explained from the analytical construction of Appendix~\ref{app-a}, where we showed that the ground state of a typical stoquastic Hamiltonian can be approximated with $\ket{\bm{+}}=\ket{+}^{\otimes N}$. Our numerical analysis reinforces this fact, as reflected in a very small value of $\mathrm{I}_{2}$ near $p_{\mathrm{ns}}=0$.
As the Hamiltonian becomes increasingly nonstoquastic, $|\bm{+}\rangle$ becomes a less reliable approximation to the ground state, which therefore delocalizes over the $\{|+\rangle,|-\rangle\}^{\otimes N}$ basis, leading to a growing $\mathrm{I}_{2}$.

The data for $\mathrm{I}_2/N$ shown in Fig.~\ref{fig:IPR} collapse cleanly across system sizes, especially at small and at large $p_\mathrm{ns}$, indicating that the unnormalized participation entropy scales linearly with system size. This result moreover suggests that the fractal dimension $\mathrm{D}_2$ nearly vanishes at $p_\mathrm{ns}=0$ (localized system) while it assumes a value characteristic of multifractality near the non-stoquastic limit.


\begin{figure}
\includegraphics[scale=0.2325]{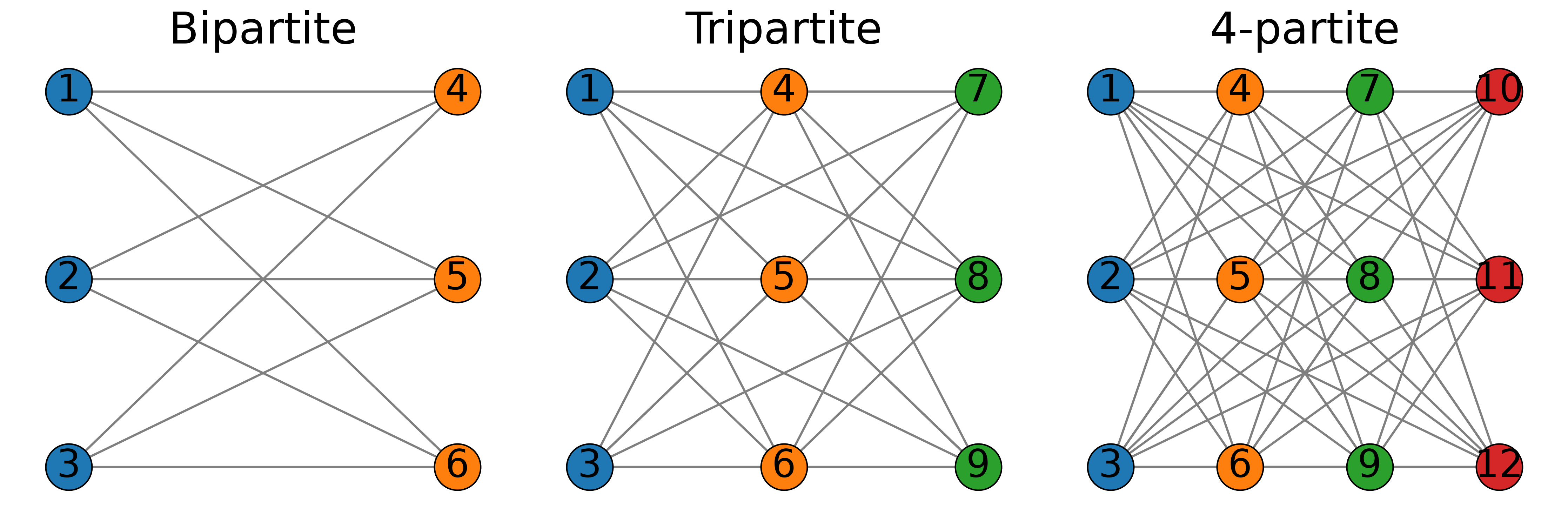}
\caption{Multipartite graphs with two, three, and four partitions (from left to right). Vertices belonging to the same partition are assigned the same color, while vertices from different partitions are partly connected. The vertices are labeled sequentially within each graph.}
\label{fig:graphs}
\end{figure}

\emph{Sign problem in Graph states.---}Here, we prove our result on the sign structure of graph states. Every stabilizer state is the ground state of a Hamiltonian built from its own stabilizer generators, $H=-\sum_{i}g_{i}$, and graph states are a special case. Curing the sign problem of such a Hamiltonian therefore amounts to removing the signs of the state itself.
While this is a restricted setting, it permits us to extend the class of ground states where the quantum resources and sign show a less clearly correlated behavior than for the generic nonstoquastic Hamiltonians studied numerically.

\textit{Graph states} are associated with a graph $G=(V,E)$, where $V$ and $E$ denote the sets of vertices respectively edges. Each vertex represents a qubit, while each edge represents a $CZ$ entangling operation. The associated state is defined as~\cite{hein2006entanglementgraphstatesapplications}
\begin{equation}
|G\rangle =
\prod_{(i,j)\in E} CZ_{ij}\,
|+\rangle^{\otimes |V|},
\end{equation}
where $|+\rangle = ({|0\rangle+|1\rangle})/{\sqrt{2}}$ (see Fig.~\ref{fig:graphs} for an illustration of various graphs).
We aim to prove the following result, already stated in the main text:
\begin{result}
(a) All bipartite graph states in $\mathcal{H}^{\otimes N}$ are intrinsically sign problem-free.
(b) $t$-partite graph states with $t\geq 3$ have an intrinsic sign problem when the local transformations allowed to cure the sign problem are restricted to the set $\{\mathbb{I}_2, H\}^{\otimes N}$.
\end{result}
\begin{proof}
In the computational basis, the Calderbank--Shor--Steane (CSS) stabilizer states are the only stabilizer states with all positive amplitudes. Hence, the problem of sign curing can be translated into finding whether there exists a local Clifford transformation that takes the given graph state to an arbitrary CSS state. Note that the CSS states have an efficient representation in terms of either $X$-type or $Z$-type stabilizers. On the other hand, for a graph $G=(V, E)$ with $N=|V|$, the graph state $|G\rangle$ has stabilizer generators $K_v = X_v \prod_{u\in N(u)} Z_u$, one for each vertex $v\in V$. In tableau form, each generator has exactly one $X$ entry and several $Z$-entries with all phases being $+1$. If $|G\rangle$ is a bipartite graph state, the vertices of its corresponding graph $G$ can be colored as $V=A\cup B$. Then, by applying Hadamard operations on all qubits in $A$, one can transform the stabilizer generators as
\begin{eqnarray}
 K_v&\to& Z_v \prod_{u\in N(u)} Z_u \;\text{ for }u\in A   \;\text{ and }\nonumber\\
K_v&\to& X_v\prod_{u\in N(u)} X_u  \;\text{ for }u\in B,
\end{eqnarray}
which correspond to the stabilizer generators of the CSS states. Moreover, the converse is also true: every graph state which is curable using the transformations from the set $\{H, \mathbb{I}\}^{\otimes N}$ will be a bipartite state. In contrast, when the given graph state corresponds to a $t$-partite graph for some $t\geq 3$, no combination of Hadamard and identity operators can cure the sign problem.
\end{proof}

The following remark is an implication of the above result:
\begin{remark}
If a Hamiltonian can be made sign-free through local Hadamard transformations, then it can not have a multipartite graph state (up to local Hadamard transformations) as its ground state.
\end{remark}

For illustration, exact counts of the stabilizer states, graph states, and bipartite graph states are given in Table~\ref{tab:state_counts}. Closed-form expressions are known for the first two~\cite{aaronson2004improved}: the number of stabilizer states on $N$ qubits is $\mathcal{N}_{\mathrm{stab}}(N)=2^{N}\prod_{\ell=1}^{N}\left(2^{\ell}+1\right)$, while the number of graph states is $\mathcal{N}_{\mathbf{G}}(N)=2^{N(N-1)/2}$. At large qubit numbers, the Hadamard-curable (bipartite) graph states form a vanishing fraction of all graph states, suppressed superexponentially in $N$.

\begin{table}[ht]
\centering
\caption{Counts for small numbers of qubits $N$ of the number of stabilizer states, graph states, and of graph states that can be made sign-free by Hadamard rotations (equivalent to bipartite graph states). The fraction of Hadamard-curable graph states is superexponentially small.}
\label{tab:state_counts}
\begin{tabular}{|c|c|c|c|}
\hline
\textbf{$N$} & \ $\mathcal{N}_{\text{stab}}$ & $\mathcal{N}_{\mathbf{G}}$ & \textbf{\# Intrinsic Sign-free G} \\
\hline
1 & $6$ & $1$ & 1 \\
2 & $60$ & $2$ & 2 \\
3 & $1080$ & $8$ & 7 \\
4 & $36{,}720$ & $64$ & 41 \\
5 & $2{,}423{,}520$ & $1{,}024$ & 376 \\
6 & $315{,}057{,}600$ & $32{,}768$ & 5177 \\
7 & $81{,}284{,}860{,}800$ & $2{,}097{,}152$ &103{,}237 \\
8 & $41{,}780{,}418{,}451{,}200$ & $268{,}435{,}456$ & 2{,}922{,}446 \\
\hline
\end{tabular}
\end{table}

\clearpage
\onecolumngrid
\let\addcontentsline\ORIGaddcontentsline  
\setcounter{secnumdepth}{3}               

\setcounter{equation}{0}
\setcounter{figure}{0}
\setcounter{table}{0}
\setcounter{section}{0}
\setcounter{theorem}{0}
\setcounter{proposition}{0}

\renewcommand{\theequation}{S\arabic{equation}}
\renewcommand{\thefigure}{S\arabic{figure}}
\renewcommand{\thetable}{S\arabic{table}}
\renewcommand{\thesection}{S\arabic{section}}
\renewcommand{\thesubsection}{S\arabic{section}.\arabic{subsection}}
\renewcommand{\thetheorem}{S\arabic{theorem}}
\renewcommand{\theproposition}{S\arabic{proposition}}

\begin{center}
  \textbf{\large Supplemental Material}\\[4pt]
  \textbf{Quantum resources and the onset of the quantum Monte Carlo sign problem}\\[4pt]
Naga Dileep Varikuti, Leela Ganesh Chandra Lakkaraju, and Philipp Hauke
\end{center}
\vspace{6pt}

\addtocontents{toc}{\protect\setcounter{tocdepth}{1}}
\setcounter{tocdepth}{2}
\vspace{-1.5em}
\tableofcontents
\vspace{1em}

\section{Quantum resource quantifiers}

In this appendix, we briefly outline the quantum-information–theoretic quantities that we use to probe the complexity of many-body quantum states, the entanglement entropy and stabilizer entropy. These quantities, and the IPR discussed in the end matter, characterize different aspects of the complexity underlying the quantum state.

\subsection{Bipartite Entanglement entropy}
The bipartite entanglement of a pure state is suitably quantified by the von Neumann entropy. For a state $|\psi_{AB}\rangle$ shared between two subsystems, it reads
\begin{eqnarray}
S_{\mathrm{vN}}(|\psi_{AB}\rangle) = -\mathrm{Tr}\left(\rho_{A}\log\rho_{A}\right),
\,\,\, \text{with} \,\,\, \rho_{A} = \mathrm{Tr}_{B}\left(|\psi\rangle\langle\psi|\right).
\end{eqnarray}
Operationally, $S_{\mathrm{vN}}$ counts the number of Bell pairs that can be asymptotically distilled from the state under local operations and classical communication~\cite{PhysRevA.53.2046}. The entanglement entropy vanishes if and only if the state is a product state across the cut, and a nonzero value therefore signals non-local correlations between $A$ and $B$. Entanglement has been identified as a key resource across quantum metrology~\cite{giovannetti2004quantum, giovannetti2006quantum, RevModPhys.89.035002, hauke2016measuring, RevModPhys.90.035005}, teleportation~\cite{bennett1993teleporting}, and error correction~\cite{shor1995scheme, RevModPhys.87.307, roffe2019quantum}. It is equally one of the principal obstacles to classical simulation of quantum many-body systems: only when its growth remains sufficiently weak, the quantum system can be simulated efficiently by matrix product states~\cite{verstraete2008matrix, schollwock2011density, perez2006matrix}. Interestingly, recent works have shown that entanglement has surprising connections with quantum chaos through out-of-time ordered correlators~\cite{varikuti2022out, PhysRevLett.120.040402}

\subsection{Stabilizer R\'enyi entropy}
Stabilizer states form a special class of quantum states that possess an
efficient classical representation. Let $\mathcal{G}_{N}$ denote the group of all $N$-qubit Pauli strings. A pure state $|\psi\rangle$ is called a stabilizer state if there exists a subgroup $\mathcal{S} \subset \mathcal{G}_{N}$ with $|\mathcal{S}| = 2^N$ such that $|\psi\rangle$ is stabilized by every element of $\mathcal{S}$, i.e., $P|\psi\rangle = |\psi\rangle$ for all $P \in \mathcal{S}$ \cite{gottesman1998heisenberg}. In other words, $|\psi\rangle$ is a simultaneous $+1$ eigenstate of all Pauli operators belonging to the stabilizer group $\mathcal{S}$. The set of unitary operators that preserve the Pauli group under conjugation forms the Clifford group. Acting with Clifford unitaries on computational basis states generates the class of stabilizer states. A key property of such dynamics is that they admit efficient classical simulation \cite{gottesman1998heisenberg, aaronson2004improved}. For a general quantum state $|\psi\rangle$, the deviation from the stabilizer manifold can be quantified using the stabilizer R\'enyi entropy of order $q$ \cite{leone2022stabilizer}, defined as
\begin{eqnarray}
\mathcal{M}_{q} = \dfrac{1}{1-q}\log_{2}\left( \sum_{p=1}^{2^{2N}}\dfrac{|\bra{\psi}p_{i}\ket{\psi}|^{2q} }{2^{2N}} \right),
\end{eqnarray}
where $q\geq 0$. In this work, we consider $q=2$, the second stabilizer R\'enyi entropy, given by
\begin{equation}
\mathcal{M}_{2} = -\log_{2}\left( 1-\mathcal{M}_{\mathrm{linear}} \right),
\end{equation}
where
\begin{align}\label{eq:nonstabpowerdef}
\mathcal{M}_{\mathrm{linear}}(|\psi\rangle)
= 1-2^N\sum_{i=0}^{2^{2N}-1}\frac{1}{2^{2N}}\langle\psi |P_i|\psi\rangle^4
= 1-2^N\mathrm{Tr}\left[ Q\left(|\psi\rangle\langle\psi|\right)^{\otimes 4} \right],
\end{align}
and $Q=\left( \dfrac{1}{2^{2N}}\sum_{i=0}^{2^{2N}-1}P_i^{\otimes 4} \right)
$ can be interpreted as a projector acting on $\mathcal{H}^{\otimes 4}$, satisfying $Q^2 = Q$. Stabilizer entropy is a faithful measure of nonstabilizerness as it vanishes for all pure stabilizer states and is invariant under the action of Clifford operators.  It has been shown recently that $\mathcal{M}_{2}$ is a monotone for the nonstabilizerness of pure states \cite{leone2024stabilizer}. This quantity has attracted considerable interest in the context of quantum many-body systems~\cite{leone2022stabilizer, santra2025complexitytransitionschaoticquantum, Magni2025quantumcomplexity, tarabunga2024nonstabilizerness, santra2025quantumresourcesnonabelianlattice, tarabunga2023many, 91hd-8k7g, piroli_longrange_magic, capecci2025role, robin2026quantumcomplexitynewdirections, Viscardi_2026, odavic2024stabilizer, haug2023quantifying, Varikuti2026impactofclifford, catalano2024magic}.
Moreover, the stabilizer R\'enyi entropy provides a bound on the number of T-gates needed to describe the quantum system of interest \cite{leone2022stabilizer}, which in turn implies exponential hardness of representing the state using classical methods based on stabilizer tableaus~\cite{gottesman1997stabilizercodesquantumerror}.

\section{Ground state of random $k$-local stoquastic Hamiltonians}
\label{app-a}

In this appendix we construct the approximate ground state at $p_{\mathrm{ns}}=0$ analytically. It is exact for the averaged Hamiltonian, and for typical $H$, the fluctuations about it are suppressed exponentially in $k$, so that the construction becomes increasingly accurate as $k\rightarrow N$.
Following Eq.~(\ref{ham-main}) of the main text, we consider random stoquastic Hamiltonians of the following form with open boundary conditions in one spatial dimension:
\begin{equation}
  H= \sum_{j=0}^{N-k}h_j\;, \text{ where }\; [h_j]_{m, n\neq m}\leq 0\,.
\end{equation}
Here, $h_j$s are random $k$-local terms acting on $k$-nearest neighbor lattice sites. For each $h_j$, we initially draw its matrix elements independently from a Gaussian distribution \(\mathcal{N}(0,1)\). We then set the signs of all off-diagonal elements such that they are non-positive. The resulting off-diagonal elements therefore follow a half-Gaussian distribution.

In the following, we first determine the ground state of the averaged Hamiltonian $\mathbb{E}[H]$. We then argue that this state provides a good approximation to the true ground state of a typical stoquastic Hamiltonian $H$ by demonstrating that the fluctuations in it when acted upon by $H$ are sufficiently small.

\subsection{Ground state of $\mathbb{E}[H]$}

For a typical stoquastic Hamiltonian $H$, the diagonal elements of each $k$-local term have zero mean, and the off-diagonal elements have a fixed, non-vanishing mean $S$. One can then write the averaged Hamiltonian $\mathbb{E}[H]$ as follows:
\begin{equation}
\mathbb{E}[H]=S\sum_{j=0}^{N-k}\left(J_{k}-\mathbb{I}\right)_{j},
\qquad \mathrm{where }\;\;
J_{k}=\sum_{m,n=0}^{2^k-1}\ket{m}\bra{n}=2^{k}\,\ket{+}\bra{+}^{\otimes k},\;\; \mathrm{and}\;\;  |+\rangle=\dfrac{|0\rangle+|1\rangle}{\sqrt{2}}
\end{equation}
The terms are thus products of single-site projectors $\ket{+}\bra{+}$ and mutually commute, so $\mathbb{E}[H]$ is diagonal in the Hadamard basis given by $\{\ket{+},\ket{-}\}^{\otimes N}$. The eigenvalues are obtained by summing the contributions from all $N-k+1$ local blocks. Each block contributes $S(2^k-1)$ if all its $k$ sites are in $\ket{+}$, and $-S$ otherwise. Consequently, the energy eigenvalues are
\begin{equation}\label{typical_eval}
E_n=S\left[n\,2^k-(N-k+1)\right],
\end{equation}
where $n$ denotes the number of contiguous blocks whose $k$ sites are all in $\ket{+}$. Since $S<0$, the energy is minimized only when $n=N-k+1$, which requires all sites to be in $\ket{+}$. Thus, the unique ground state of $\mathbb{E}[H]$ is given by
\begin{eqnarray}
|\bm{+}\rangle=\ket{+}^{\otimes N}=\dfrac{1}{\sqrt{2^{N}}}
\sum_{j=0}^{2^N-1}\ket{j}.
\end{eqnarray}
In the subsequent analysis, we use this state as a ground-state heuristic for typical stoquastic Hamiltonians. One can obtain the first excited state by flipping a single boundary spin from $\ket{+}$ to $\ket{-}$. Having access to the first excited state enables us to get the energy gap of $\mathbb{E}[H]$, which is given by
\begin{equation}\label{gapeq}
\Delta_{\mathbb{E}[H]}=|S|\,2^{k}.
\end{equation}
This is independent of $N$ and depends only on $k$. In the following subsection, we check how close $|\bm{+}\rangle$ lies to the ground state of a typical realization.

\subsection{Relative fluctuations of $H | + \rangle$}

For a state $|\psi\rangle$ to serve as a good approximation to the true ground state of a typical stoquastic Hamiltonian, it should satisfy the following: (i) its amplitudes in the computational basis should be positive, and (ii) the fluctuations of the elements $\bra{i}H\ket{\psi}$ across the stoquastic ensemble, for any $i\in\{0,1\}^{N}$, should be sufficiently small. The latter condition implies that a typical random Hamiltonian acts approximately homogeneously on the initial state, with different realizations producing nearly the same action up to small fluctuations. In addition, we require those fluctuations to be smaller than the energy gap of the Hamiltonian, so that the true ground state remains perturbatively related to $|\psi\rangle$. Here, we show that $\ket{\bm{+}}$ provides a good approximation of the true ground state of a typical stoquastic Hamiltonian. To verify, we consider the following:
\begin{eqnarray}\label{IHPSI1}
H\ket{\bm{+}} &=& \left(\sum_{j=0}^{N-k}h_{j}\right)|+\rangle^{\otimes N}
\nonumber\\
&=& \frac{1}{\sqrt{2^{k}}}\sum_{j=0}^{N-k}\left[\sum_{m=0}^{2^k-1}
\left(\sum_{n}[h_{j}]_{mn}\right)
|+\rangle^{\otimes j}\ket{m}|+\rangle^{\otimes(N-j-k)}\right],
\end{eqnarray}
where $|m\rangle$ runs over the $2^{k}$ computational basis states of the block on which $h_{j}$ acts. The uniform superposition is therefore preserved only if the bracketed row sums are the same for every $m$. We check this by considering, for an arbitrary computational basis state $\ket{i}$ with $i\in\{0,1\}^{N}$, the following:
\begin{eqnarray}
\langle i|H|\bm{+}\rangle &=& \dfrac{1}{\sqrt{2^{N}}}\sum_{j=0}^{N-k}
\left([h_{j}]_{m_{j}m_{j}}+\sum_{n\neq m_{j}}[h_{j}]_{m_{j}n}\right),
\end{eqnarray}
where $m_{j}=i_{j}i_{j+1}\cdots i_{j+k-1}$ is the $k$-bit substring of $i$ on the sites acted on by $h_{j}$.
Averaging over the distribution of the elements of the Hamiltonian, which washes out any dependence on $\ket{i}$, yields
\begin{eqnarray}\label{IHPSI2}
\mathbb{E}_{[h_j]_{mn}}\left( \dfrac{\langle i |H|\bm{+}\rangle}{\langle i|\bm{+}\rangle} \right) = S\; (N-k+1) (2^k-1),
\end{eqnarray}
where $S=\mathbb{E}([h_j]_{m, n})$. Therefore, the action of $H$ on $\ket{\bm{+}}$ remains uniform in the computational basis up to the first moments of the elements of the stoquastic Hamiltonian ensemble. Note that the ratio $\bra{i}H\ket{\bm{+}}/\langle i|\bm{+}\rangle$ gives the local energy associated with the configuration $\ket{i}$. If $\ket{\bm{+}}$ were an exact eigenstate of $H$, this ratio would equal the corresponding eigenvalue for every $i$. Its fluctuations therefore directly quantify the extent to which $\ket{\bm{+}}$ fails to be an eigenstate, in units of energy, making it directly comparable to the gap. Moreover, dividing by $\langle i|\bm{+}\rangle=2^{-N/2}$ removes the normalization of $\ket{\bm{+}}$, which would otherwise introduce a factor $2^{-N/2}$ in the amplitude, independent of the Hamiltonian. Note also that the averaged energy eigenvalue in the above expression coincides with the ground state energy we obtained in the previous subsection for $\mathbb{E}[H]$ [see Eq.~(\ref{typical_eval})].

While the above statement that $H$ acts uniformly on $\ket{\bm{+}}$ is true on average, a single random Hamiltonian will not generally act so uniformly. How far it typically deviates can be assessed through the fluctuations of $\bra{i}H\ket{\bm{+}}$ over the stoquastic ensemble. These fluctuations must be measured against the gap $\Delta_{\mathbb{E}[H]}$ of $\mathbb{E}[H]$, since it is their ratio that determines how strongly the excited states of $\mathbb{E}[H]$ are mixed into the true ground state, and hence how far it departs from $\ket{\bm{+}}$.
\begin{eqnarray}
\mathbb{E}\left( \langle i|H|\ket{\bm{+}}^2 \right)&=& \dfrac{1}{2^N}\mathbb{E}\left[ \sum_{j}\sum_{j'}\left( [h_j]_{ii} +\sum_{i\neq i'} [h_j]_{ii'}  \right) \left([h_{j'}]_{ii} +\sum_{i\neq i''} [h_{j'}]_{i, i''}  \right)  \right]\nonumber\\
&=&\dfrac{1}{2^N}\left[ \sum_{j}\mathbb{E}\left( [h_{j}]^2_{ii} \right) + \sum_{j, j'}\mathbb{E}\left( \sum_{(i', i'')\neq i} [h_j]_{i, i'}[h_{j'}]_{i, i''} \right)  \right]\nonumber\\
&=&\dfrac{(N-k+1)}{2^N}\left[ 1+(N-k)(2^k-1)^2S^2 +(2^k-1)M + (2^k-1)(2^k-2)S^2 \right],
\end{eqnarray}
where $M=\mathbb{E}([h_j]^2_{i, i'})$ for $i\neq i'$.
The variance therefore is
\begin{eqnarray}
\mathrm{Var}\left( \dfrac{\langle i|H\ket{\bm{+}}}{\langle i|\bm{+}\rangle} \right)
&=& 2^{N}\,\left(\mathbb{E}\left( \langle i|H\ket{\bm{+}}^2 \right) - \left[\mathbb{E}\left( \langle i|H\ket{\bm{+}} \right)\right]^2 \right)\nonumber\\
&=& (N-k+1)\left[ 1+ (2^k-1)(M-S^2) \right]\, .
\end{eqnarray}

Following the above equation and Eq.~(\ref{gapeq}), the relative fluctuations in $\bra{i}H\ket{\bm{+}}$ can be written as

\begin{eqnarray}
 \mathrm{RF} &=& \dfrac{\sqrt{2^N\mathrm{Var}_{H}(\bra{i}H\ket{\bm{+}}})}{\Delta_{\mathbb{E}[H]}}= \dfrac{\sqrt{(N-k+1)(1+(2^k-1)(M-S^2))}}{|S|2^k}.
\end{eqnarray}
Note that $\mathrm{RF}$ is invariant under the shift $H\rightarrow H+c\,\mathbb{I}$, which leaves the ground state unchanged: both the numerator and the gap are unaffected, since every amplitude is
displaced by the same constant. Any faithful measure must share this invariance, and $\mathrm{RF}$ accordingly captures how close the ground state of a typical random $k$-local stoquastic Hamiltonian lies to
$\ket{\bm{+}}$.

When $2^k\gg 1$, it then follows that
\begin{eqnarray}
\mathrm{RF}\approx \dfrac{\sqrt{(N-k+1)(M-S^2)}}{|S|2^{k/2}}
\end{eqnarray}
That is, for large $k$, the approximation of the true ground state by $\ket{+}$ becomes exponentially better. In contrast, for a fixed $k$, $\mathrm{RF}_{\mathrm{reg}}$ increases with $N$ as $\sim\sqrt{N-k+1}$. That is, increasing $N$ while keeping $k$ fixed takes the true ground state away from $|+\rangle^{\otimes N}$. Therefore, for some sufficiently large $k$, one can expect that the ground state of a random $k$-local stoquastic Hamiltonian remains perturbatively connected to the state $|\bm{+}\rangle$.

\section{Additional numerical results on entanglement and nonstabilizerness}

\begin{figure*}[t]
    \centering
    \includegraphics[width=\textwidth]{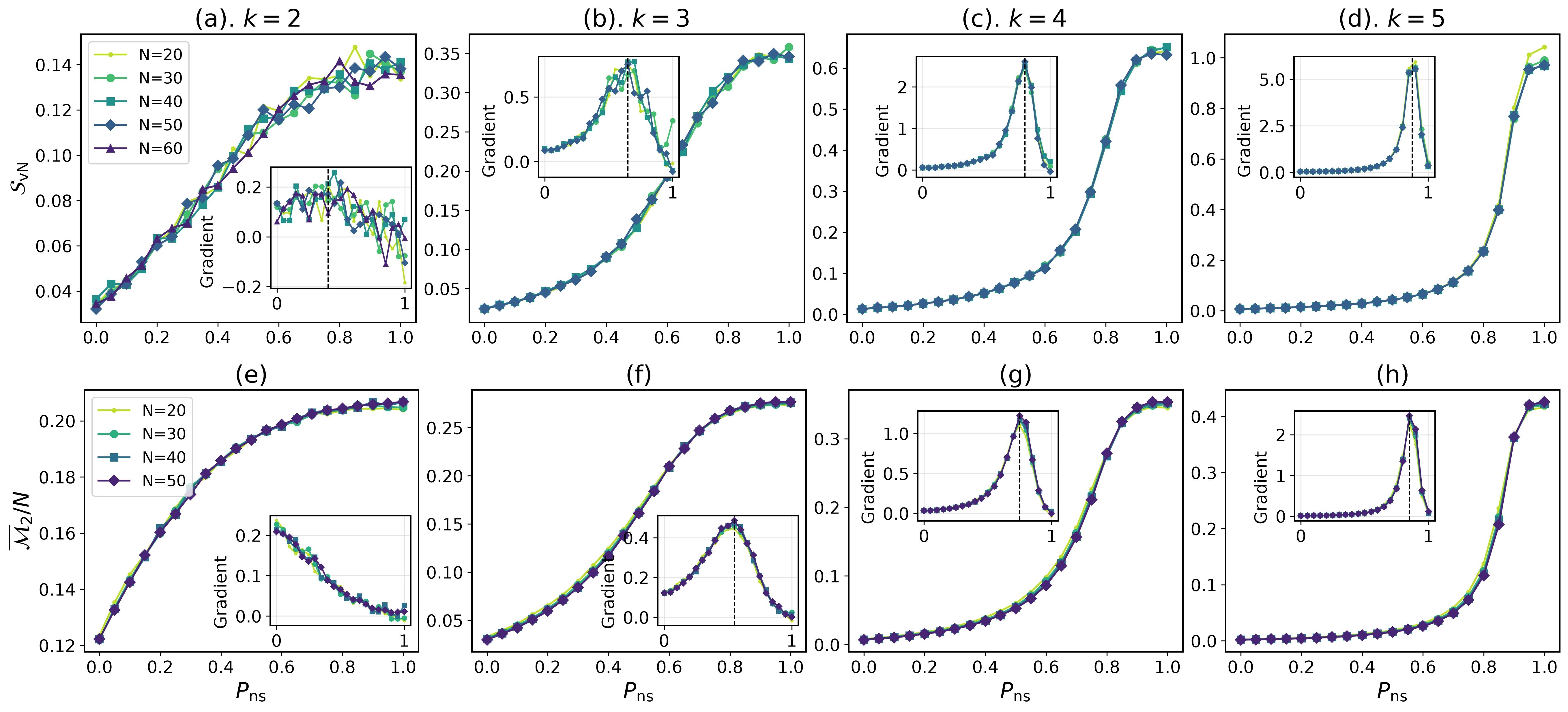}
    \caption{Quantum resources in the ground state of random nonstoquastic Hamiltonians as a function of the nonstoquasticity parameter $p_{\mathrm{ns}}$, which interpolates between the stoquastic limit ($p_{\mathrm{ns}} = 0$) and the GOE limit ($p_{\mathrm{ns}} = 1$). Data for $k = 2$, $3$, $4$, and $5$, respectively (the $k = 3$ results, panels (b) and (f), appear also in the main-text Fig.~\ref{fig:coll}).
    Top panels (a)--(d): bipartite entanglement, quantified by the von Neumann entanglement entropy.
    Bottom panels (e)--(h): nonstabilizerness, quantified by the stabilizer R\'enyi entropy, for the same values of $k$. Each data point is averaged over $\sim 10^{3}$ independent realizations of the random Hamiltonian. The insets show the derivative of the corresponding resource with respect to $p_{\mathrm{ns}}$. For entanglement, the peaks occur near $p^{\star}_{\mathrm{ns}}\approx 0.4, 0.65, 0.8,$ and $0.875$, while for non-stabilizerness, they occur near $p^{\star}_{\mathrm{ns}}\approx 0, 0.55, 0.75,$ and $0.85$, corresponding to $k=2, 3, 4,$ and $5$, respectively. The peaks of the derivative become taller and narrower, and shift to larger $p_{\mathrm{ns}}$, as $k$ increases. For the nonstabilizerness, the maxima of the derivative are displaced to smaller $p_{\mathrm{ns}}$ relative to the entanglement entropy. This effect is most prominent for $k = 2$, where the derivative of nonstabilizerness is largest at $p_{\mathrm{ns}} = 0$ and decreases monotonically.}
    \label{fig:res_tot}
\end{figure*}

In this section, we present additional numerical results on the bipartite entanglement entropy and the nonstabilizerness; see Fig.~\ref{fig:res_tot}.
We consider $k = 2$, $3$, $4$, and $5$, reproducing also the $k = 4$ case, which already appears in the main text, to allow for a direct comparison.
Each data point is obtained by averaging over $\sim 10^{3}$ independent realizations of the random Hamiltonian.
The insets display the respective derivatives with respect to $p_{\mathrm{ns}}$, which permits us to identify the region where the entanglement entropy and nonstabilizerness change most rapidly.

The top panels (a)--(d) show the bipartite entanglement in the ground state of the random nonstoquastic Hamiltonians, quantified by the von Neumann entanglement entropy, as $p_{\mathrm{ns}}$ is varied from $0$ (the stoquastic limit) to $1$ (the GOE limit).
As $k$ increases, the peaks of its derivative become taller and narrower, which indicates that the growth of entanglement becomes increasingly abrupt and concentrated in a smaller window of $p_{\mathrm{ns}}$. The position of the peak also shifts to larger $p_\mathrm{ns}$ as $k$ increases. Further, the dependence on $k$ is qualitatively different at the two ends of the interpolation. Close to the stoquastic limit, the entanglement entropy is increasingly suppressed as $k$ grows, so that ground states of higher-$k$ Hamiltonians remain only weakly entangled over a wider range of $p_{\mathrm{ns}}$. Close to the GOE limit, the trend is reversed, and the entanglement entropy is larger for higher $k$. Taken together, these observations indicate that increasing $k$ sharpens the crossover from weakly to strongly entangled ground states.

The bottom panels (e)--(h) show the corresponding nonstabilizerness, quantified by the stabilizer R\'enyi entropy.
Again, the ground states near the stoquastic limit exhibit decreasing nonstabilizerness as $k$ grows, while the behavior is inverted near the GOE limit, mirroring the trend observed for the entanglement in the same limit.
For $k = 3, 4$ and $5$, the derivative exhibits a well-defined peak, located at $p_{\mathrm{ns}} \approx 0.55$, $\approx 0.75$ and $\approx 0.85$, respectively. The case $k = 2$ behaves qualitatively differently: the derivative shows no peak in the interior of the interval, but instead attains its maximum at $p_{\mathrm{ns}} = 0$ and appears to decrease monotonically as $p_{\mathrm{ns}}$ increases. In other words, for $k = 2$ the nonstabilizerness grows most rapidly as soon as an arbitrarily small amount of nonstoquasticity is introduced, and its growth slows down steadily thereafter. This is in contrast to the behavior of entanglement, which shows a peak near $p_{\mathrm{ns}}\approx 0.4$. With increasing $k$, the peaks for both resources become successively closer.

Interestingly, the locations of the maxima of the nonstabilizerness derivative are consistently shifted to smaller $p_\mathrm{ns}$ values to the left with respect to their counterparts in the entanglement case, indicating that nonstabilizerness responds to the onset of nonstoquasticity earlier than entanglement does. Thus, although both resources ultimately grow with nonstoquasticity, they do not do so at the same rate, and nonstabilizerness can be generated substantially before significant entanglement builds up.

\section{Analysis of the energy gap}
\begin{figure*}
\includegraphics[scale=0.45]{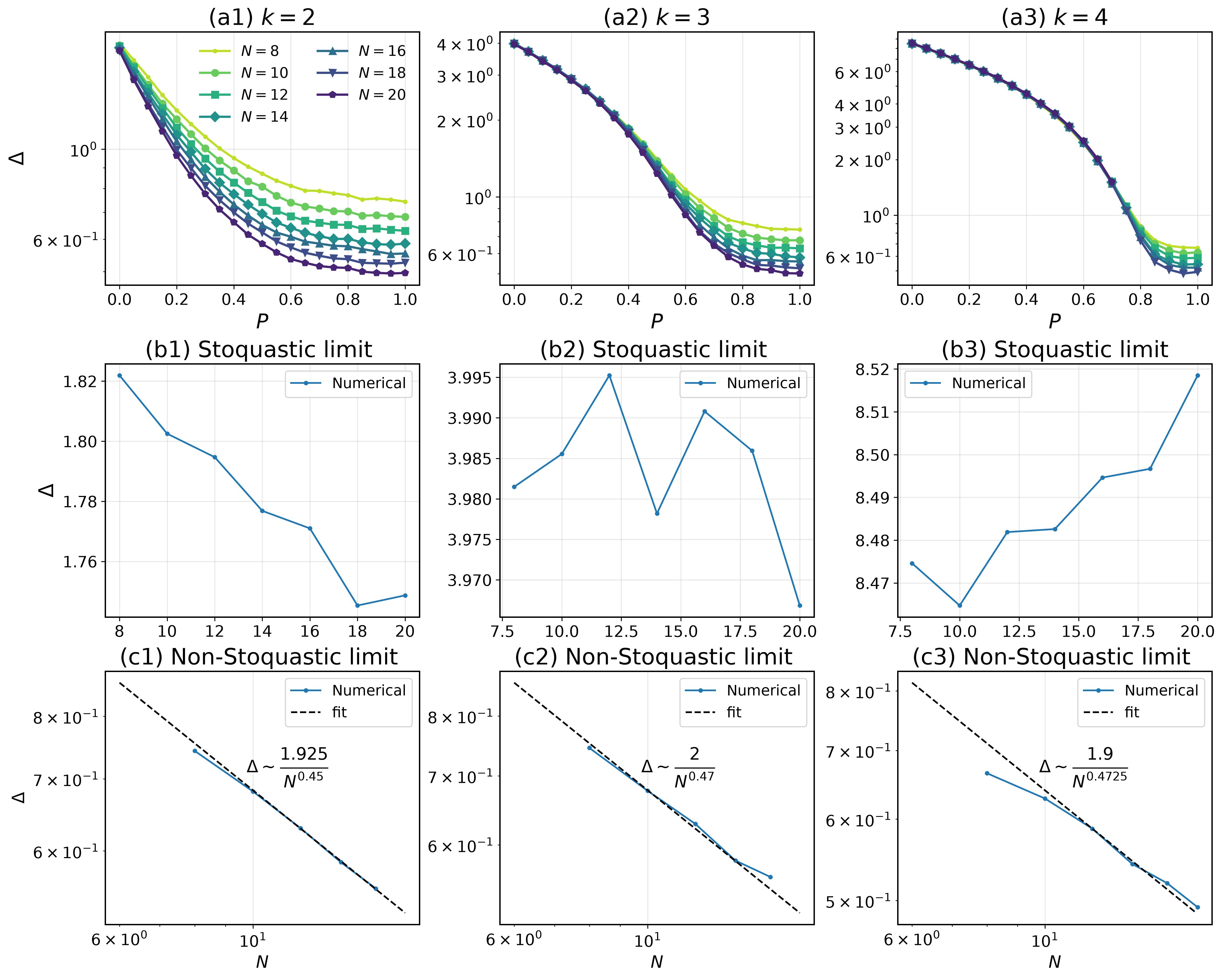}
    \caption{Spectral gap $\overline{\Delta}$ of random nonstoquastic Hamiltonians as a function of system size $N$. The columns correspond to $k = 2$, $3$, and $4$, from left to right. All data are averaged over $\sim 10^{3}$ independent realizations of the random Hamiltonian. (a1)--(a3) Averaged gap versus the degree of nonstoquasticity $p_{\mathrm{ns}}$. Towards the stoquastic limit ($p_{\mathrm{ns}} \to 0$), the curves for different $N$ nearly coincide, whereas the averaged gap decreases consistently as $p_{\mathrm{ns}} \to 1$. (b1)--(b3) Averaged gap as a function of $N$ in the fully stoquastic regime ($p_{\mathrm{ns}}=0$), showing that the gap remains finite and nearly independent of system size. (c1)--(c3) Averaged gap as a function of $N$ in the fully nonstoquastic limit ($p_{\mathrm{ns}} = 1$), on a log-log scale. The dashed lines are power-law fits $\overline{\Delta} \sim N^{-\alpha}$ with $\alpha \approx 0.45$, $0.47$, and $0.4725$ for $k = 2$, $3$, and $4$, respectively.}
\label{fig:gap}
\end{figure*}

In this section, we analyze the spectral gap of the random nonstoquastic Hamiltonians, defined as the energy difference between the first excited state and the ground state, as the degree of nonstoquasticity is tuned. Figure~\ref{fig:gap} shows the numerical results for different system sizes $N$. The three columns of the figure correspond to $k = 2$, $3$, and $4$, respectively, and all results are averaged over $\sim 10^{3}$ independent realizations of the random Hamiltonian.

Panels (a1)--(a3) show the averaged gap $(\overline{\Delta})$ as a function of $p_{\mathrm{ns}}$. Towards the stoquastic limit, the curves corresponding to different system sizes nearly coincide, indicating that in this regime the gap is essentially independent of $N$. This behavior is confirmed in panels (b1)--(b3), where the averaged gap is plotted as a function of $N$ for a fixed $p_{\mathrm{ns}}=0$. For $k = 2$, the gap decreases only very slightly with $N$, whereas for $k = 3$ it fluctuates weakly around an approximately constant value, while for $k=4$ it increases slightly with $N$, with no discernible systematic dependence on system size. Interestingly, the range of $p_{\mathrm{ns}}$ over which the curves for different $N$ collapse agrees closely with the region identified in our analysis of the quantum resources, where the gradients of the entanglement and the nonstabilizerness separate the effectively stoquastic regime from the nonstoquastic one. The spectral and resource-based diagnostics therefore point consistently to the same crossover. Since the gap remains finite and does not close with increasing system size in this regime, the corresponding Hamiltonians are expected to be amenable to efficient simulation via quantum Monte Carlo methods.

As $p_{\mathrm{ns}}$ increases towards $1$, $\overline{\Delta}$ decreases consistently, and its dependence on the system size becomes increasingly pronounced. Near the fully nonstoquastic limit at $p_{\mathrm{ns}} = 1$, the gap exhibits an approximate power-law decay with $N$, as shown in the lower panels (c1)--(c3). The numerical data yields $\overline{\Delta} \sim N^{-0.45}$ for $k = 2$, $\overline{\Delta} \sim N^{-0.47}$ for $k = 3$, and $\overline{\Delta} \sim N^{-0.4725}$ for $k = 4$, so the decay becomes somewhat steeper as $k$ increases. This closing of the gap with system size, together with the enhanced resource content discussed above, signals that for larger $k$ values, the ground states in the strongly nonstoquastic regime become progressively harder to access with classical simulation techniques.

\section{Additional details on the projector Quantum Monte Carlo method and simulation accuracy}

\begin{figure*}[ht]
    \centering
    \includegraphics[scale=0.125]{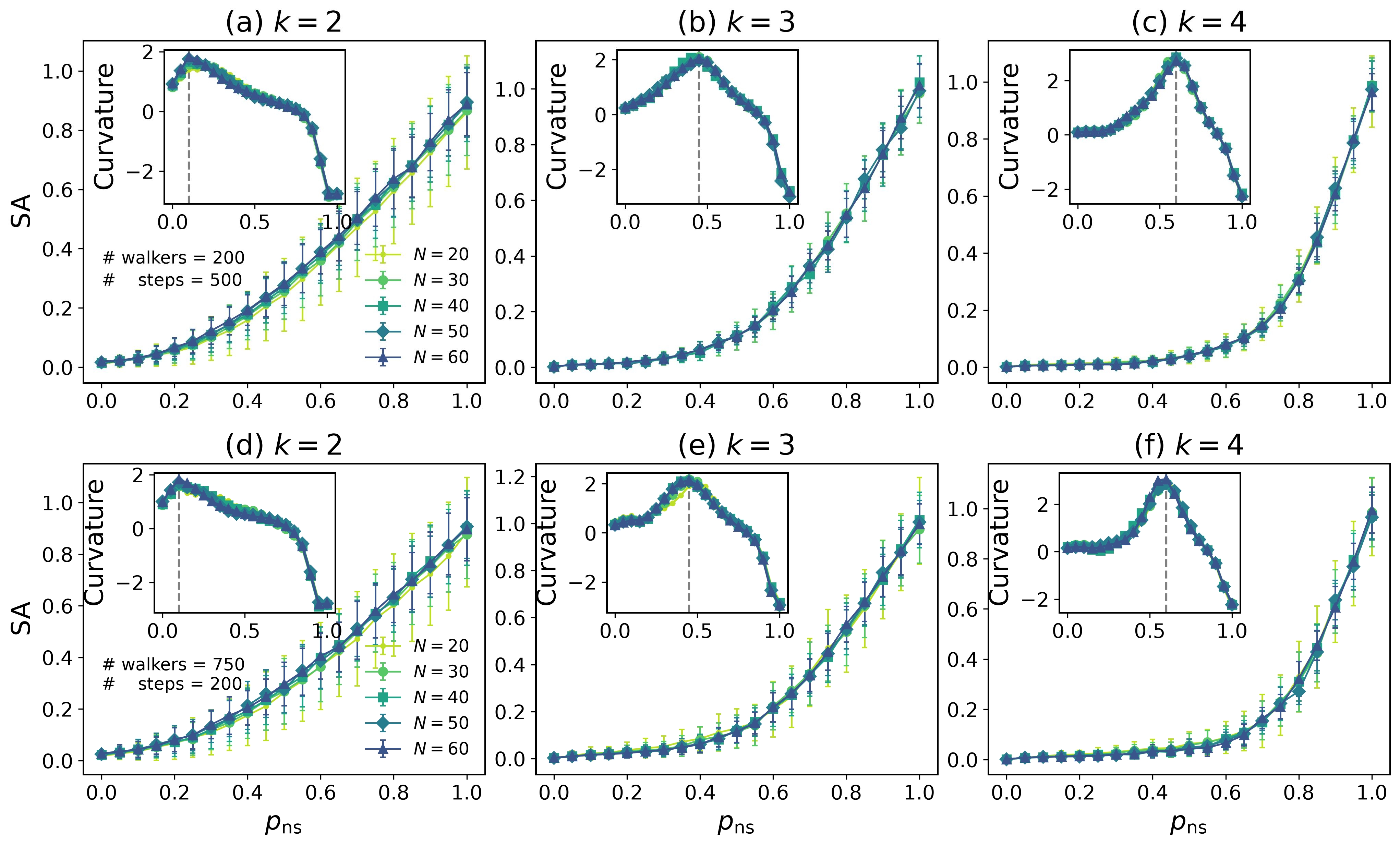}
    \caption{Simulation accuracy of the PQMC method in estimating the ground-state energy,
    measured against the DMRG reference, for random Hamiltonians with $k=2$, $3$ and $4$, system sizes $N=20$--$60$, and two different classical budgets. The top row shows the results for (walkers, steps) $=(200,500)$, the bottom row for $(750,200)$. Markers denote the mean over $\sim O(10^2)$ random instances and error bars the standard deviation across the ensemble. Across all panels, the SA remains small initially and rises steeply as $p_{\mathrm{ns}}$ is increased, with the onset of the rise shifting to larger $p_{\mathrm{ns}}$ for increasing $k$, while the curves for different $N$ collapse onto one another. The insets show the curvature of the data,
whose peaks roughly identify the crossover from the easy to the hard-to-simulate regime. The curvature peaks near $p_{\mathrm{ns}}\approx 0.1$, $0.45$ and $0.6$ for $k=2$, $3$, and $4$ respectively, and its position remains insensitive to the choice of classical budget. Throughout, we take the uniform superposition $|\bm{+}\rangle$ as the trial state, which is also the exact ground state of $\mathbb{E}[H]$ in the stoquastic limit.}
    \label{fig:PQMC_diff}
\end{figure*}

In this section, we first outline the basic projector quantum Monte Carlo (PQMC) algorithm used for estimating the ground-state energy of a random symmetric Hamiltonian, and then illustrate the corresponding numerical results for the random nonstoquastic Hamiltonians considered in this work. Given a Hamiltonian $H$ with ground state $|\phi_{0}\rangle$, any initial state $|\psi_{\mathrm{init}}\rangle$ with $\langle\phi_{0}|\psi_{\mathrm{init}}\rangle\neq0$ can be projected onto the ground state by imaginary-time evolution~\cite{Becca_Sorella_2017},
\begin{equation}
|\phi_{0}\rangle=\lim_{\beta\rightarrow\infty}
\frac{e^{-\beta H}|\psi_{\mathrm{init}}\rangle}
{\big\|\,e^{-\beta H}|\psi_{\mathrm{init}}\rangle\,\big\|}\;,
\end{equation}
since the contribution of excited states is suppressed by the factor $e^{-\beta(E_{n}-E_{0})}$ for $n\geq 1$.
In practice, $e^{-\beta H}$ is implemented as $e^{-\beta H} \approx \left( \mathbb{I}-\tau H\right)^r$, where $ \beta = r\tau$,  with a small time step $\tau$. For stoquastic Hamiltonians, the matrix elements of $e^{-\beta H}$ are nonnegative, allowing them to be interpreted (up to normalization) as transition probabilities. This enables stochastic sampling of imaginary-time paths, from which ground-state expectation values, such as the energy, can be estimated. For more details on the implementation of QMC methods, refer to~\cite{Becca_Sorella_2017}. For completeness, we provide the core algorithm of the PQMC in Algorithm~\ref{alg:pqmc}.
The sign problem in PQMC arises when contributions from walkers with opposite signs cancel each other, causing the energy estimator to be divided by a small denominator and thereby amplifying statistical noise. Algorithm~\ref{alg:pqmc} illustrates that this occurs when the off-diagonal matrix elements $H_{ij}$ of the Hamiltonian have strongly fluctuating signs. However, our results in the main text show that the simulation accuracy does not trivially deteriorate with increasing degree of nonstoquasticity of the Hamiltonian, but rather correlates strongly with the growth of entanglement, nonstabilizerness, and the average sign of the ground state.

We now corroborate these findings with additional numerical details on the simulation accuracy of the PQMC against the reference DMRG to estimate the ground-state energy. The corresponding results are presented in Fig.~\ref{fig:PQMC_diff} for $k=2$, $3$, and $4$, and system sizes $N=20$--$60$, considering two different classical budgets: $(N_w, T)=(200, 500)$ in the top row and $(750, 200)$ in
the bottom row, with $\Delta\tau=0.01$ in both cases.

The data for different $N$ appear to collapse onto a single curve in all the panels. Since $E_{\mathrm{DMRG}}$ is extensive, this suggests that the absolute error $|E_{\mathrm{QMC}}-E_{\mathrm{DMRG}}|$ grows linearly with $N$. Moreover, the SA clearly separates into two regimes. Near the stoquastic limit, it stays small and nearly constant, implying that the PQMC estimates the ground-state energy reliably even when a considerable fraction of the off-diagonal elements has already turned positive. Beyond a crossover, the SA rises steeply, and as
$p_{\mathrm{ns}}$ approaches $1$, it approaches unity, where $E_{\mathrm{QMC}}$ nearly vanishes due to the cancellation between the positive and the negative walkers.

To identify the crossover region, we compute the curvature of the mean SA, softened with a Gaussian filter of width $\sigma=0.1$ to suppress the fluctuations arising from the numerical differentiation. Interestingly, the curvature exhibits a clear peak in all the cases considered (see insets of Fig.~\ref{fig:PQMC_diff}). The peaks appear near $p_{\mathrm{ns}}\approx 0.1$, $0.45$, and $0.6$ for $k=2$, $3$, and $4$ respectively, indicating that the crossover shifts to larger $p_\mathrm{ns}$ with increasing $k$, consistent with the behavior of entanglement and nonstabilizerness. Finally, the positions of the curvature peaks remain insensitive to the redistribution of the classical budget between the walkers and the projection steps, suggesting that the crossover is a property of the model rather than of
the simulation parameters.

\begin{algorithm}[H]
\caption{Projector Quantum Monte Carlo (PQMC) with branching walkers}
\label{alg:pqmc}
\begin{algorithmic}[1]

\State \textbf{Input:} Hamiltonian $H$ (real symmetric), time step $\Delta\tau$, number of walkers $N_w$, projection steps $T$, energy shift $E_T$.

\State Initialize walker positions $\{i_w\}_{w=1}^{N_w}$ uniformly over the computational basis, with signs $s_w = +1$.

\For{$t = 1$ to $T$}

    \State Initialize empty walker list $\mathcal{W}_{\text{new}}$

    \For{each walker $w$ located at basis state $i$ with sign $s_w$}

        \State Extract the nonzero elements $\{H_{ji}\}$ of column $i$

        \State For each connected state $j$, compute
        \[
        G_{ji} =
        \begin{cases}
        1 - \Delta\tau\,(H_{ii}-E_T), & j = i, \\
        -\Delta\tau\, H_{ji}, & j \neq i.
        \end{cases}
        \]

        \State Compute the normalization weight
        \[
        W_i = \sum_j |G_{ji}|
        \]

        \State Sample the next state $j$ with probability
        $
        p_{ji} = \dfrac{|G_{ji}|}{W_i}
        $ and update the sign, $s_w \gets s_w \, \mathrm{sgn}(G_{ji})$

        \State Branch the walker into
        $
        m = \lfloor W_i + \eta \rfloor, \quad \eta \sim U(0,1)
        $
        copies, and add $m$ walkers at state $j$ with sign $s_w$ to $\mathcal{W}_{\text{new}}$

    \EndFor

    \State Replace the walker population: $\mathcal{W} \gets \mathcal{W}_{\text{new}}$

    \State Measure the energy,
    \[
    E^{(t)} = \frac{\sum_{w} s_w E_L(i_w)}{\sum_{w} s_w},
    \qquad
    E_L(i) = \sum_{j} H_{ji}
    \]

\EndFor

\State \textbf{Output:} Ground-state energy estimate $\overline{E}$, obtained by averaging $E^{(t)}$ over the steps after the initial burn-in.

\end{algorithmic}
\end{algorithm}

\section{Sign complexity of Haar-random real states}
\label{app-sign}

In this section, we evaluate the sign complexity $\mathcal{A}$ for Haar-random states with real-valued amplitudes. Writing $\psi_{i}=\langle i|\psi\rangle$, we consider
\begin{equation}
X=\sum_{i=0}^{2^{N}-1}\psi_{i}\,\mathrm{sign}(\psi_{i})=\sum_{i}\psi_{i}|\psi_{i}|\;,
\end{equation}
so that $\mathcal{A}=1-|X|$. For a sign-problem-free state, $X=1$, leading to
$\mathcal{A}=0$. Cancellations between amplitudes of opposite signs, in contrast, drive $\mathcal{A}$ towards unity.

A Haar-random state with real amplitudes can be thought of as a point picked uniformly at random from the unit sphere in $d=2^{N}$ dimensions. The Haar measure over this sphere is invariant under the transformation $\psi\rightarrow-\psi$, under which $X\rightarrow-X$, and therefore
$\overline{X}=0$. We are however interested in obtaining $|X|$. Since the amplitudes of the real random vectors are Gaussian distributed, $\overline{|X|}$ can be obtained from the second moment given by $\overline{X^2}$:
\begin{equation}\label{x2}
X^{2}=\sum_{i}\left(\psi_{i}|\psi_{i}|\right)^{2}
+\sum_{i\neq j}\psi_{i}|\psi_{i}|\,\psi_{j}|\psi_{j}|\;.
\end{equation}
Following Ref.~\cite{collins2006integration}, we obtain
\begin{equation}
\overline{\psi_i\psi_j\psi_k\psi_l}=\frac{\delta_{ij}\delta_{kl}+\delta_{ik}\delta_{jl} +\delta_{il}\delta_{jk}}{2^N(2^N+2)}.
\end{equation}
Setting all indices equal, the average of the first term on the right-hand side of Eq.~(\ref{x2}) gives
\begin{equation}
\overline{\psi_i^4}=\frac{3}{2^N(2^N+2)}.
\end{equation}
In contrast, the second term, when averaged over the random states, vanishes. As a result, $\overline{X^2}=3/(2^N+2)$.

Since $X$ is a sum of $d$ weakly correlated contributions, it is Gaussian
distributed for large $d$. The mean modulus of a centered Gaussian is
$\sqrt{2/\pi}$ times its standard deviation. This gives
\begin{equation}
\overline{\mathcal{A}}=1-\sqrt{\frac{2\,\overline{X^{2}}}{\pi}}
=1-\sqrt{\frac{6}{\pi\left(2^{N}+2\right)}}\;,
\end{equation}
As the above equation shows, for Haar-random states with real amplitudes, the sign complexity approaches unity exponentially with system size.

\end{document}